\documentclass[12pt]{article}

\usepackage{graphicx}
\usepackage{amsmath}
\usepackage{subfig}
\usepackage{epstopdf}
\usepackage{datetime}

\newdate{date}{02}{07}{2012}
\date{\displaydate{date}}

\begin{document}

\title{The theory of Hawking radiation in laboratory analogues}

\author{Scott J Robertson\\[1mm]
Dipartimento di Fisica,\\
Universit\`{a} degli Studi di Pavia, Via Bassi 6, 27100 Pavia, Italy\\[2mm]
email: scott.robertson@unipv.it
}

\maketitle

\begin{abstract}
Hawking radiation, despite being known to theoretical physics for nearly forty years, remains elusive and undetected. It also suffers, in its original context of gravitational black holes, from practical and conceptual difficulties. Of particular note is the trans-Planckian problem, which is concerned with the apparent origin of the radiation in absurdly high frequencies. In order to gain better theoretical understanding and, it is hoped, experimental verification of Hawking radiation, much study is being devoted to laboratory systems which use moving media to model the spacetime geometry of black holes, and which, by analogy, are also thought to emit Hawking radiation. These analogue systems typically exhibit dispersion, which regularizes the wave behaviour at the horizon at the cost of a more complicated theoretical framework. This tutorial serves as an introduction to Hawking radiation and its analogues, developing the moving medium analogy for black holes and demonstrating how dispersion can be incorporated into this generalized framework.
\end{abstract}

\newpage

\section{Introduction\label{sub:Introduction}}

Black holes -- so-called because of the apparent impossibility of escape from them -- are not entirely black. That was the intriguing claim made by Hawking \cite{Hawking-1974,Hawking-1975} nearly forty years
ago. Examining the behaviour of quantum fields in the vicinity of a black hole, he showed that, far from being emission-free, it should emit a steady flux of thermal radiation, with a temperature proportional to $\kappa$, the gravitational field strength at the event horizon:
\begin{equation}
k_{B}T=\frac{\hbar\kappa}{2\pi c}=\frac{\hbar c^{3}}{8\pi GM}\,.\label{eq:black_hole_temperature}
\end{equation}
With this remarkable result, Hawking completed a thermodynamic treatment of black holes that had begun with Bekenstein \cite{Bekenstein-1973,Bekenstein-1974} and has since been continued by many others, e.g.,\cite{Hartle-Hawking-1976,Unruh-1976,Damour-Ruffini-1976, Sanchez-1978,Fredenhagen-Haag-1990,Parikh-Wilczek-2000}. This field brings together the normally disparate areas of gravity, quantum theory and thermodynamics; a glance at the various fundamental constants appearing in Eq. (\ref{eq:black_hole_temperature}) makes this fusion clear. Thirst for understanding of the underlying connections between these mighty realms of physics provides ample motivation for the study of what has come to be known as Hawking radiation.

A pre-requisite for any such study must be the acknowledgement that Hawking radiation is not without its own problems, both practical and conceptual. On the practical side, the predicted temperature -- at least in the gravitational context in which it was first derived -- is virtually untestable. A solar mass black hole would, according to Eq. (\ref{eq:black_hole_temperature}), have a temperature of about $10^{-6}\,\mathrm{K}$ -- six orders of magnitude smaller than the temperature of the cosmic microwave background (CMB). Any radiation from the black hole would be drowned out by the CMB. Therefore, experimental verification of black hole radiation would seem to require an extremely light black hole, orders of magnitude lighter than the Sun.  Such ``micro'' black holes may have formed early in the life of the universe \cite{Carr-Hawking-1974}, or they may be formed today in the high-energy collisions at the Large Hadron Collider \cite{Dimopoulos-Landsberg-2001,Giddings-Thomas-2002}.  Such tiny black holes would, according to Eq. (\ref{eq:black_hole_temperature}), have a very high temperature and quickly evaporate in a burst of radiation; however, if they do exist, they have so far escaped detection.

Conceptually, there is the trans-Planckian problem \cite{Jacobson-1991,Brout-et-al-Primer}, which has to do with the validity of the derivation of Hawking radiation.  Let us briefly explain the problem here; it is discussed mathematically in \S\ref{sub:Event_horizon}.  In most derivations, the spacetime is assumed to collapse, as in a star collapsing to form a black hole \cite{Hawking-1974,Hawking-1975,Brout-et-al-Primer}. Modes of the quantum vacuum are incident from infinity, propagating through the collapsing spacetime and out to infinity again, experiencing a gravitational redshift as they climb out of the ever-deepening gravitational well. The steady thermal flux seen at late times can be traced back to those vacuum modes which \textit{just} manage to escape the event horizon, slowed and redshifted to greater and greater degrees. Thus, any low-frequency mode seen in the late-time thermal spectrum can be traced back to an incident vacuum mode of ever-increasing frequency; indeed, to an \textit{exponentially} increasing frequency! The frequencies of these incident modes very quickly exceed the Planck scale \cite{Brout-et-al-Primer}, widely believed to be a fundamental quantum limit. We cannot justify the use of quantum field theory at such scales -- yet it appears that Hawking radiation is dependent on the existence of these initial trans-Planckian frequencies in order to generate the final low frequencies at which the radiation should be observed \cite{Jacobson-1991}. Whatever is the correct physics, can we be sure that it will preserve Hawking radiation?

These difficulties can be tackled by appealing to artificial event horizons, or physical systems possessing horizons analogous to those of gravitational black holes \cite{ArtificialBlackHoles,Schutzhold-Unruh,LivingReview}. This idea was first proposed by Unruh \cite{Unruh-1981}, who found that perturbations of a stationary background fluid flow behave just as a scalar field in Lorentzian spacetime \cite{Unruh-1981,Visser-1993}. In particular, if the flow velocity crosses the speed of sound, the surface where it does so is entirely analogous to a black-hole event horizon, and on quantizing the perturbation field, one predicts analogue Hawking radiation in such a system. Of course, this model is subject to the same trans-Planckian problem as the gravitational one. In reality, however, the trans-Planckian problem is avoided by the ubiquitous phenomenon of \textit{dispersion} \cite{Jacobson-1991}. That is, the behaviour of waves changes at different scales by mechanisms which are better-understood than quantum gravity. For example, the discreteness of atoms or molecules places a fundamental limit on the wavelength of sound waves. It was later shown that, even after taking high-frequency dispersion into account, Hawking radiation is still predicted, with the same temperature as Hawking's dispersionless model \footnote{Assuming dispersion and the ``surface gravity'' are not too strong, as we shall see in later chapters.} \cite{Unruh-1995,Brout-et-al-1995}. The conclusion is that the trans-Planckian problem is a mathematical artifact, while Hawking radiation exists quite independently of the physics at the high-energy scale. This discovery has prompted a great deal of interest in a range of black-hole analogue systems: in Bose-Einstein condensates (BECs) \cite{Garay-et-al-2000,Garay-et-al-2001,Barcelo-Liberati-Visser-2001-arXiv,Barcelo-Liberati-Visser-2001,Giovanazzi-et-al-2004}, in ultracold fermions \cite{Giovanazzi-2005}, in superfluid helium \cite{Jacobson-Volovik-1998,Volovik-1999,Fischer-Volovik-2001}, in water \cite{Schutzhold-Unruh-2002}, in electromagnetism \cite{Reznik-1997,Schutzhold-Unruh-2005}, and in optics \cite{Leonhardt-Piwnicki-1999,Leonhardt-Piwnicki-2000,Leonhardt-2002}.  (See \S\ref{sub:Overview} for an overview of current research.)  These black-hole analogies might not teach us about quantum gravity directly, though they can demonstrate the ways in which the Hawking spectrum might change in response to new physics at the Planck scale.  Perhaps more importantly, they offer a chance to study Hawking radiation as a general phenomenon related not so much to gravitation as to the restless nature of the quantum vacuum; any insight into the origin of this vacuum radiation \cite{Unruh-1977,Jacobson-1996,Schutzhold-Unruh-2008} is to be welcomed.

This tutorial aims to give an introduction to the theory of Hawking radiation in analogue systems.  Given the wide variety of possible analogue systems, the tutorial aims to be as general as possible.  However, analogue systems will obey different wave equations with a wide variety of dispersion relations.  The framework is here laid mainly within the context of acoustic waves in fluids, in the spirit of Unruh's original insight \cite{Unruh-1981,Unruh-1995}.  This point of view has the advantage of simplicity, in the intuitive nature of the system, its general relativistic form (see \S\ref{sub:Schwarzschild_metric_moving_medium}) and its straightforward generalization to dispersive media (see \S\ref{sub:The-wave-equation}).

The tutorial falls naturally into two distinct parts.  Part \ref{part:Dispersionless_model} is concerned with the analogy between the black hole spacetime, as generally understood in the gravitational context, and the spacetime corresponding to a moving medium.  The latter point of view is more general, leading to the wealth of analogous systems briefly mentioned above.  We then go on to derive Hawking radiation in this generalized spacetime, paying close attention to the important steps and ingredients in the derivation.  In Part \ref{part:Dispersive_model} we generalize the theory further by allowing the medium to be dispersive, and so of greater applicability to experimentally realizable systems.  Hawking radiation is re-derived, but only important differences from the previous derivation are highlighted.  Since there is no exact analytic expression for the Hawking spectrum in the presence of dispersion, we consider numerical and analytical techniques for its calculation.  Finally, we examine the results of these calculations for a simple case.

\newpage

\part{Dispersionless model\label{part:Dispersionless_model}}

\section{The black hole spacetime as a moving medium\label{sub:Schwarzschild_metric_moving_medium}}

\subsection{The Schwarzschild metric}

General Relativity identifies gravity with curvature of spacetime, described by the spacetime metric $ds^{2}$ \cite{Misner-Thorne-Wheeler}. It has a unique spherically symmetric vacuum metric: the Schwarzschild metric, which in coordinates $\left(t_{S},r,\theta,\phi\right)$ takes the form
\begin{equation}
ds^{2}=\left(1-\frac{r_{S}}{r}\right)c^{2}dt_{S}^{2}-\left(1-\frac{r_{S}}{r}\right)^{-1}dr^{2}-r^{2}d\Omega^{2}\,,
\label{eq:Schwarzschild_metric}
\end{equation}
where $d\Omega^{2}=d\theta^{2}+\sin^{2}\theta\, d\phi^{2}$ is the angular line element, $t_{S}$ is the Schwarzschild time coordinate and $r_{S}=2GM/c^{2}$ is the Schwarzschild radius. This describes, for example, the spacetime exterior to a star or planet with relatively low rotation rate. As $r_{S}/r\rightarrow0$, the Schwarzschild metric approaches the flat Minkowski metric, so the coordinates $\left(t_{S},r,\theta,\phi\right)$ correspond to the usual spherical coordinates of flat spacetime for an observer at infinity. However, Eq. (\ref{eq:Schwarzschild_metric}) contains two singularities, at $r=0$ and $r=r_{S}$. Since the Schwarzschild metric is valid only in vacuum, these singularities are relevant only when the entirety of the mass is confined to a radius smaller than $r_{S}$, in which case it will inevitably collapse to a single point of infinite density at $r=0$.  Such objects are called \textit{black holes}.  The point $r=0$ is a genuine singularity of Schwarzschild spacetime \cite{Misner-Thorne-Wheeler}, and we shall not be concerned with it here. It is the surface $r=r_{S}$ -- the \textit{event horizon} -- that is of interest to us.

Let us briefly review the effects of the event horizon by examining light trajectories, or null curves, with $ds^{2}=0$. For simplicity, we shall consider only radial trajectories, so we also set $d\Omega^{2}=0$. This leaves us with a differential equation for the radial null curves:
\begin{equation}
\frac{dt_{S}}{dr}=\pm\frac{1}{c}\left(1-\frac{r_{S}}{r}\right)^{-1}\,.
\label{eq:Schwarzschild_radial_null_curves}
\end{equation}
Far from the Schwarzschild radius, where $r\gg r_{S}$, $\left|dt_{S}/dr\right|\rightarrow1/c$, so that light behaves just as it does in flat spacetime. However, as we approach the Schwarzschild radius, $\left|dt_{S}/dr\right|$ diverges in such a way that light takes longer and longer to travel any distance, and, if travelling towards the event horizon, can never reach it in a finite time $t_{S}$. As far as the Schwarzschild time $t_{S}$ is concerned, the event horizon is infinitely far away.

\subsection{The Painlev\'e-Gullstrand-Lema\^itre metric}

Despite this peculiar behaviour, the event horizon is not a genuine singularity of Schwarzschild spacetime \cite{Misner-Thorne-Wheeler}; it simply appears as such in the coordinates $\left(t_{S},r\right)$, in terms of which the two regions separated by the event horizon are infinitely far apart. This problem can be addressed by following in the footsteps of Painlev\'{e} \cite{Painleve-1921}, Gullstrand \cite{Gullstrand-1922} and Lema\^{i}tre \cite{Lemaitre-1933}, defining a new time
\begin{equation}
t=t_{S}+2\frac{\sqrt{r_{S}r}}{c}+\frac{r_{S}}{c}\ln\left(\frac{\sqrt{r/r_{S}}-1}{\sqrt{r/r_{S}}+1}\right)\,,
\label{eq:Lemaitre_time}
\end{equation}
which, when substituted in Eq. (\ref{eq:Schwarzschild_metric}), yields
the transformed metric
\begin{equation}
ds^{2}=c^{2}dt^{2}-\left(dr+\sqrt{\frac{r_{S}}{r}}c\, dt\right)^{2}-r^{2}d\Omega^{2}\,.
\label{eq:Lemaitre_metric}
\end{equation}
In the coordinates $\left(t,r,\theta,\phi\right)$, the metric clearly
has no singularity at $r=r_{S}$.  At constant $t$, the spatial metric is precisely that
of Minkowski space, and  the distance to the event horizon is always finite.
Notice that, while the definition of $t$, Eq. (\ref{eq:Lemaitre_time}),
is applicable only when $r>r_{S}$, the metric of Eq. (\ref{eq:Lemaitre_metric})
is easily extendable to all values of $r$ greater than zero. The
new coordinate $t$ has opened up a previously inaccessible region
of the spacetime. Keeping $t$ fixed while decreasing $r$, we see
from Eq. (\ref{eq:Lemaitre_time}) that, as we approach the Schwarzschild
radius, we must have $t_{S}\rightarrow\infty$ to compensate for the
divergence of the logarithm. Thus, with respect to our original coordinates,
the transformation to the coordinates $\left(t,r\right)$ is accompanied
by an extension of the spacetime into the infinite future. Our discussion
of light trajectories anticipated this: since an ingoing light ray
approaches the horizon at an infinitely slower rate (with respect
to $t_{S}$), crossing the horizon requires $t_{S}\rightarrow\infty$.

Let us take a moment to interpret the metric (\ref{eq:Lemaitre_metric}).
Again, we shall consider only radial trajectories, setting $d\Omega^{2}=0$.
The key point to note is that, if $dr/dt=-c\sqrt{r_{S}/r}$, the metric
reduces to $ds^{2}=c^{2}dt^{2}$.  Since
this condition clearly maximises $ds^{2}$, these trajectories are
geodesics, and $t$ measures proper time along them. It is as though
space consists of a (Galilean) fluid \cite{Jacobson-1999,Hamilton-Lisle-2008}, flowing inwards with velocity $-c\sqrt{r_{S}/r}$
to converge on the point $r=0$. The geodesics just defined are those
which are stationary with respect to this fluid. They define a locally inertial frame
which we shall call the \textit{co-moving frame}, and in this frame -- i.e., with respect to the fluid -- the speed of
light is $c$.  At the Schwarzschild radius, the fluid flows inward
with speed $c$; anything that falls beneath this radius, no matter its velocity
with respect to the spacetime fluid, will inexorably
be dragged towards the centre at $r=0$. This view is reinforced by
looking at the radial null curves in the coordinates $\left(t,r\right)$.
Setting $ds^{2}=0$ and $d\Omega^{2}=0$, we find two possible trajectories
for light:
\begin{alignat}{1}
\frac{dt}{dr}=\frac{1}{c}\left(1-\sqrt{\frac{r_{S}}{r}}\right)^{-1}\,,\qquad & \mathrm{or}\qquad\frac{dt}{dr}=-\frac{1}{c}\left(1+\sqrt{\frac{r_{S}}{r}}\right)^{-1}\,.
\label{eq:Lemaitre_radial_null_curves}
\end{alignat}
The forms of these trajectories near the point $r=r_{S}$ are shown
in Figure \ref{fig:Lemaitre-Null-Curves}. The second possibility represents
rays propagating \textit{with} the fluid; the total velocity
is $-c-c\sqrt{r_{S}/r}$, the sum of the light's velocity and the
fluid's velocity. It is perfectly regular at the horizon, and tilts
over as it propagates, travelling faster and faster as it moves to
ever smaller radii. The first possibility represents rays propagating \textit{against} the fluid, having a total velocity of $c-c\sqrt{r_{S}/r}$.
This is not regular at the horizon, nor should it be: there, the competing
velocities of the fluid and the light exactly cancel, giving a total
velocity of zero. Rays at higher radii will have a positive total
velocity, and will eventually escape to infinity; rays at smaller
radii will have a negative total velocity, unable to overcome the
fluid flow, and will propagate inwards to $r=0$.

\subsection{A general metric for moving media}

This analogy with a moving medium forms the basis of artificial black
holes and event horizon analogies. We may simply replace $-c\sqrt{r_{S}/r}$
in Eq. (\ref{eq:Lemaitre_metric}) with the more general velocity
profile $V\left(x\right)$ to obtain, in $1+1$-dimensional spacetime,
the metric\begin{equation}
ds^{2}=c^{2}dt^{2}-\left(dx-V\left(x\right)dt\right)^{2}\,,\label{eq:generalized_Lemaitre_metric}\end{equation}
where now $c$ is to be interpreted as the velocity with respect to
the medium in question (not necessarily the speed of light).
The co-moving frame, then, is the frame in which this medium is at rest, and in which the wave speed is exactly $c$.
By contrast, the frame with coordinates $x$ and $t$ shall henceforth be called the \textit{lab frame}.
We shall always assume the medium to be left-moving in the lab frame, so that $V<0$.

\begin{figure}
\includegraphics[width=0.8\columnwidth]{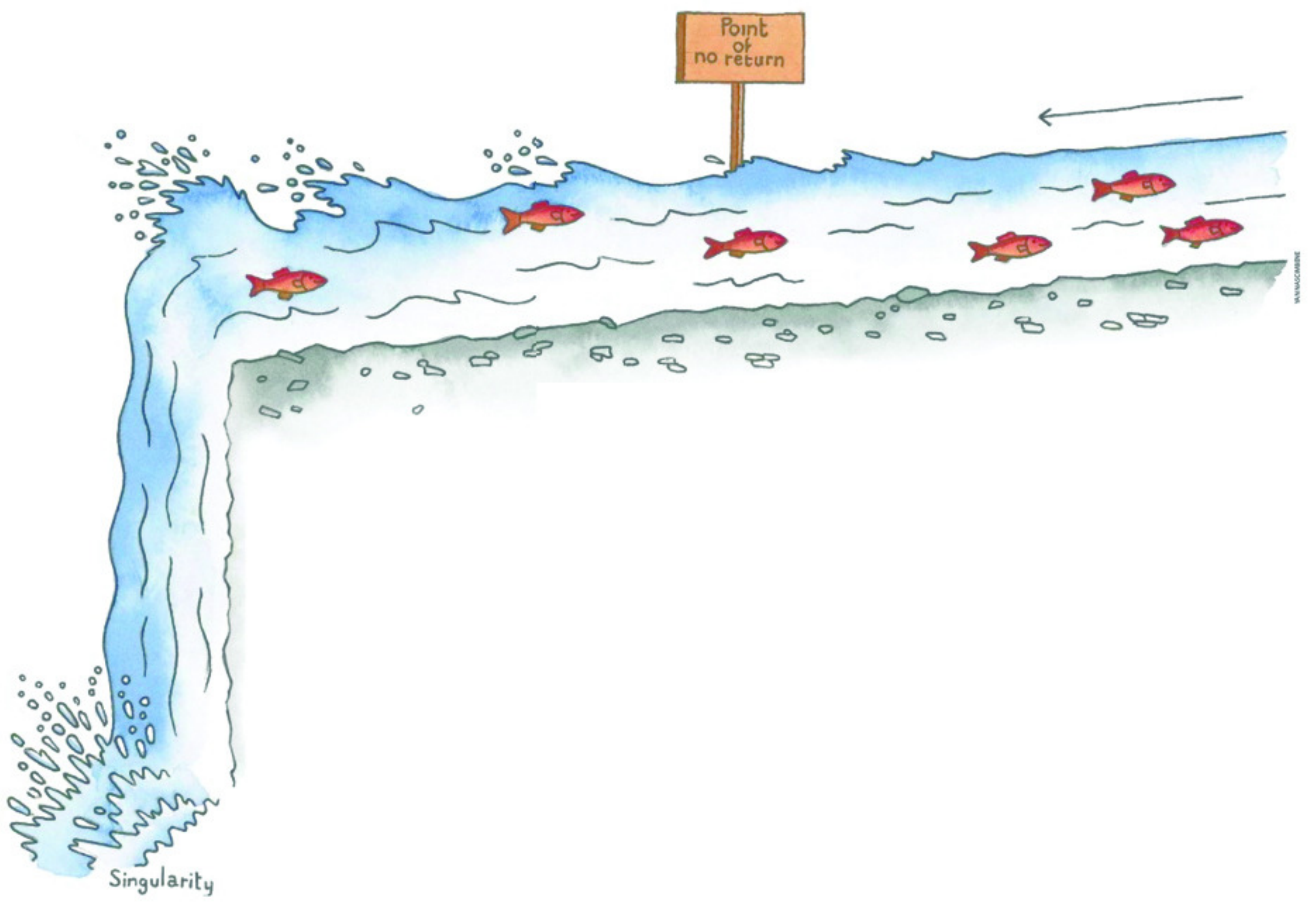}
\caption[\textsc{Black hole horizon in a river}]{\textsc{Black hole horizon in a river}: At the {}``Point of no return'',
the flow speed of the water is exactly equal to the maximum speed
of the fish. Any fish that travel further downstream than this point
will be dragged inexorably towards the waterfall. [Illustration by Yan Nascimbene, from ``Black Holes and the Information Paradox'', L. Susskind, \textit{Scientific American}, April 1997]\label{fig:Fishy-Black-Hole}}
\end{figure}

From now on, we shall consider the general metric (\ref{eq:generalized_Lemaitre_metric}),
which need not be gravitational in origin.  It may, for example, be applied to a system so far removed
from astrophysics as a river flowing towards a waterfall \cite{Hamilton-Lisle-2008}, so that
the flow speed increases in the direction of flow, as illustrated
in Figure \ref{fig:Fishy-Black-Hole}. Imagine this river is populated
by fish who can swim only up to a maximum speed $c$ with respect to the water. Then
the above metric suffices to describe the trajectories of fish in
this river. Fish who are far from the waterfall, where the current
is low, are free to swim around as they please, experiencing no significant
resistance in either direction. However, as the current increases,
there may be a point at which $\left|V\right|=c$. As the fish approach
this point, they will find it increasingly difficult to swim back
upstream; passing this point, motion upstream is impossible, for the
current is so strong that the fish, no matter how hard or in which
direction they swim with respect to the water, are doomed to be swept
over the waterfall. The point where $\left|V\right|=c$ is the event
horizon, and the trajectories of fish swimming at exactly the speed
$c$ are analogous to the trajectories of light near a black hole
horizon \footnote{Slightly more realistically, it is the trajectories of waves in the water -- sound waves \cite{Unruh-1981,Visser-1993,Jacobson-1999} or even surface waves \cite{Rousseaux-et-al-2008,Rousseaux-et-al-2010,Weinfurtner-et-al-2011} -- which respect the black hole analogy.  It is most common in the literature to make reference to sound waves, and for this reason the flow velocity is usually classified as \textit{subsonic} ($|V|<c$) or \textit{supersonic} ($|V|>c$).  These terms are adopted in this tutorial.
\label{foot:sonic}}.

\subsection{White holes\label{sub:White_holes}}

\begin{figure}
\includegraphics[width=0.8\columnwidth]{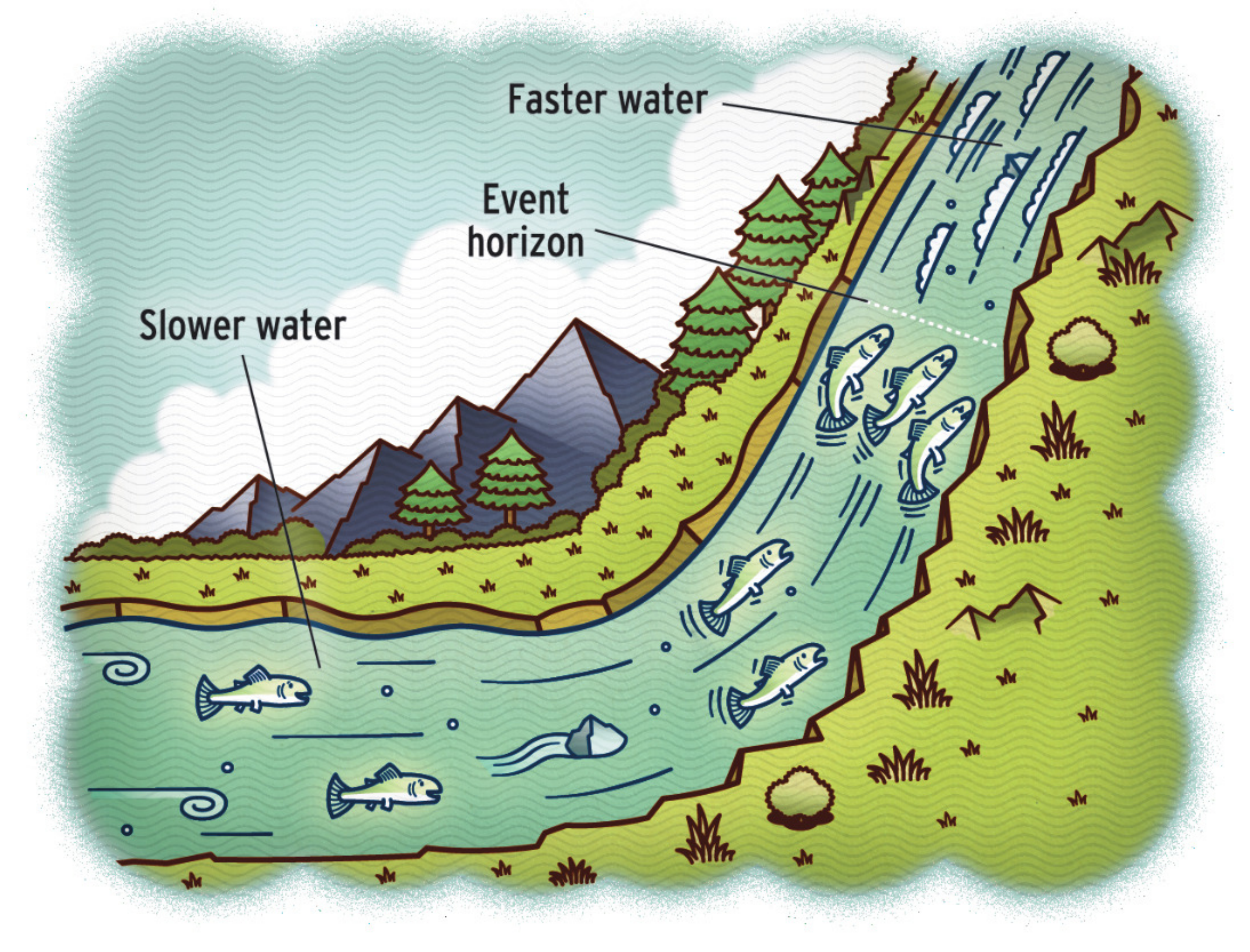}
\caption[\textsc{White hole horizon in a river}]{\textsc{White hole horizon in a river}: As in Figure \ref{fig:Fishy-Black-Hole},
the event horizon is the point at which the flow speed is exactly
equal to the maximum speed of the fish. Here, however, the flow speed
\textit{decreases} in the direction of flow, so that the fish may
swim arbitrarily close to the event horizon, but may never cross it. [Illustration by Peter Hoey, from ``Test of Hawking's Prediction on the Horizon with Mock 'White Hole' '', A. Cho, \textit{Science} \textbf{319}, 1321 (2008)]\label{fig:Fishy-White-Hole}}
\end{figure}

Now consider a slightly different scenario, portrayed in Figure \ref{fig:Fishy-White-Hole}.
Here, water is flowing \textit{from} a waterfall, so that its flow is initially very fast and slows
as it travels.  There is also a point here at which $\left|V\right|=c$, but it behaves in a qualitatively
different way from that in Figure \ref{fig:Fishy-Black-Hole}.  Fish far from the waterfall, where the
current is low, may come and go as they please; but as they travel
upstream, they will find it increasingly difficult to continue, and
must come to a complete standstill exactly at the horizon.  Whereas before the fish could
not \textit{escape} from the region beyond the horizon, now they find they cannot \textit{enter} it.
This is a white hole spacetime.  In more precise terms,
the supersonic flow (see footnote \ref{foot:sonic}) leads \textit{away} from the horizon in a black hole spacetime,
but \textit{towards} it in a white hole spacetime.  Equivalently, we may think of the white hole
as the time-reversed black hole: in the spacetime metric (\ref{eq:generalized_Lemaitre_metric}), we
substitute $V\rightarrow-V$ to transform between them.  This one-to-one correspondence between black and white holes persists in the presence of dispersion (see, e.g., Appendix D of \cite{Macher-Parentani-2009-ii}).

The white hole is of limited use in astrophysics. It \textit{is} encoded
in Schwarzschild spacetime, from which it is derived by taking the negative square roots in Eqs. (\ref{eq:Lemaitre_time})-(\ref{eq:Lemaitre_metric}),
but it extends the spacetime into the infinite past. Therefore, its
validity requires that the Schwarzschild metric is valid for $t_{S}\rightarrow-\infty$,
which is seldom the case, as black holes are believed to form from
gravitational collapse. However, this is specific to the Schwarzschild case, and there
is no fundamental restriction on the existence of white holes in the
more general case of moving media.  Indeed, many experimental setups achieve sub- or supersonic flow only over a limited region, so that a black hole-white hole pair is formed.  White holes have thus found relevance in the field of analogue Hawking radiation.

One point of particular interest and controversy is the stability or otherwise of spacetimes containing white holes.  While black holes are always found to be stable \cite{Leonhardt-et-al-2003,Barcelo-et-al-2006}, white holes were found in \cite{Leonhardt-et-al-2003} to be intrinsically unstable, and, conversely, to be stable in \cite{Mayoral-et-al-2011}.  A systematic numerical study in \cite{Barcelo-et-al-2006} found that the stability or otherwise of white holes depends crucially on the boundary conditions imposed, though the nature of the physically appropriate boundary conditions remains for now a moot point (see also \S 5.2.4 of \cite{LivingReview} for a concise description of this problem and the issues involved).  Another spacetime showing signs of instability is the black hole-white hole pair.  Periodic geometries including such a pair are found to induce narrow instability ``fingers'' in the parameter space \cite{Garay-et-al-2000,Garay-et-al-2001,Jain-et-al-2007}.  Better understood, however, is the dynamical instability of the black hole-white hole pair in a non-periodic configuration, in which the inner region acts as a resonant cavity: the so-called \textit{black hole laser} \cite{Corley-Jacobson-1999,Leonhardt-Philbin-2008,Coutant-Parentani-2010,Finazzi-Parentani-2010}.

For simplicity, in the following we shall restrict our attention to black hole configurations, but white holes can be treated in an analogous fashion (see \cite{Macher-Parentani-2009} for an analysis and comparison of black holes and white holes).

\subsection{Dispersion\label{sub:Dispersion}}

There is one caveat to the generalized metric (\ref{eq:generalized_Lemaitre_metric}) for moving media, but it is a serious one.  Assuming for simplicity a massless field, then $c$, a constant parameter, is the speed of waves with respect to the medium \textit{no matter their frequency or wavevector}.  The medium thus described is \textit{dispersionless}: the wave speed is absolutely fixed.  In some ways, this is a good thing: the wave equation is simpler; the general solution is more tractable; the positions of any horizons are defined absolutely.  But, as we shall soon see, it is precisely the absence of dispersion\footnote{More precisely, it is absence of \textit{high-frequency} dispersion that is problematic.  Thus massive fields, though they are dispersive at low frequencies and would allow us to retain the equivalence between medium and Lorentzian spacetime, do not resolve the trans-Planckian issue.  In any case, they would restrict us to media with a very specific dispersion, whereas realistic media yield complicated dispersion relations and inevitably break the medium-spacetime equivalence.} that leads to the trans-Planckian problem\footnote{Although dispersive profiles are also problematic if the asymptotic velocity vanishes; see ``Conceptual issues'' of \S\ref{sub:Overview} and \S VI of \cite{Corley-Jacobson-1996}.}.  So, while the generalized metric (\ref{eq:generalized_Lemaitre_metric}) may free us from the specific case of a gravitational black hole, it does not rid us of this fundamental conceptual issue.  In the case of gravity, quantum theory is thought likely to modify things near the Planck scale, introducing high-frequency dispersive effects into General Relativity; but a quantum theory of gravity is as yet unknown.  In real media that may be used in experiments, however, dispersive behaviour is generally well-understood, so that it may be possible to incorporate it into the theory of Hawking radiation.  We examine how this can be done in Part \ref{part:Dispersive_model}.

\section{The wave equation and its solutions\label{sub:Acoustic_field_wave_equation}}

\subsection{Deriving the wave equation}

Having now generalized the metric of a Schwarzschild black hole to that of a moving medium,
let us study the behaviour of fields in a ($1+1$)-dimensional spacetime described by the metric (\ref{eq:generalized_Lemaitre_metric}).
For simplicity, we shall assume a massless scalar field; in the context
of a real fluid, for example, such a field arises from small perturbations in the
background flow (see \cite{Unruh-1981} for a detailed derivation).
From the outset, we shall treat the field as complex, for even though we later (in \S \ref{sub:k-representation}) constrain it to be real, its decomposition into complex-valued modes is fundamental to the quantization of the field and, ultimately, to the Hawking process itself.

Let us begin with the Principle of Least Action: the field
$\phi\left(t,x\right)$ varies from one configuration to another in
such a way that the action is an extremum (usually a minimum).
The action is the integral
\begin{equation}
S=\int\int dx\, dt\, L\left(\phi,\phi^{\star},\partial_{t}\phi,\partial_{t}\phi^{\star},\partial_{x}\phi,\partial_{x}\phi^{\star}\right),
\label{eq:action}
\end{equation}
so the physics of the model is completely contained in the Lagrangian density, $L$.
An extremum of the action is found by infinitesimally varying the
fields $\phi$ and $\phi^{\star}$ and their derivatives,
then setting the resulting variation in $S$ to zero. This yields
the Euler-Lagrange equation,
\begin{equation}
\frac{\partial L}{\partial\phi^{\star}}-\frac{\partial}{\partial t}\left(\frac{\partial L}{\partial\left(\partial_{t}\phi^{\star}\right)}\right)-\frac{\partial}{\partial x}\left(\frac{\partial L}{\partial\left(\partial_{x}\phi^{\star}\right)}\right)=0\,.
\label{eq:Euler-Lagrange}
\end{equation}
The Lagrangian density for a massless scalar field is \cite{Birrell-Davies}
\begin{equation}
L=\frac{1}{2}\sqrt{-g}g^{\mu\nu}\partial_{\mu}\phi^{\star}\partial_{\nu}\phi\,,
\label{eq:scalar_massless_Lagrangian}
\end{equation}
which is simply the covariant form of the corresponding Lagrangian density in flat space; $g_{\mu\nu}$ is the metric tensor, $g$ is its determinant, and
$g^{\mu\nu}$, with raised indices, is its inverse.
From the metric (\ref{eq:generalized_Lemaitre_metric}),
we find that the Lagrangian density is
\begin{equation}
L=\frac{1}{2c}\left(\left|\left(\partial_{t}+V\partial_{x}\right)\phi\right|^{2}-c^{2}\left|\partial_{x}\phi\right|^{2}\right)\,,
\label{eq:scalar_massless_Lagrangian_moving_fluid}
\end{equation}
and plugging this into the Euler-Lagrange equation (\ref{eq:Euler-Lagrange}) yields the wave
equation
\begin{equation}
\left(\partial_{t}+\partial_{x}V\right)\left(\partial_{t}+V\partial_{x}\right)\phi-c^{2}\partial_{x}^{2}\phi=0\,.
\label{eq:dispersionless_acoustic_wave_equation}
\end{equation}
The partial derivatives act on everything to their right, including factors of $V$.

\subsection{General solution: the $u$- and $v$-branches\label{sub:General_solution}}

A general solution to the wave equation (\ref{eq:dispersionless_acoustic_wave_equation})
is easily found. We define new variables $u$ and $v$ as follows:\begin{alignat}{1}
u=t-\int^{x}\frac{dx^{\prime}}{c+V\left(x^{\prime}\right)}\,,\qquad & v=t+\int^{x}\frac{dx^{\prime}}{c-V\left(x^{\prime}\right)}\,.\label{eq:defining_U_and_V}\end{alignat}
On substitution in the metric (\ref{eq:generalized_Lemaitre_metric}), we find that the metric in the coordinates $\left(u,v\right)$ takes the form
\begin{equation}
ds^{2} = \left(c^{2}-V^{2}(x)\right)\,du\,dv\,,
\label{eq:U_V_metric}
\end{equation}
which in turn, via the Lagrangian (\ref{eq:scalar_massless_Lagrangian}) and the Euler-Lagrange equation (\ref{eq:Euler-Lagrange}), leads to the wave equation
\begin{equation}
\partial_{u}\partial_{v}\phi=0\,.
\label{eq:U_V_wave_equation}
\end{equation}
So, in the absence of horizons, $\phi$
is simply a sum of two arbitrary functions, one a function of $u$
only, the other a function of $v$ only:\begin{eqnarray}
\phi & = & \phi_{u}\left(u\right)+\phi_{v}\left(v\right)\nonumber \\
 & = & \phi_{u}\left(t-\int^{x}\frac{dx^{\prime}}{c+V\left(x^{\prime}\right)}\right)+\phi_{v}\left(t+\int^{x}\frac{dx^{\prime}}{c-V\left(x^{\prime}\right)}\right)\,.\label{eq:general_solution}\end{eqnarray}
In the co-moving frame, $\phi_{u}$ is right-moving (counter-propagating -- recall we take $V<0$) while $\phi_{v}$
is left-moving (co-propagating). The fact that the two functional forms maintain their
shapes is a consequence of the absence of dispersion: all wave components
have the same velocity, $c$, with respect to the fluid, and so does
the waveform as a whole. The only ambiguity is in the direction of
travel; thus the solution splits into a right-moving and a left-moving
part.

Note that the metric (\ref{eq:U_V_metric}) is related to the metric of a stationary medium (in which $V$ is identically zero) simply by multiplication by a coordinate-dependent prefactor.  This is a \textit{conformal transformation}, and its usefulness stems from the invariance of the Lagrangian (\ref{eq:scalar_massless_Lagrangian}) -- and consequently of the wave equation (\ref{eq:U_V_wave_equation}) -- under such a transformation.  This observation corroborates a well-known theorem that all curved two-dimensional spaces are \textit{conformally flat} (i.e., related to flat space via a conformal transformation); see, e.g., \cite{Leonhardt-Philbin-Invisibility}.

\subsection{Event horizon\label{sub:Event_horizon}}

In the vicinity of a horizon where $V=-c$, inspection of Eqs. (\ref{eq:defining_U_and_V}) shows that $u$ diverges as the horizon is approached.  So the range $u \in \left(-\infty,\infty\right)$ applies only to one side of the horizon, and the coordinate pair $(u,v)$ does not cover the entire spacetime.  To include the region on the other side of the horizon requires the introduction of an additional $u$-coordinate, and, consequently, an extra function in the general solution (\ref{eq:general_solution}).

Let us suppose that the horizon is situated at the origin, $x=0$;
and, moreover, that the derivative of $V$ is non-zero there. Then
we may approximate the flow velocity to first-order in $x$ as follows:\begin{equation}
V\left(x\right)\approx-c+\alpha x\,.\label{eq:linearized_flow_velocity}\end{equation}
where $\alpha>0$ for a black hole horizon.  (A white hole horizon would have $\alpha<0$.)

\begin{figure}
\includegraphics[width=0.8\columnwidth]{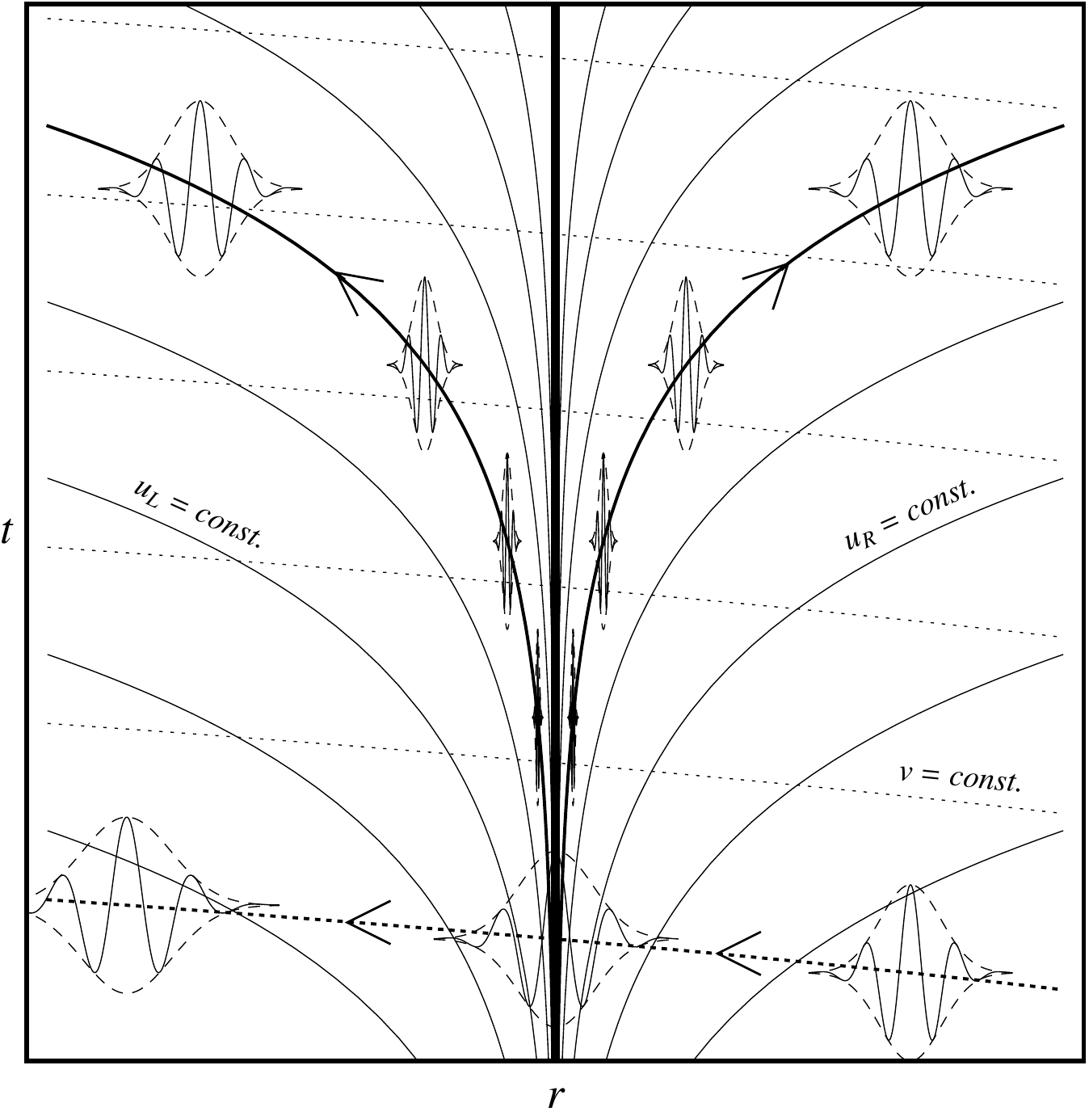}
\caption[\textsc{Radial wave trajectories near a black hole horizon}]{\textsc{Radial wave trajectories near a black hole horizon}: Space-time
diagram of light trajectories in the metric of Eq. (\ref{eq:Lemaitre_metric}), where the thick centre line represents the event horizon $r=r_{S}$ and, in this near-horizon region, the flow velocity profile is approximately linear: $V(r)\approx -c\,+\,\alpha (r-r_{S})$.
The dotted lines show trajectories of co-propagating waves (where the coordinate $v$, defined in the second of Eqs. (\ref{eq:defining_U_and_V}), is constant), and these obey the second of Eqs. (\ref{eq:Lemaitre_radial_null_curves});
these experience nothing unusual at the horizon, so that a wavepacket on the $v$-branch can pass through the horizon completely unhindered.  In contrast, the solid lines show
the trajectories of counter-propagating waves (where the coordinate $u$, from the first of Eqs. (\ref{eq:defining_U_and_V}), is constant), and these obey the first of Eqs. (\ref{eq:Lemaitre_radial_null_curves});
as these are propagated further back in time, they come to a standstill at the horizon.  The $u$-branch is thus split into two pieces: $u_{R}$ and $u_{L}$ (defined by Eq. (\ref{eq:U_horizon})), its restrictions to the right- and left-hand regions, respectively.  Both are right-moving with respect to the background fluid flow, but only the $u_{R}$-modes are able to propagate to the right in the lab frame, whereas the $u_{L}$-modes are dragged to the left.  Since the $u_{L}$ and $u_{R}$ geodesics bunch together at the horizon, any wavepacket traced back in time will become arbitrarily thin, with arbitrarily short wavelength; this is the trans-Planckian problem.\label{fig:Lemaitre-Null-Curves}}
\end{figure}

Equation (\ref{eq:general_solution}) states that counter-propagating waves are described by the arbitrary function $\phi_{u}\left(u\right)$, where $u$ is given by
\begin{equation}
u\,=\,t-\int^{x}\frac{dx^{\prime}}{c+V\left(x^{\prime}\right)} \,\approx\, t-\frac{1}{\alpha}\log\left(\frac{\alpha}{c}\left|x\right|\right)\,,
\label{eq:U_horizon}
\end{equation}
the second equality holding in the vicinity of the horizon. It is evident from Eq. (\ref{eq:U_horizon}) that a horizon is located at $x=0$, for the space is divided into two separate regions, with $x=0$ marking the boundary between them. To see this, imagine first that we have a wavepacket centred at a certain value of $u$, and that this is located, at a certain time, at a positive value of $x$; such wavepackets are illustrated in Figure \ref{fig:Lemaitre-Null-Curves}.  If $t$ increases, then $\log\left(\alpha\left|x\right|/c\right)/\alpha$ must increase by exactly the same amount, and so $\left|x\right|$ increases and the wavepacket moves to the right. Similarly, if we trace the wavepacket back in time by decreasing $t$, $\log\left(\alpha\left|x\right|/c\right)/\alpha$ must also decrease by the same amount, and $x$ decreases; the wavepacket has come from the left. But the logarithm diverges to $-\infty$ at the origin; this means that, no matter how far back in time we look, $\log\left(\alpha\left|x\right|/c\right)/\alpha$ can be decreased by a corresponding amount \textit{without $x$ ever becoming negative}.  The wavepacket must have originated arbitrarily close to the event horizon, moving very slowly forwards at first, picking up speed the further it travels.  The longer the horizon has existed, the closer to it the wavepacket must have been at the moment of formation; furthermore, the \textit{thinner} it must have been, since the lines of constant $u$ bunch together ever closer as they approach the horizon.  As we trace it back into the asymptotic past, a given wavepacket at a given position must have arisen from a wavepacket arbitrarily close to the horizon, of arbitrary thinness and composed of arbitrarily short wavelengths.  This is precisely the trans-Planckian problem.

Exactly the same analysis holds in the region to the left of the horizon, as also shown in Fig. \ref{fig:Lemaitre-Null-Curves}$\left(b\right)$.

We shall always take $V$ to be constant in time, so that, mathematically at least, any horizons must have existed since the infinite past (though this is a subtle issue, and we shall return to it in \S\S\ref{sub:In_and_out_modes} and \ref{sub:Conceptual_issues}).  In this case, \textit{all} wavepackets, when traced backwards in time, are found to have originated arbitrarily close to the horizon and can never have been in the region on the opposite side.  Since a complete description of the field must apply throughout the entire space, we conclude that there must be two distinct $u$-coordinates, each localized to one side of the event horizon.  Let us denote them $u_{R}$ and $u_{L}$.  In the definition (\ref{eq:defining_U_and_V}), these are distinguished by the lower bounds of the integral over $x$, that for $u_{R}$ being in the right-hand region and that for $u_{L}$ in the left-hand region.  Then the coordinates $u_{R}$ and $u_{L}$ both range from $-\infty$ to $+\infty$, but cover only their respective halves of the spacetime, as seen in Fig. \ref{fig:Lemaitre-Null-Curves}.  We thus see that the existence of a horizon -- or, equivalently here, of one subsonic and one supersonic region -- results in three independent solutions of the wave equation ($u_{R}$, $u_{L}$ and $v$) rather than the usual two ($u$ and $v$).  $u_{R}$ and $u_{L}$ are both of $u$-character since both describe waves travelling to the right \textit{in the co-moving frame of the fluid}.  The difference between them is in the lab frame: $u_{L}$ applying in the supersonic region where the fluid flow is faster than the wave speed, the $u_{L}$-waves are seen to be dragged to the left in the lab frame; whereas, since $u_{R}$ applies in the subsonic region where the fluid flow is less than the wave speed, the $u_{R}$-waves still manage to move to the right in the lab frame, albeit at a reduced speed.

The requirement of two $u$-coordinates in addition to a single $v$-coordinate to completely describe the spacetime answers an immediate objection that might be raised against the possibility of Hawking radiation in $1+1$-dimensional spacetime: namely, the conformal flatness of all two-dimensional spacetimes, mentioned previously in \S\ref{sub:General_solution}.  At first sight, this seems to suggest that all two-dimensional spacetimes are conformally \textit{equivalent}, thus precluding the occurrence of Hawking radiation in some of them (those with horizons) given its absence in others (without horizons).  But conformal flatness does not imply conformal equivalence if there exist non-equivalent spacetimes which are flat nonetheless.  This is precisely the case here: whereas the horizonless spacetime is conformally equivalent to the two-dimensional plane, the existence of the horizon splits the spacetime into two regions, \textit{each} of which is conformally equivalent to the two-dimensional plane, these planes being ``stitched together'' at the horizon \cite{Leonhardt-Philbin-Invisibility}.  Therefore, the flat equivalents of spacetimes with and without horizons are topologically distinct; they are \textit{not} conformally equivalent to each other, and the absence of Hawking radiation in one does not imply its absence in the other.

\section{Field modes and quantization\label{sec:Field_modes_and_quantization}}

\subsection{Stationary modes and the dispersion relation\label{sub:Stationary_modes}}

The Lagrangian density (\ref{eq:scalar_massless_Lagrangian_moving_fluid}) is invariant under time translation; that is, performing a small shift in the time coordinate, $t\rightarrow t+\Delta t$, leaves $L$ unaffected.  This implies \cite{Noether-1918} the existence of \textit{stationary modes} of the form\begin{equation}
\phi\left(t,x\right)=e^{-i\omega t}\phi_{\omega}\left(x\right)\,.\label{eq:stationary_solution}\end{equation}
On quantization, these correspond to energy eigenstates or quasiparticles \cite{Birrell-Davies}.  Equation (\ref{eq:general_solution}) indicates that, in the absence of an event horizon, the spatial part of
the solution takes the form\begin{eqnarray}
\phi_{\omega}\left(x\right) & = & C_{u}\,\exp\left(i\omega\int^{x}\frac{dx^{\prime}}{c+V\left(x^{\prime}\right)}\right)+C_{v}\,\exp\left(-i\omega\int^{x}\frac{dx^{\prime}}{c-V\left(x^{\prime}\right)}\right)\nonumber \\
 & = & C_{u}\,\phi_{\omega}^{u}\left(x\right)+C_{v}\,\phi_{\omega}^{v}\left(x\right)\,,\label{eq:stationary_mode_U_V}\end{eqnarray}
 whereas, accounting for the considerations of \S\ref{sub:Event_horizon}, in the presence of an event horizon (at $x=x_{H}$) we have instead
 \begin{eqnarray}
 \phi_{\omega}\left(x\right) & = & C_{u_{R}}\,\exp\left(i\omega\int^{x}_{x_{R}} \frac{dx^{\prime}}{c+V\left(x^{\prime}\right)}\right)\,\theta\left(x-x_{H}\right)+C_{u_{L}}\,\exp\left(i\omega\int^{x}_{x_{L}} \frac{dx^{\prime}}{c+V\left(x^{\prime}\right)}\right)\,\theta\left(x_{H}-x\right) \nonumber \\
 & & \qquad \qquad \qquad \qquad \qquad \qquad \qquad +C_{v}\,\exp\left(-i\omega\int^{x}\frac{dx^{\prime}}{c-V\left(x^{\prime}\right)}\right)\nonumber \\
 & = & C_{u_{R}}\,\phi_{\omega}^{u_{R}}\left(x\right)\,\theta\left(x-x_{H}\right)+C_{u_{L}}\,\phi_{\omega}^{u_{L}}\left(x\right)\,\theta\left(x_{H}-x\right)\nonumber \\
 & & \qquad \qquad \qquad \qquad \qquad \qquad \qquad +C_{v}\,\phi_{\omega}^{v}\left(x\right)\,,\label{eq:stationary_mode_U_V_horizon}\end{eqnarray}
 where Heaviside step functions ($\theta(x)=0$ for $x<0$ and $1$ for $x>0$) have been included to clarify the restricted domains of $\phi_{\omega}^{u_{R}}$ and $\phi_{\omega}^{u_{L}}$.
Like the general solution (\ref{eq:general_solution}), the general stationary mode for a given frequency splits into two independent parts, one counter-propagating and one co-propagating
\footnote{Being stationary modes, the wave envelope does not move at all.  A way of thinking about the wave velocity is to consider a wavepacket, strongly peaked at the given values of $\omega$ and $k$.  Then the wavepackets corresponding to the $u$- and $v$-modes will travel in opposite directions with respect to the fluid.  Wavepackets are included in Fig. \ref{fig:Lemaitre-Null-Curves}.}, which we henceforth refer to as $u$- and $v$-modes, respectively; and, in the presence of a horizon, the $u$-branch splits further into modes localised to either the subsonic or supersonic region.

Defining the local wavevector
$k\left(x\right)$ such that $\phi_{\omega}^{u/v}\left(x\right)=\exp\left(i\int^{x}dx^{\prime}\, k_{\omega}^{u/v}\left(x^{\prime}\right)\right)$,
we see that\begin{alignat}{1}
k_{\omega}^{u}\left(x\right)=\frac{\omega}{c+V\left(x\right)}\,,\qquad & k_{\omega}^{v}\left(x\right)=-\frac{\omega}{c-V\left(x\right)}\,,\label{eq:U_V_wavevectors}\end{alignat}
relations summarized by the single equation\begin{equation}
\left(\omega-Vk\right)^{2}=c^{2}k^{2}\,.\label{eq:Doppler_formula}\end{equation}
This formula is very intuitive once we recognise that the lab frequency $\omega$ is
related to the co-moving frequency $\omega_{\mathrm{cm}}$ via the Doppler formula: $\omega=\omega_{\mathrm{cm}}+Vk$
\footnote{The ``spacetime fluid'' itself is Galilean with respect to the lab frame -- the spatial coordinate is transformed according to $dx^{\prime}=dx-Vdt$ in Eq. (\ref{eq:generalized_Lemaitre_metric}) -- which is why the Doppler formula appears in Galiliean form.  This is in contrast to relativistic motion occurring \textit{within} the fluid \cite{Hamilton-Lisle-2008}.}.
Equation
(\ref{eq:Doppler_formula}), then, is simply $\omega_{\mathrm{cm}}^{2}=c^{2}k^{2}$
-- the dispersion relation in the co-moving frame. Comparing with Eqs.
(\ref{eq:U_V_wavevectors}), we see that\begin{equation}
\omega_{\mathrm{cm}}=\omega-Vk=\begin{cases}
ck & \mathrm{for}\: u\mathrm{-modes}\\
-ck & \mathrm{for}\: v\mathrm{\mathrm{-modes}}\end{cases}\,.\label{eq:free-fall-freq_u-and-v}\end{equation}
The function $\pm ck$ will be referred to as the dispersion profile or curve.  In this dispersionless case it is simply proportional to $k$, but general dispersion profiles are more complicated, and will be considered in Part \ref{part:Dispersive_model}.

\begin{figure}
\includegraphics[width=0.8\columnwidth]{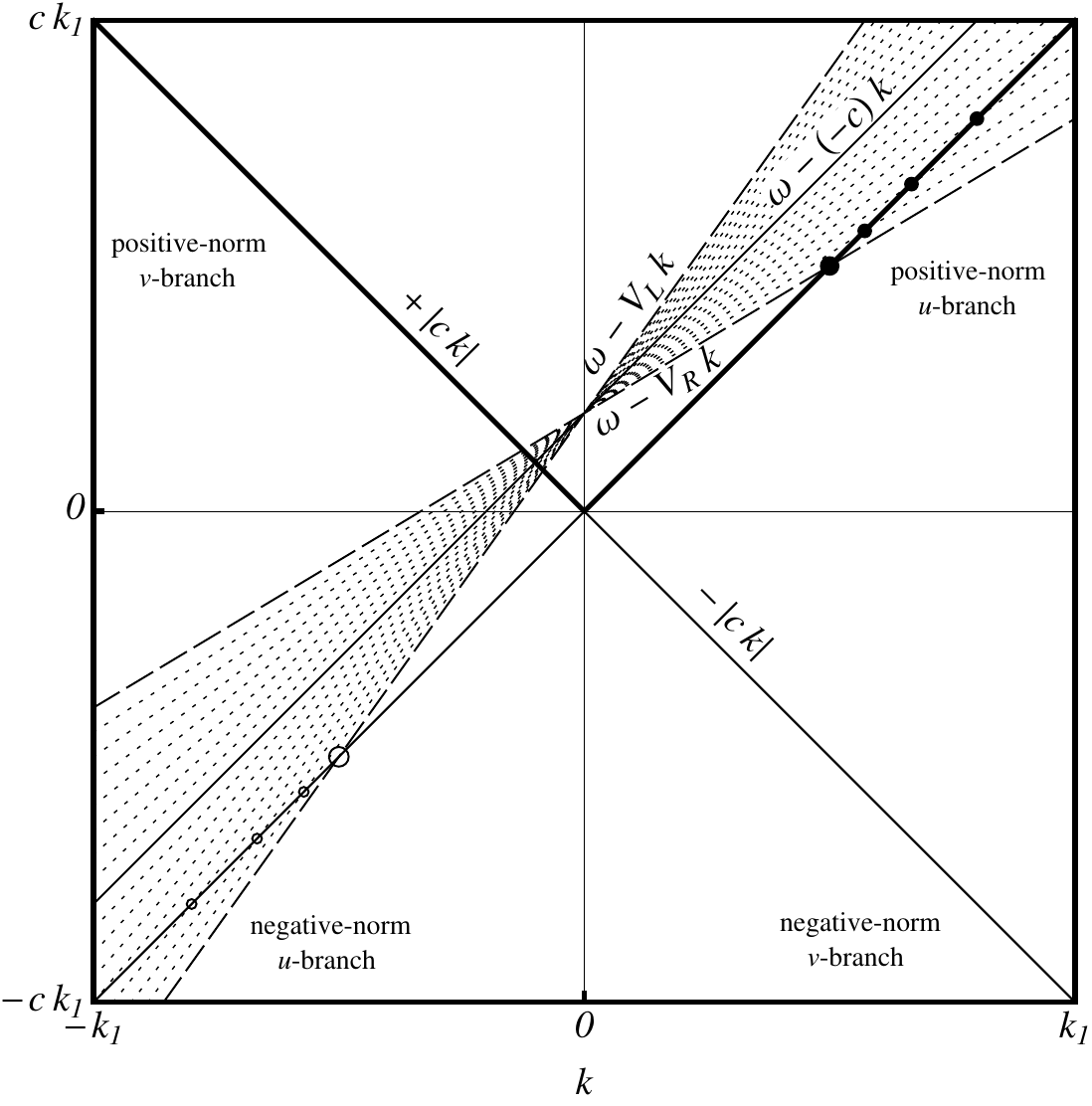}
\caption[\textsc{Dispersion profile in the co-moving frame}]{\textsc{Dispersion profile in the co-moving frame}:
In the rest frame of the medium, the co-moving frequency obeys the simple relation $\omega_{\mathrm{cm}}^{2}=c^{2}k^{2}$.  This splits into four sectors: the $u$- and $v$-branches are determined by the velocity in the co-moving frame, $\omega_{\mathrm{cm}}/k=\pm c$ (see Eq. (\ref{eq:free-fall-freq_u-and-v})), and correspond to the diagonal lines $ck$ and $-ck$, respectively; the positive- and negative-norm branches are determined by the sign of the co-moving frequency itself (see Eq. (\ref{eq:free-fall-freq_pos-and-neg})), and therefore correspond to $|ck|$ (the thicker line on the upper-half of the plot) and $-|ck|$ (on the lower-half of the plot), respectively.  The co-moving frequency is related to the conserved lab-frame frequency via the Doppler formula $\omega_{\mathrm{cm}}=\omega-Vk$, plotted here as dashed lines for the asymptotic velocities $V_{R}$ and $V_{L}$ and as dotted lines for the intervening velocities.  The possible wavevector solutions occur at the intersections of the two curves, so that their evolution with changing velocity (i.e. with changing slope of the line $\omega-Vk$) can be traced.  Critical behaviour occurs as the velocity passes through $-c$, at which point the line $\omega-Vk$ is parallel to the $u$-branch dispersion curve: it crosses the $u$-branch on the positive-norm side when $V>-c$, but on the negative-norm side when $V<-c$.  As $V\rightarrow -c$, the wavevector on either the positive- or negative-norm side diverges -- this is the trans-Planckian problem.
\label{fig:Dispersion_comoving-frame}}
\end{figure}

Equation (\ref{eq:free-fall-freq_u-and-v}) -- which states that the co-moving frequency must both satisfy the dispersion relation in the co-moving frame and be related to the lab frequency via the Doppler formula -- can be solved graphically from one of two possible viewpoints:
\begin{itemize}
\item{\textsc{Co-moving frame}: The dispersion profile in the rest frame of the fluid is plotted alongside the co-moving frequency, which is given by the straight line $\omega - V k$.  The points of intersection occur at the possible wavevector solutions.  In this picture, variation of the velocity $V$ does not alter the dispersion profile, but the slope of the Doppler curve, whose $y$-intercept is equal to $\omega$ and is therefore to be kept constant.  An illustration, with $V$ varying between sub- and supersonic values, is given in Figure \ref{fig:Dispersion_comoving-frame}.}
\item{\textsc{Lab frame}: The lab frequency $\omega$, which is constant and appears as a straight horizontal line, is plotted alongside $V k \pm c k$, the dispersion profile as viewed from the lab frame in which the fluid is not at rest.  Again, the points of intersection give the possible wavevector solutions.  In this picture, variation of $V$ alters the dispersion curve, causing it to tilt on a dispersion diagram.  The equivalent of Figure \ref{fig:Dispersion_comoving-frame} as viewed from the lab frame is shown in Figure \ref{fig:Dispersion_lab-frame}.}
\end{itemize}

\subsection{Scalar product\label{sub:Scalar_product}}

The Lagrangian $L$ is also invariant under phase rotation:
$\phi\rightarrow\phi\, e^{i\alpha}$, where $\alpha$ is a real constant.
For any two solutions $\phi_{1}$ and $\phi_{2}$ of the wave equation (\ref{eq:dispersionless_acoustic_wave_equation}), this symmetry in the phase implies \cite{Noether-1918} conservation of a scalar quantity -- the \textit{scalar product} \cite{Birrell-Davies} -- defined as
\begin{eqnarray}
\left(\phi_{1},\phi_{2}\right) & = & i\int_{-\infty}^{+\infty}dx\left\{ \phi_{1}^{\star}\left(\partial_{t}+V\partial_{x}\right)\phi_{2}-\phi_{2}\left(\partial_{t}+V\partial_{x}\right)\phi_{1}^{\star}\right\} \nonumber \\
 & = & i\int_{-\infty}^{+\infty}dx\left\{ \phi_{1}^{\star}\pi_{2}-\phi_{2}\pi_{1}^{\star}\right\} \,,
 \label{eq:scalar_product}
 \end{eqnarray}
where the canonical momentum is
\begin{equation}
\pi=\frac{\partial L}{\partial\left(\partial_{t}\phi^{\star}\right)}=\left(\partial_{t}+V\partial_{x}\right)\phi\,.
\label{eq:canonical_momentum}
\end{equation}
The definition (\ref{eq:scalar_product}) of the scalar product directly implies the following relation between complex conjugate solutions:
\begin{equation}
\left(\phi_{1}^{\star},\phi_{2}^{\star}\right) = -\left(\phi_{1},\phi_{2}\right)^{\star} \,. \label{eq:scalar_product_cc}
\end{equation}
The scalar product of a solution with itself is called the \textit{norm}.  Note that this norm is \textit{not} positive-definite: if $\phi$ solves the wave equation (\ref{eq:dispersionless_acoustic_wave_equation}) and has positive norm, then the complex-conjugate solution $\phi^{\star}$ -- which also solves the real wave equation (\ref{eq:dispersionless_acoustic_wave_equation}) -- must, according to Eq. (\ref{eq:scalar_product_cc}), have negative norm.

The scalar product (\ref{eq:scalar_product}) only makes sense for a complex field, vanishing when $\phi$ is real.  However, even when the \textit{total} field is real, it makes sense to decompose it into complex components, and these will evolve in such a way that their scalar products are conserved.

\subsection{Field decomposition: the $k$-representation\label{sub:k-representation}}

Quantization of the field requires expressing it as a sum over eigenstates, which are then quantized individually.
For now, let us assume that $V$ is constant in $x$; then $L$ is invariant under spatial translations and $k$ is also a conserved quantity \cite{Noether-1918}.  We may then work with the states of constant $k$, which correspond to momentum eigenstates, and for which Fourier analysis assures completeness on the entire real line.  This might seem a trivial case, but it will turn out to be very instructive for the generalization to inhomogeneous velocity profiles.

\subsubsection{Orthonormal basis}

The canonical momentum operator $\left(\partial_{t}\,+\,V\partial_{x}\right)$, defined in Eq. (\ref{eq:canonical_momentum}), is the derivative with respect to time in the co-moving frame: when acting on a stationary mode, it multiplies it by $-i\,\omega_{\mathrm{cm}} = \mp i\,c\,k$, where the minus (plus) sign corresponds to $u$- ($v$-) modes.  $V$ being constant, the stationary modes are simply plane waves $\exp(ikx-i\omega t)$, whose scalar products (see Eq. (\ref{eq:scalar_product})) are
\begin{eqnarray}
\left(e^{ik_{1}x-i\omega^{u}(k_{1})t},e^{ik_{2}x-i\omega^{u}(k_{2})t}\right) & = & 4\pi\, c\,k_{1}\,\delta\left(k_{1}-k_{2}\right)\,,\nonumber \\
\left(e^{ik_{1}x-i\omega^{v}(k_{1})t},e^{ik_{2}x-i\omega^{v}(k_{2})t}\right) & = & -4\pi\, c\,k_{1}\,\delta\left(k_{1}-k_{2}\right)\,, \nonumber \\
\left(e^{ik_{1}x-i\omega^{u}(k_{1})t},e^{ik_{2}x-i\omega^{v}(k_{2})t}\right) & = & 0\,.
\label{eq:norm_u-v-modes_k-repn}
\end{eqnarray}
(Since $V$ is constant, there is no event horizon and hence no splitting of the $u$-branch into $u_{R}$ and $u_{L}$.)  That is, the plane waves are mutually orthogonal, and, utilizing Eqs. (\ref{eq:free-fall-freq_u-and-v}), their norms are $4\pi\,\omega_{\mathrm{cm}}\,\delta\left(k_{1}-k_{2}\right)$.  Although we can renormalize these modes using an appropriate prefactor, the occurrence of one mode and one complex-conjugate mode in the definition (\ref{eq:scalar_product}) of the scalar product ensures that we can never change the \textit{sign} of the norm. Therefore, the sign of the norm is equal to the sign of the co-moving frequency, and the dispersion relation (\ref{eq:Doppler_formula}) leads also to the classification
\begin{equation}
\omega_{\mathrm{cm}} = \omega - Vk = \begin{cases}
|c k| & \mathrm{for}\:\mathrm{positive-norm}\:\mathrm{modes}\\
-|c k|& \mathrm{for}\:\mathrm{negative-norm}\:\mathrm{modes}\end{cases}\,.\label{eq:free-fall-freq_pos-and-neg}\end{equation}
Combining classifications (\ref{eq:free-fall-freq_u-and-v}) and (\ref{eq:free-fall-freq_pos-and-neg}), we see that positive-norm modes are those which:
\begin{itemize}
\item if $k>0$, are $u$-modes;
\item if $k<0$, are $v$-modes;
\end{itemize}
while for negative-norm modes, this correspondence is reversed.  (In Figs. \ref{fig:Dispersion_comoving-frame} and \ref{fig:Dispersion_lab-frame}, the splitting of the dispersion profile into the $u$- and $v$-branches, and into the positive- and negative-norm branches, is labelled explicitly.)  Thus positive-norm modes correspond to the intuitive notion that the sign of $k$ determines the direction of travel (with respect to the medium); and the negative-norm modes, through complex conjugation, correspond to precisely the opposite.

The orthonormal $k$-mode basis, then, is the set of modes
\begin{eqnarray}
\phi_{k}^{u}\left(x,t\right) & = & \frac{1}{\sqrt{4\pi\,|c k|}} \exp \left(ikx - i\omega^{u}(k)t\right) \, \equiv \, \phi_{k}^{u}\left(x\right) e^{-i\omega^{u}(k)t}\,,\nonumber\\
\phi_{k}^{v}\left(x,t\right) & = & \frac{1}{\sqrt{4\pi\,|c k|}} \exp \left(ikx - i\omega^{v}(k)t\right) \, \equiv \, \phi_{k}^{v}\left(x\right) e^{-i\omega^{v}(k)t}\,,
\label{eq:normalized_k-modes}
\end{eqnarray}
normalized to $\pm\,\delta(k-k^{\prime})$ where the sign is determined as above, and where $\omega^{u/v}(k)=(V\pm c)k$.

\subsubsection{Field decomposition}

According to Fourier analysis, the set of all real wavevectors form a complete set in position-space.  The separate $u$- and $v$-branches found here are needed to describe the full time-dependence of the field, the wave equation (\ref{eq:dispersionless_acoustic_wave_equation}) being second-order in time.  So the general solution of the wave equation can be written in the form
\begin{equation}
\phi \left(x,t\right) = \int_{-\infty}^{+\infty} dk\,\left[ a^{u}(k) \phi^{u}_{k} \left(x\right)e^{-i\omega^{u}(k)t} + a^{v}(k) \phi^{v}_{k} \left(x\right)e^{-i\omega^{v}(k)t} \right]\,,
\label{eq:k_mode_decomposition}
\end{equation}
where $a^{u}(k)$ and $a^{v}(k)$ are complex-valued functions of $k$.  It is useful to separate the positive- and negative-norm modes in the integral, as follows:
\begin{eqnarray}
\phi \left(x,t\right) & = & \int_{-\infty}^{0} dk\,\left[ a^{v}(k) \phi^{v}_{k}\left(x\right)e^{-i\omega^{v}(k)t} + a^{v\star}(k) \phi^{v\star}_{k}\left(x\right)e^{i\omega^{v}(k)t} \right] \nonumber \\
& & \qquad \qquad + \int_{0}^{\infty} dk\,\left[ a^{u}(k) \phi^{u}_{k}\left(x\right)e^{-i\omega^{u}(k)t} + a^{u\star}(k) \phi^{u\star}_{k}\left(x\right)e^{i\omega^{u}(k)t} \right]\,,
\label{eq:k_mode_decomposition_p-n-norm}
\end{eqnarray}
where, for simplicity, the coefficients of complex-conjugate modes have themselves been written as complex conjugates so that the \textit{total} field is real.  The first term in each of the integrands, written without a $^{\star}$, is a positive-norm mode; its complex conjugate, written with a $^{\star}$, has negative norm.  If the total field $\phi$ is known, then the orthonormality properties of the modes allow extraction of the coefficients using the scalar product:
\begin{alignat}{1}
a^{u}(k)=\left(\phi_{k}^{u},\phi\right)\,,\qquad & a^{u\star}(k)=-\left(\phi_{k}^{u\star},\phi\right)\,,\nonumber \\
a^{v}(k)=\left(\phi_{k}^{v},\phi\right)\,,\qquad & a^{v\star}(k)=-\left(\phi_{k}^{v\star},\phi\right)\,.\label{eq:field_expansion_coefficients}
\end{alignat}

\subsubsection{Quantization}

In the form (\ref{eq:k_mode_decomposition_p-n-norm}), the field is readily quantized via the usual methods of Quantum Field Theory (QFT) \cite{Birrell-Davies}.  We promote the real-valued field variable $\phi$ to the Hermitian operator $\hat{\phi}$, the coefficients $a^{u/v}(k)$ to operators $\hat{a}_{k}^{u/v}$ and their complex conjugates $[a^{u/v}(k)]^{\star}$ to the Hermitian conjugate operators $[\hat{a}_{k}^{u/v}]^{\dagger}$:
\begin{eqnarray}
\hat{\phi} \left(x,t\right) & = & \int_{-\infty}^{0} dk\,\left[ \hat{a}^{v}_{k} \phi^{v}_{k}\left(x\right)e^{-i\omega^{v}(k)t} + \hat{a}^{v\dagger}_{k} \phi^{v\star}_{k}\left(x\right)e^{i\omega^{v}(k)t} \right] \nonumber \\
& & \qquad \qquad + \int_{0}^{\infty} dk\,\left[ \hat{a}^{u}_{k} \phi^{u}_{k}\left(x\right)e^{-i\omega^{u}(k)t} +\hat{a}^{u\dagger}_{k} \phi^{u\star}_{k}\left(x\right)e^{i\omega^{u}(k)t} \right]\,.
\label{eq:k_mode_decomposition_quantized}
\end{eqnarray}
Similarly, using the definition (\ref{eq:canonical_momentum}), the canonical momentum operator is
\begin{eqnarray}
\hat{\pi} \left(x,t\right) & = & \int_{-\infty}^{0} dk\,\left[ \hat{a}^{v}_{k} \pi^{v}_{k}\left(x\right)e^{-i\omega^{v}(k)t} + \hat{a}^{v\dagger}_{k} \pi^{v\star}_{k}\left(x\right)e^{i\omega^{v}(k)t} \right] \nonumber \\
& & \qquad \qquad + \int_{0}^{\infty} dk\,\left[ \hat{a}^{u}_{k} \pi^{u}_{k}\left(x\right)e^{-i\omega^{u}(k)t} +\hat{a}^{u\dagger}_{k} \pi^{u\star}_{k}\left(x\right)e^{i\omega^{u}(k)t} \right]\,.
\label{eq:k_mode_decomposition_quantized_momentum}
\end{eqnarray}
The relations (\ref{eq:field_expansion_coefficients}) may be similarly quantized, yielding expressions for the mode operators in terms of the full field operators:
\begin{eqnarray}
\hat{a}_{k}^{u} & = & i\int_{-\infty}^{+\infty}dx\left\{ \phi_{k}^{u\star}\left(x\right)e^{i\omega^{u}(k)t}\hat{\pi}\left(x,t\right)-\pi_{k}^{u\star}\left(x\right)e^{i\omega^{u}(k)t}\hat{\phi}\left(x,t\right)\right\} \,,\nonumber \\
\hat{a}_{k}^{u\dagger} & = & -i\int_{-\infty}^{+\infty}dx\left\{ \phi_{k}^{u}\left(x\right)e^{-i\omega^{u}(k)t}\hat{\pi}\left(x,t\right)-\pi_{k}^{u}\left(x\right)e^{-i\omega^{u}(k)t}\hat{\phi}\left(x,t\right)\right\} \,,
\label{eq:mode_operators}
\end{eqnarray}
with analogous relations for the $v$-mode operators.  The quantization procedure is completed by imposing the canonical commutation relations
\footnote{We work in natural units where $\hbar=1$.}
\begin{alignat}{1}
\left[\hat{\phi}\left(x,t\right),\hat{\pi}\left(x^{\prime},t\right)\right]=i\,\delta\left(x-x^{\prime}\right)\,,\qquad & \left[\hat{\phi}\left(x,t\right),\hat{\phi}\left(x^{\prime},t\right)\right]=\Big[\hat{\pi}\left(x,t\right),\hat{\pi}\left(x^{\prime},t\right)\Big]=0\,.
\label{eq:canonical_commutation_relations}
\end{alignat}
Finally, from these it can be shown that the mode operators satisfy the Bose commutation relations
\begin{equation}
\left[\hat{a}_{k}^{u},\hat{a}_{k^{\prime}}^{u\dagger}\right]=\left[\hat{a}_{k}^{v},\hat{a}_{k^{\prime}}^{v\dagger}\right]=\delta\left(k-k^{\prime}\right)\,,
\label{eq:Bose_commutation_relations}
\end{equation}
all other commutators being zero.  Thus we see that $\hat{a}^{u/v}_{k}$ and $\hat{a}^{u/v\dagger}_{k}$ -- the quantum amplitude operators multiplying modes of positive and negative norm -- are bosonic annihilation and creation operators, respectively, for the $u$- or $v$-mode with wavevector $k$.

This last point is of paramount importance, and worth emphasizing: positive- and negative-norm modes correspond to bosonic annihilation and creation operators, respectively\footnote{Had we considered a complex rather than a real field, the field operator $\hat{\phi}$ would not be Hermitian and the annihilation and creation operators appearing in its decomposition would not be Hermitian conjugates of each other.  They would then correspond to \textit{different} particles, i.e., particles and antiparticles.  Here, with a Hermitian field operator, we might say that we are considering particles which are their own antiparticles.}.  Notice that, although the notation was chosen with foresight, we did not make any mathematical assumptions about the mode operators.  Their bosonic property (\ref{eq:Bose_commutation_relations}) follows from the norms of the modes they multiply and the imposition of the canonical commutation relations (\ref{eq:canonical_commutation_relations}).  The correspondence between norm and bosonic quality represents a generalization to moving media of QFT in flat space.  There, it is standard to separate the complex-conjugate positive- and negative-\textit{frequency} modes, finding as here that the corresponding mode operators obey the Bose commutation relations (\ref{eq:Bose_commutation_relations}).  What we have just shown is that it is the sign of the \textit{norm} which determines these mode operators, and this happens to be the same as the sign of the frequency \textit{in the co-moving frame} -- \textit{not} the sign of the frequency in the lab frame.  Worded this way, this point seems obvious, since for the constant-velocity profile we could have transformed into the co-moving frame, in which case the co-moving and lab frequencies would have coincided and flat-space QFT would have been applicable.  Ultimately, however, it is the norm of the modes that enters the derivation of Eqs. (\ref{eq:Bose_commutation_relations}), and this will continue to be true for inhomogeneous profiles, where the co-moving frequency is no longer so well-defined and the norm emerges as the true fundamental quantity.

The quantization procedure above is admittedly rather abstract.  An analogous procedure, from a physically grounded viewpoint, is performed in condensed matter theory using the Bogoliubov theory of dilute Bose gases; see, for example, reference \cite{Castin-2000}.

\subsection{Field decomposition: the $\omega$-representation\label{sub:w-representation}}

Plane waves (\ref{eq:normalized_k-modes}) of constant $k$ are valid when $V$ is constant -- even if it is not constant everywhere.  They are especially useful in the constant-velocity asymptotic regions, where they correspond to quasiparticles that can be measured at infinity.  An exact solution for a stationary inhomogeneous velocity profile will be expressible as a sum over plane waves -- of equal lab frequency $\omega$ -- in the asymptotic regions.  This can be interpreted as a mixing or coupling between the various plane waves.

Therefore, it is much more useful to consider $\omega$ as the integration variable when summing the modes \cite{Macher-Parentani-2009}, since this groups together all plane waves that can be mixed together by an inhomogeneous flow.  First, we transform from the $k$-representation to the $\omega$-representation when $V$ is constant; we then consider the generalization to inhomogeneous velocity profiles.

\subsubsection{Homogeneous flow}

For reasons that will become clear, let us begin with the $k$-representation that does \textit{not} separate positive- and negative-norm modes:
\begin{equation}
\hat{\phi}\left(x,t\right)=\int_{-\infty}^{+\infty}dk\,\left[ \hat{a}_{k}^{v}\phi_{k}^{v}\left(x\right)e^{-i\omega^{v}(k) t}+\hat{a}_{k}^{u}\phi_{k}^{u}\left(x\right)e^{-i\omega^{u}(k) t}\right]\,,
\label{eq:k-repn_no-norm}
\end{equation}
where, consistently with Eqs. (\ref{eq:k_mode_decomposition}) and (\ref{eq:k_mode_decomposition_p-n-norm}), we have $\phi_{-k}^{u}\left(x\right)=\phi_{k}^{u\star}\left(x\right)$ and $\hat{a}_{-k}^{u}=\hat{a}_{k}^{u\dagger}$, and similar identities for the $v$-modes.  Recall from \S \ref{sub:k-representation} that the $u$-modes with $k>0$ and the $v$-modes with $k<0$ have positive norm.

We wish to express the field operator (\ref{eq:k-repn_no-norm}) in the equivalent form
\begin{equation}
\hat{\phi}\left(x,t\right)=\int_{-\infty}^{+\infty} d\omega \, \left[ \hat{a}_{\omega}^{v}\phi_{\omega}^{v}\left(x\right)e^{-i\omega t}+\hat{a}_{\omega}^{u}\phi_{\omega}^{u}\left(x\right)e^{-i\omega t} \right] \,.
\label{eq:w-repn_no-norm}
\end{equation}
This requires a redefining of the modes and their corresponding operators.  Firstly, we note that, in order for the quantization procedure to carry over exactly, the $\omega$-representation of the modes and operators should be normalized with respect to $\omega$:
\begin{alignat}{1}
\left(\phi_{\omega}^{u},\phi_{\omega^{\prime}}^{u}\right)=\delta\left(\omega-\omega^{\prime}\right)\,,\qquad & \left[\hat{a}_{\omega}^{u},\hat{a}_{\omega^{\prime}}^{u\dagger}\right]=\delta\left(\omega-\omega^{\prime}\right)\,,
\label{eq:w_mode_normalization}
\end{alignat}
and similarly for the $v$-modes. Now, the Dirac $\delta$ function $\delta\left(\omega-\omega^{\prime}\right)\,=\,|dk/d\omega|\,\delta\left(k-k^{\prime}\right)$, while the differential $d\omega=|d\omega/dk|\,dk$.  So, if the modes and operators in the $k$- and $\omega$-representations are related via \cite{Macher-Parentani-2009}
\begin{alignat}{1}
\phi_{\omega}^{u}=\sqrt{\left|\frac{dk^{u}}{d\omega}\right|}\phi_{k^{u}\left(\omega\right)}\,,\ \ \  & \phi_{\omega}^{v}=\sqrt{\left|\frac{dk^{v}}{d\omega}\right|}\phi_{k^{v}\left(\omega\right)}\,,\label{eq:modes_k2w}\\
\hat{a}_{\omega}^{u}=\sqrt{\left|\frac{dk^{u}}{d\omega}\right|}\hat{a}_{k^{u}\left(\omega\right)}\,,\ \ \  & \hat{a}_{\omega}^{v}=\sqrt{\left|\frac{dk^{v}}{d\omega}\right|}\hat{a}_{k^{v}\left(\omega\right)}\,,
\label{eq:operators_k2w}
\end{alignat}
we find that the normalization conditions (\ref{eq:w_mode_normalization}) are satisfied \textit{and} that the integrand of the $\omega$-representation of the field (\ref{eq:w-repn_no-norm}) transforms exactly into the integrand of the $k$-representation (\ref{eq:k-repn_no-norm}), since the factors of $|d\omega/dk|$ cancel.  So the orthonormal basis in the $\omega$-representation is the set of modes
\begin{eqnarray}
\phi_{\omega}^{u}\left(x,t\right) & = & \frac{1}{\sqrt{4\pi\,|c\,k^{u}\left(\omega\right)\,v_{g}\left(k^{u}(\omega)\right)|}} \exp \left(ik^{u}(\omega)x - i\omega t\right) \equiv \phi_{\omega}^{u}\left(x\right)e^{-i\omega t}\,,\nonumber\\
\phi_{\omega}^{v}\left(x,t\right) & = & \frac{1}{\sqrt{4\pi\,|c\,k^{v}\left(\omega\right)\,v_{g}\left(k^{v}(\omega)\right)|}} \exp \left(ik^{v}(\omega)x - i\omega t\right) \equiv \phi_{\omega}^{v}\left(x\right)e^{-i\omega t}\,,
\label{eq:normalized_w-modes}
\end{eqnarray}
normalized to $\pm\,\delta\left(\omega-\omega^{\prime}\right)$ according to the sign of the norm.  We have defined $v_{g}(k)=d\omega/dk$, which equals $\omega/k$ in the absence of dispersion, so that $k(\omega)\,v_{g}(k(\omega))$ appearing in Eqs. (\ref{eq:normalized_w-modes}) is simply equal to $\omega$.  (In the dispersive case, $v_{g}$ is the group velocity; see \S\ref{sub:The-wave-equation}.)

As before, we would like to separate the $\omega$-modes of the field operator (\ref{eq:w-repn_no-norm}) into positive- and negative-norm modes, i.e., into those terms corresponding to annihilation and creation of quasiparticles.  We have already seen in \S\ref{sub:k-representation} that the sign of the norm is simply related to the wavevector $k$ -- but there is a non-trivial aspect in its relation to frequency.  This can be clearly seen in the dispersion diagrams of Figs. \ref{fig:Dispersion_comoving-frame} and \ref{fig:Dispersion_lab-frame}.  For $\omega>0$, the solution on the $v$-branch always has positive norm; the $u$-mode, on the other hand, has positive norm when the flow is subsonic, and \textit{negative} norm when the flow is supersonic.  Restricting our attention to the $u$-part of the field, we have
\begin{equation}
\hat{\phi}^{u}\left(x,t\right)  =
\begin{cases}
\int_{0}^{\infty} d\omega \Big[ \hat{a}^{u}_{\omega} \phi_{\omega}^{u}\left(x\right)e^{-i\omega t} + \hat{a}^{u\dagger}_{\omega} \phi_{\omega}^{u\star}\left(x\right)e^{i\omega t} \Big] & \mathrm{for}\: |V| < c \\
\int_{0}^{\infty} d\omega \Big[ \hat{a}^{u}_{-\omega} \phi_{-\omega}^{u}\left(x\right)e^{i\omega t} + \hat{a}^{u\dagger}_{-\omega} \phi_{-\omega}^{u\star}\left(x\right)e^{-i\omega t} \Big] & \mathrm{for}\: |V| > c
\end{cases}\,,
\end{equation}
where, as before, the first terms of the integrands are the positive-norm modes and their complex conjugates have negative norm.  We emphasize the crucial fact that, comparing subsonic with supersonic flow, $u$-modes of equal lab frequency have opposite norm.

\subsubsection{Inhomogeneous flow}

There is nothing particularly significant about the sign of the frequency when the flow is truly homogeneous, for we could easily shift between subsonic and supersonic flow via a simple change of reference frame.  There is no gain in transforming to the $\omega$-representation, for $k$ itself is a conserved quantity, and mode mixing cannot occur.

For an \textit{in}homogeneous flow, however, the significance is profound.  There is a preferred reference frame in which the flow profile is time-independent, and in which $\omega$ is conserved.  It is in this frame that the $\omega$-representation is useful, in which the linking of modes with equal frequency has physical meaning.  And if, in this frame, the asymptotic flow velocities are one subsonic and the other supersonic -- if, in other words, an event horizon is present -- then, through the switching of norm induced by this transition, positive-norm modes on one side can couple to negative-norm modes on the other.

We know the form of the $u$-modes from \S\ref{sub:Stationary_modes}, and we have seen in \S\ref{sub:Event_horizon} that we must include two distinct modes, one corresponding to each side of the event horizon.  Taking account of the norm switching as $V$ varies between sub- and supersonic values, the spatially distinct \textit{positive-norm} $u$-modes can, for $\omega>0$, be written
\begin{eqnarray}
\phi_{\omega,R}^{u}\left(x,t\right) & = & \theta\left(x\right)\frac{1}{\sqrt{4\pi c \, \omega}}\exp\left(-i\omega u_{R}\right)\,,\nonumber\\
\phi_{-\omega,L}^{u}\left(x,t\right) & = & \theta\left(-x\right)\frac{1}{\sqrt{4\pi c \, \omega}}\exp\left(i\omega u_{L}\right)\,,\label{eq:spatially_distinct_u-modes}
\end{eqnarray}
where $u_{R}$ and $u_{L}$ are the $u$-variables defined on the right- and left-hand sides (i.e., the subsonic and supersonic regions), respectively.  Since those of the same frequency can mix, we can form linear combinations of one of the modes in Eqs. (\ref{eq:spatially_distinct_u-modes}) with the complex conjugate of the other.  To normalize such a linear combination, we simply employ the linearity of the scalar product:
\begin{multline*}
\left(\alpha\phi_{\omega_{1},R}^{u}+\beta\phi_{-\omega_{1},L}^{u\star},\alpha\phi_{\omega_{2},R}^{u}+\beta\phi_{-\omega_{2},L}^{u\star}\right)\\
=\alpha^{\star}\alpha\left(\phi_{\omega_{1},R}^{u},\phi_{\omega_{2},R}^{u}\right)+\alpha^{\star}\beta\left(\phi_{\omega_{1},R}^{u},\phi_{-\omega_{2},L}^{u\star}\right)+\beta^{\star}\alpha\left(\phi_{-\omega_{1},L}^{u\star},\phi_{\omega_{2},R}^{u}\right)+\beta^{\star}\beta\left(\phi_{-\omega_{1},L}^{u\star},\phi_{-\omega_{2},L}^{u\star}\right)\\
=\left(\left|\alpha\right|^{2}-\left|\beta\right|^{2}\right)\delta\left(\omega_{1}-\omega_{2}\right)\,.
\end{multline*}
So, if $\left|\alpha\right|^{2}-\left|\beta\right|^{2}=1$, the linear
combination $\alpha\phi_{\omega,R}^{u}+\beta\phi_{-\omega,L}^{u\star}$
is normalized with positive norm; it is also automatically orthogonal
to any such combination involving the modes
$\phi_{-\omega,L}^{u}$ and $\phi_{\omega,R}^{u\star}$, or modes of a different frequency, since the
individual modes are orthogonal amongst themselves. Therefore, any set of modes
\begin{alignat}{1}
\phi_{\omega,1}^{u}=\alpha_{\omega,1}\phi_{\omega,R}^{u}+\beta_{\omega,1}\phi_{-\omega,L}^{u\star}\,,\qquad & \phi_{\omega,2}^{u}=\alpha_{\omega,2}\phi_{-\omega,L}^{u}+\beta_{\omega,2}\phi_{\omega,R}^{u\star}\,,
\label{eq:generalized_U_modes}
\end{alignat}
where
\begin{equation}
\left|\alpha_{\omega,j}\right|^{2}-\left|\beta_{\omega,j}\right|^{2}=1\,,
\label{eq:normalization_of_U_modes}
\end{equation}
form a complete set of orthonormal, positive-norm $u$-modes.  The various possibilities correspond to different ways of characterizing the quasiparticles, and they are made possible by the horizon's splitting of the $u$-branch into two independent parts.

\section{Hawking radiation\label{sec:Hawking_radiation}}

\subsection{In- and out-modes\label{sub:In_and_out_modes}}

Despite the limitless possibilities that the freedom in the choice of $u$-modes allows, very few of them
are useful. We should deal only with ``natural'' sets of modes that correspond
to possible measurements. Consider the simplest example: $\alpha_{\omega,j}=1$,
$\beta_{\omega,j}=0$, which is simply the set of localized
modes given in Eqs. (\ref{eq:spatially_distinct_u-modes}).
These correspond to single outgoing wavepackets.
A quasiparticle travelling to the right with frequency $\omega$
is precisely an excitation of the mode $\phi_{\omega,R}^{u}$; similarly,
a quasiparticle travelling to the left with frequency $-\omega$ is an excitation
of $\phi_{-\omega,L}^{u}$. They differ in this respect from all other
linear combinations, which correspond to two outgoing wavepackets
rather than just one. In light of this property, the set of modes
$\phi_{\omega,R}^{u}$ and $\phi_{-\omega,L}^{u}$ are termed \textit{out-modes}:
those which correspond to a single outgoing wave in the asymptotic future.
The $v$-modes $\phi_{\omega}^{v}$ also have this property, and are therefore the out-modes of the $v$-branch.

Complementing the out-modes are the \textit{in-modes}, those corresponding to a single ingoing wave in the asymptotic past.  The $v$-modes $\phi_{\omega}^{v}$ are such modes, so that, on the $v$-branch, the in-modes are equal to the out-modes; this is a consequence of the regular behaviour of the $v$-coordinate at the horizon.  In contrast, as we saw in \S\ref{sub:Event_horizon}, the $u$-modes, in both the subsonic and supersonic regions, drift away from the horizon.  How, then, can we possibly form an \textit{ingoing} $u$-mode?  One possible answer (see \S\ref{sub:Conceptual_issues} for another) is that, in the asymptotic past, we are able to form ingoing waves if the event horizon is not present there.  That is, we may assume that the horizon has not been forever present, but was formed at some instant in the past, before which the flow was everywhere sub- or supersonic.  In this initial, horizonless spacetime, the $u$-modes are well-behaved and are precisely the in-modes we are looking for \footnote{The $u$-modes of the horizonless spacetime would also be out-modes if this spacetime were not to form an event horizon; but, since the out-modes are by nature defined with respect to the asymptotic future, the true out-modes are those corresponding to the presence of the horizon.}.  There being no horizon at this initial stage, there is no divergence of the wavevector at any point, so that -- although it will vary from point to point if $V$ is inhomogeneous -- the \textit{sign} of the wavevector can never change.  Positive-norm in-modes are thus composed entirely of positive wavevectors, and similarly the negative-norm modes of negative wavevectors.

We are here interested only in the long-time, stationary state brought about by the final presence of the horizon, and not in any transient effects reliant on the details of its formation.  It is convenient to refer only to the final spacetime with horizon present, and so, to find the form of the in-modes that must be used, they must first be propagated from the initial spacetime to the final one.  Since the precise evolution is unimportant, we may consider simply an instantaneous change of the velocity profile.  Immediately after this change, the spatial form of the in-modes is precisely as in the initial horizonless spacetime -- but they will no longer be stationary.  We may, however, form linear combinations of them to find stationary in-modes appropriate to the final spacetime, so long as we form such combinations only from modes \textit{of the same norm}; the reason for this is made clear in \S\ref{sub:Spontaneous_creation}.  As already noted, this requires that positive-norm modes be formed only from positive wavevectors.  This implies in turn that the positive-norm modes, if analytically continued onto the complex $x$-plane, must be analytic in the upper half of this plane \cite{Damour-Ruffini-1976,Brout-et-al-Primer,Brout-et-al-1995}.  Recalling from Eq. (\ref{eq:U_horizon}) that, in the vicinity of the horizon, the coordinates $u_{R}$ and $u_{L}$ go as $\log\left((\alpha/c) |x|\right)/\alpha$, traversing the horizon $x=0$ on the upper-half $x$-plane analytically connects these two coordinates with the addition of an imaginary part: $u_{R}\rightarrow u_{L} + i\pi/\alpha$ and $u_{L}\rightarrow u_{R} - i\pi/\alpha$.  Applying these continuations to Eqs. (\ref{eq:spatially_distinct_u-modes}), we find that the $u$-modes on either side of the horizon also analytically connect to each other, but, upon exponentiation, the relative imaginary component of the coordinates translates into a relative \textit{amplitude} between the right- and left-hand $u$-modes.  This relative amplitude determines the $\alpha$ and $\beta$ coefficients of Eqs. (\ref{eq:generalized_U_modes}), and after normalizing according to Eq. (\ref{eq:normalization_of_U_modes}), we find that the stationary positive-norm in-modes of the $u$-branch are, for $\omega>0$,
\begin{eqnarray}
\phi_{\omega,R}^{u,\mathrm{in}} & = & \frac{1}{\sqrt{2\,\sinh\left(\frac{\pi\omega}{\alpha}\right)}}\left(e^{\frac{\pi\omega}{2\alpha}}\phi_{\omega,R}^{u,\mathrm{out}}+e^{-\frac{\pi\omega}{2\alpha}}\phi_{-\omega,L}^{u,\mathrm{out}\star}\right)\,,\nonumber \\
\phi_{-\omega,L}^{u,\mathrm{in}} & = & \frac{1}{\sqrt{2\,\sinh\left(\frac{\pi\omega}{\alpha}\right)}}\left(e^{\frac{\pi\omega}{2\alpha}}\phi_{-\omega,L}^{u,\mathrm{out}}+e^{-\frac{\pi\omega}{2\alpha}}\phi_{\omega,R}^{u,\mathrm{out}\star}\right)\,,\label{eq:in-modes_out-modes}
\end{eqnarray}
where $\phi_{\omega,R}^{u,\mathrm{out}}$ and $\phi_{-\omega,L}^{u,\mathrm{out}}$ are simply the localised modes of Eqs. (\ref{eq:spatially_distinct_u-modes}) with their out-mode character made explicit.  The negative-norm in-modes are simply the complex conjugates $\phi_{\omega,R}^{u,\mathrm{in}\star}$ and $\phi_{-\omega,L}^{u,\mathrm{in}\star}$.

Despite the mathematical elegance of this derivation of the stationary in-modes, it is physically obscure.  It is therefore of value to consider briefly the evolution of wavepackets, as is usually done in more direct physical treatments of the problem \cite{Hawking-1974,Hawking-1975,Brout-et-al-Primer} -- the stationary modes of Eqs. (\ref{eq:in-modes_out-modes}) may then be thought of as linear combinations of wavepackets.  Those wavepackets located far from the horizon at the time of its formation are essentially unaffected by its appearance, except for a possible change of direction due to the transition between subsonic and supersonic flow.  This accounts for the regularity of the in-modes away from the horizon.  For those wavepackets located near the horizon at the time of its formation, there will be a marked shift in velocity and a marked change in its wavevector as it slowly moves away from the horizon, so that the in-modes exhibit the divergence in wavevector at the horizon characteristic of the $u$-modes.  The wavepackets we have so far considered evolve into outgoing wavepackets on one side of the horizon, and therefore behave just as the out-modes do.  The crucial difference, then, for the in-modes is that there are wavepackets which actually cross the horizon at the time of its formation.  Such wavepackets will then evolve as two ``disconnected'' pieces -- but they are not \textit{quite} disconnected, because in the past they are analytically connected to each other.  This is the origin of the analytic continuation between the right- and left-hand sides across the horizon, which is characteristic of the in-modes.  Given the logarithmic divergence of the phase at the horizon, this analytic continuation can only be of the form (\ref{eq:in-modes_out-modes}) or their complex conjugates, depending on whether analyticity is imposed on the upper or lower half-plane.  Straightforward algebra then shows that these two possibilities correspond to the positive- and negative-norm modes, respectively.

In-modes and out-modes form two distinct and natural sets of orthonormal $u$-modes, and we are free to decompose the $u$-part of the field operator in terms of either. Together with
the $v$-modes, they form a complete set of orthonormal modes that
solve the wave equation (\ref{eq:dispersionless_acoustic_wave_equation}).
Therefore, the total field operator may be written:
\begin{eqnarray}
\negthickspace\negthickspace\negthickspace\hat{\phi}\left(t,x\right) & = & \int_{0}^{\infty}d\omega\left\{ \hat{a}_{\omega,R}^{u,\mathrm{in}}\phi_{\omega,R}^{u,\mathrm{in}}\left(t,x\right)+\hat{a}_{-\omega,L}^{u,\mathrm{in}}\phi_{-\omega,L}^{u,\mathrm{in}}\left(t,x\right)+\hat{a}_{\omega}^{v}\phi_{\omega}^{v}\left(t,x\right)+\mathrm{h.c.}\right\} \nonumber \\
 & = & \int_{0}^{\infty}d\omega\left\{ \hat{a}_{\omega,R}^{u,\mathrm{out}}\phi_{\omega,R}^{u,\mathrm{out}}\left(t,x\right)+\hat{a}_{-\omega,L}^{u,\mathrm{out}}\phi_{-\omega,L}^{u,\mathrm{out}}\left(t,x\right)+\hat{a}_{\omega}^{v}\phi_{\omega}^{v}\left(t,x\right)+\mathrm{h.c.}\right\}\,\,\,
\label{eq:field_in-out-mode_decomposition}
\end{eqnarray}
where $\mathrm{h.c.}$ stands for Hermitian conjugate, containing the negative-norm modes and creation operators.
Substituting the transformations (\ref{eq:in-modes_out-modes}) into the expressions (\ref{eq:field_in-out-mode_decomposition}) for the total field operator, we find the corresponding transformation between the mode operators:
\begin{eqnarray}
\hat{a}_{\omega,R}^{u,\mathrm{out}} & = & \frac{1}{\sqrt{2\,\sinh\left(\frac{\pi\omega}{\alpha}\right)}}\left(e^{\frac{\pi\omega}{2\alpha}}\,\hat{a}_{\omega,R}^{u,\mathrm{in}}+e^{-\frac{\pi\omega}{2\alpha}}\,\hat{a}_{-\omega,L}^{u,\mathrm{in}\dagger}\right)\,,\nonumber \\
\hat{a}_{-\omega,L}^{u,\mathrm{out}} & = & \frac{1}{\sqrt{2\,\sinh\left(\frac{\pi\omega}{\alpha}\right)}}\left(e^{\frac{\pi\omega}{2\alpha}}\,\hat{a}_{-\omega,L}^{u,\mathrm{in}}+e^{-\frac{\pi\omega}{2\alpha}}\,\hat{a}_{\omega,R}^{u,\mathrm{in}\dagger}\right)\,.\label{eq:operator_transformation}
\end{eqnarray}
It stands to reason that, since the modes of the in- or out-bases are formed from both positive- and negative-norm modes of the other, so the mode operators of one basis should combine both annihilation and creation operators of the other.

\subsection{Spontaneous creation\label{sub:Spontaneous_creation}}

Inequivalence of incoming and outgoing modes is not an exotic phenomenon in and of itself.  Any scattering process exhibits this kind of behaviour: an incoming wavepacket or particle is deflected, often into several other wavepackets of particles, each of which can be considered a single outgoing mode.  Normally, however, this is a process of \textit{conversion}: the incoming particles are converted into outgoing particles, such that the sum of the rates of outgoing particles -- all contributing with the same sign -- is equal to the rate of ingoing particles.  Decrease the rate of incoming particles to zero, and the rate of outgoing particles will likewise vanish.  We might say that all these particles have the same sign of norm.

The difference here is that the converted particles have different signs of norm, leading to the normalization condition $|\alpha|^{2}-|\beta|^{2}=1$.  Since the norm is always conserved, a part-conversion of the incoming modes into negative-norm modes must result in an increase in the amount of positive-norm modes.  This process is, at least in part, an \textit{amplification} \cite{Leonhardt-QO,Unruh-2011}.  For oppositely-normed particles equal numbers of both must be added to the system to keep the overall norm constant.  Moreover, at the quantum level, an amplifier adds particles to the system \textit{even when there are no incident particles} \cite{Caves-1982}.

Mathematically, the vacuum state is that in which no modes are excited: it is the absence of quasiparticles. Since one cannot annihilate an excitation from the vacuum state, it must vanish when acted on by any annihilation operator.  The vacuum is thus defined as the eigenstate of all annihilation operators with eigenvalue zero \cite{Birrell-Davies}:
\begin{equation}
\hat{a}\left|0\right\rangle =0\qquad\forall\:\hat{a}\,.
\label{eq:definition_of_vacuum}
\end{equation}
Clearly, this depends on the particular set of annihilation operators, and hence on the particular basis of modes \cite{Fulling-1973}; in particular, we can define both an \textit{in}-vacuum and an \textit{out}-vacuum, according to whether the vacuum state has zero eigenvalue for all in-annihilation operators of out-annihilation operators.  That said, if a mode transformation only mixes modes of the same norm, then the corresponding transformation between the mode operators will expand annihilation operators only in terms of other annihilation operators, and the state with zero eigenvalue for one set of these operators will also have zero eigenvalue with respect to the other: their vacuum states are identical.  This is why, when forming the stationary final-state in-modes in \S\ref{sub:In_and_out_modes}, we could only combine modes of equal norm: this ensures that the in-vacuum state defined by the late-time in-modes is exactly the same as that defined by the initial in-modes.  Similarly, if positive-norm in-modes were found to scatter only into positive-norm out-modes, then the vacua defined by the in- and out-modes would be identical, and an absence of incoming particles would lead to an absence of outgoing particles.

On the other hand, if scattering into opposite-norm modes, and hence mixing of annihilation and creation operators, takes place -- as it clearly does in Eqs. (\ref{eq:in-modes_out-modes}) and (\ref{eq:operator_transformation}) -- then the in-vacuum is not equal to the out-vacuum, and the absence of incoming particles \textit{must} lead to some presence of outgoing particles \cite{Fulling-1973}!  This is the mysterious effect of spontaneous creation, which is purely quantum mechanical in origin \footnote{An analogous kind of creation takes place in the Unruh effect \cite{Unruh-1976}, where the vacuum seen by an inertial observer is found to differ from the vacuum seen by an accelerating observer, leading to the detection of a thermal bath by the latter.}.

\subsection{Radiation from an event horizon\label{sub:Radiation_from_an_event_horizon}}

We may demonstrate this particle creation explicitly using Eq. (\ref{eq:operator_transformation}).
Imposing the in-vacuum, the expectation value of right-moving outgoing quasiparticles
is found to be\begin{eqnarray}
\left\langle n_{\omega,R}^{u,\mathrm{out}}\right\rangle  & = & \left\langle 0_{\mathrm{in}}\right|\hat{a}_{\omega,R}^{u,\mathrm{out}\dagger}\hat{a}_{\omega^{\prime},R}^{u,\mathrm{out}}\left|0_{\mathrm{in}}\right\rangle \nonumber \\
 & = & \left\langle 0_{\mathrm{in}}\right|\frac{1}{\sqrt{2\,\sinh\left(\frac{\pi\omega}{\alpha}\right)}}\left(e^{\frac{\pi\omega}{2\alpha}}\hat{a}_{\omega,R}^{u,\mathrm{in}\dagger}+e^{-\frac{\pi\omega}{2\alpha}}\hat{a}_{\omega,L}^{u,\mathrm{in}}\right) \nonumber \\
& & \qquad \qquad \qquad \times \frac{1}{\sqrt{2\,\sinh\left(\frac{\pi\omega^{\prime}}{\alpha}\right)}}\left(e^{\frac{\pi\omega^{\prime}}{2\alpha}}\hat{a}_{\omega^{\prime},R}^{u,\mathrm{in}}+e^{-\frac{\pi\omega^{\prime}}{2\alpha}}\hat{a}_{\omega^{\prime},L}^{u,\mathrm{in}\dagger}\right)\left|0_{\mathrm{in}}\right\rangle \nonumber \\
 & = & \frac{1}{\sqrt{4\,\sinh\left(\frac{\pi\omega}{\alpha}\right)\,\sinh\left(\frac{\pi\omega^{\prime}}{\alpha}\right)}}e^{-\frac{\pi\left(\omega+\omega^{\prime}\right)}{2\alpha}}\left\langle 0_{\mathrm{in}}\right|\hat{a}_{\omega,L}^{u,\mathrm{in}}\hat{a}_{\omega^{\prime},L}^{u,\mathrm{in}\dagger}\left|0_{\mathrm{in}}\right\rangle \nonumber \\
 & = & \frac{1}{\sqrt{4\,\sinh\left(\frac{\pi\omega}{\alpha}\right)\,\sinh\left(\frac{\pi\omega^{\prime}}{\alpha}\right)}}e^{-\frac{\pi\left(\omega+\omega^{\prime}\right)}{2\alpha}}\left\langle 0_{\mathrm{in}}\right|\hat{a}_{\omega^{\prime},L}^{u,\mathrm{in}\dagger}\hat{a}_{\omega,L}^{u,\mathrm{in}}+\delta\left(\omega-\omega^{\prime}\right)\left|0_{\mathrm{in}}\right\rangle \nonumber \\
 & = & \frac{1}{2\,\sinh\left(\frac{\pi\omega}{\alpha}\right)}e^{-\frac{\pi\omega}{\alpha}}\,\delta\left(\omega-\omega^{\prime}\right)\nonumber \\
 & = & \frac{1}{e^{\frac{2\pi\omega}{\alpha}}-1}\,\delta\left(\omega-\omega^{\prime}\right)\,.\label{eq:thermal_spectrum}\end{eqnarray}
Remarkably, the spectrum of quasiparticles emitted is precisely a bosonic
thermal distribution with temperature $\alpha/\left(2\pi\right)$;
replacing fundamental constants,\begin{equation}
k_{B}T=\frac{\hbar\alpha}{2\pi}\,.\label{eq:Hawking_temperature}\end{equation}
The appearance of the $\delta$ function shows that Eq. (\ref{eq:thermal_spectrum}) represents
a density rather than a number. This is because the orthonormal
modes form a continuous spectrum, and are not normalized to unity
but to a delta function -- see Eqs. (\ref{eq:w_mode_normalization}).  An argument in \cite{Corley-Jacobson-1996} overcomes this problem by turning to wavepackets, and shows that the \textit{spectral flux density} -- the number of quasiparticles emitted per unit time per unit (angular) frequency interval -- is obtained by dividing the calculated expectation value by $2\pi$:
\begin{equation}
\frac{\partial^{2}N}{\partial t \, \partial \omega} = \frac{1}{2\pi}\,\frac{1}{e^{\frac{2\pi\omega}{\alpha}}-1} \, .
\label{eq:thermal_spectral_flux_density}
\end{equation}

Due to the symmetry of the transformation (\ref{eq:operator_transformation}),
the expectation value of left-moving outgoing quasiparticles is exactly equal
to the thermal spectrum of (\ref{eq:thermal_spectrum}). There is
a deeper significance to this than just symmetry, however. The transformation
of the in- and out-operators allows us to write the in-vacuum explicitly
in terms of the out-vacuum. Using the Fock basis, in which the annihilation
and creation operators behave in the standard way \cite{Birrell-Davies}:
\begin{alignat}{1}
\hat{a}_{\omega}\left|n\right\rangle _{\omega}=\sqrt{n}\left|n-1\right\rangle _{\omega}\,,\qquad & \hat{a}_{\omega}^{\dagger}\left|n\right\rangle _{\omega}=\sqrt{n+1}\left|n+1\right\rangle _{\omega}\,,
\label{eq:annihilation_creation_Fock_basis}
\end{alignat}
the in-vacuum state is given by
\begin{eqnarray}
\left|0_{\mathrm{in}}\right\rangle  & = & Z^{-\frac{1}{2}}\prod_{\omega}\sum_{n=0}^{\infty}\frac{1}{n!}\left(e^{-\frac{\pi\omega}{a}}\,\hat{a}_{\omega,R}^{u,\mathrm{out}\dagger}\,\hat{a}_{\omega,L}^{u,\mathrm{out}\dagger}\right)^{n}\left|0_{\mathrm{out}}\right\rangle \nonumber \\
 & = & Z^{-\frac{1}{2}}\prod_{\omega}\sum_{n=0}^{\infty}e^{-\frac{n\pi\omega}{a}}\left|n\right\rangle _{\omega,R}^{u,\mathrm{out}}\left|n\right\rangle _{\omega,L}^{u,\mathrm{out}}\,,
 \label{eq:in-vacuum_out-Fock-basis}
 \end{eqnarray}
where $Z$ is a normalizing prefactor defined such that $\left\langle 0_{\mathrm{in}}\right.\left|0_{\mathrm{in}}\right\rangle =1$.  (This can be checked simply by acting on Eq. (\ref{eq:in-vacuum_out-Fock-basis}) with an arbitrary in-annihilation operator, and utilizing the transformation (\ref{eq:operator_transformation}), to show that it vanishes.)
The fact that the out-creation operators appear only in $R$-$L$
pairs shows that the radiation, though it looks thermal in the right
and left sides separately, is strongly correlated \textit{between}
the two sides. Quasiparticles are emitted in pairs, one in the subsonic,
the other in the supersonic region; and a measurement of the number
of quasiparticles in any state on one side of the horizon infers that there
are equally many quasiparticles in the corresponding state on the other side.
The left and right systems, separated by the horizon, are maximally entangled \cite{Barnett-Phoenix-1989}.
This entanglement induces correlations between the left and right partners, which may prove invaluable in the eventual experimental detection of Hawking radiation \cite{Balbinot-et-al-2008,Carusotto-et-al-2008,Balbinot-et-al-2010,Mayoral-et-al-2011}; see also \S\ref{sub:Overview}.  Most recently, the entropy associated with the entangled partners has been investigated in \cite{Giovanazzi-2011,Rinaldi-2011} -- somewhat closing the circle of analogue systems by harking back to Bekenstein's insight on gravitational black holes, but using the emitted radiation as a measure of entropy rather than the event horizon area.

\section{Discussion\label{sub:Conclusion_and_discussion}}

\subsection{Summary\label{sub:Summary}}

Beginning with the generalized form of the black-hole spacetime (\ref{eq:generalized_Lemaitre_metric}), we found the wave equation (\ref{eq:dispersionless_acoustic_wave_equation}) for a massless scalar field, with the exact solution (\ref{eq:general_solution}) in terms of co- and counter-propagating components.  The counter-propagating components were found to separate at the event horizon into two distinct spatial parts.  Stationary (single-frequency) modes were derived, satisying the dispersion relation (\ref{eq:Doppler_formula}) which combines the co-moving dispersion relation with the Doppler effect.  Another conserved quantity is the scalar product (\ref{eq:scalar_product}), which gives rise to positive- and negative-norm stationary modes related by complex conjugation.  First assuming a homogeneous flow, the total field was written as a sum over the normalized stationary modes, and upon quantization it was found that positive- and negative-norm modes -- corresponding to positive and negative co-moving frequency -- are multiplied by bosonic annihilation and creation operators, respectively.  Crucially, the relation between co-moving frequency and lab frequency changes sign between subsonic and supersonic flow, and while it is the former that determines the norm, it is the latter that is conserved.  This leads, in the case of inhomogeneous flow with an event horizon, to the possibility of positive- and negative-norm mixing across the horizon.  The bases of incoming and outgoing modes are generally different in any physical system, but the mixing of positive- and negative-norm modes leads naturally to a mixing of annihilation and creation operators (see Eq. (\ref{eq:operator_transformation})), and hence to the inequality of the in- and out-vacuum states which are defined by their respective annihilation operators (see Eq. (\ref{eq:definition_of_vacuum})).  So quasiparticles are spontaneously emitted, and are found to conform to the thermal spectrum (\ref{eq:thermal_spectrum}), with temperature (\ref{eq:Hawking_temperature}) proportional to the velocity gradient at the event horizon.  Moreover, the quasiparticles emitted on each side of the horizon are maximally entangled, and emitted precisely in pairs.

The derivation has been kept quite general, in that it applies to any spacetime which can be described by the metric (\ref{eq:generalized_Lemaitre_metric}).  Of course, this includes gravitational black holes: using $V\left(r\right)=-c\sqrt{r_{S}/r}$, the velocity profile of the Schwarzschild black hole (see Eq. (\ref{eq:Lemaitre_metric})), we find that at the Schwarzschild radius $V^{\prime}\left(r_{S}\right)=c/\left(2r_{S}\right)=c^{3}/\left(4GM\right)$, and plugging into Eq. (\ref{eq:Hawking_temperature}), we have $k_{B}T=\hbar c^{3}/\left(8\pi GM\right)$ -- exactly Hawking's original formula, Eq. (\ref{eq:black_hole_temperature}).

\subsection{Conceptual issues\label{sub:Conceptual_issues}}

We identified the vacuum state as the in-vacuum, that which contains no ingoing quasiparticles; the ingoing quasiparticle states, in turn, were identified as the late-time equivalents of those corresponding to an initial horizonless spacetime.  This inevitably raises an intriguing question: how is the vacuum state of an eternal horizon -- with no horizonless spacetime in the asymptotic past -- to be defined?  There would seem to be no reason to use the modes (\ref{eq:in-modes_out-modes}) to define the vacuum state, since there can be no $u$-in modes in such a spacetime.  However, the crucial property of the modes (\ref{eq:in-modes_out-modes}) is not so much their ``ingoing'' character, but their \textit{analyticity} in the complex $x$-plane.  One can be led to the same vacuum state by a simple and intuitive argument: that the state appears as vacuum to co-moving observers crossing the horizon \cite{Unruh-1977}.  This leads again to the requirement of analyticity across the horizon and hence to the ``in''-modes (\ref{eq:in-modes_out-modes}) -- a property selected, as remarked in \S\ref{sub:In_and_out_modes}, by the ingoing criterion in virtue of their analyticity in the horizonless asymptotic past.  If the vacuum is defined differently, the modes used to define it would not be analytic at the horizon, and a freely-falling observer would observe an infinite flux of particles precisely at the horizon \cite{Unruh-1977}.  It thus appears that the state which gives rise to Hawking radiation is the only physically sensible one, and quite independent of the collapse phase.

In either the collapsing or eternal spacetime, then, Hawking radiation seems to be intimately connected with the existence of wavepackets of the quantum field smoothly connecting the two regions which, in classical general relativity, are disconnected by the event horizon.  The horizon breaks these wavepackets into two pieces, which propagate out into their respective regions \footnote{This is often interpreted as the creation of a particle-antiparticle pair via a quantum fluctuation, these then being separated by the horizon and unable to recombine \cite{Hawking-1975,Unruh-1977}.}.  Those pieces furthest from the horizon propagate outwards first, while those nearer the horizon escape later.  As time progresses, then, the Hawking particles originate from wavepackets of ever-increasing wavevector, ever-increasing thinness, and take an ever-increasing time to propagate out from the horizon.  In the dispersionless model, there exists an infinite reservoir of these trans-Planckian modes, so that the Hawking radiation can continue indefinitely (subject to eventual evaporation of the horizon, not accounted for here).  It is a remarkably counter-intuitive feature of the dispersionless model that the late-time steady-state Hawking radiation originates from an infinity of modes ``captured'' at a single point of space (and time, if collapse is included), then gradually released and redshifted so as to produce a steady thermal spectrum.

The analytic continuation giving rise to the ``in''-modes (\ref{eq:in-modes_out-modes}) is exact, even though it is derived from linearization of $V$ at the horizon (performed in \S\ref{sub:Event_horizon} to find the logarithmic phase).  This linearization is easily justified: as discussed in \S\ref{sub:Event_horizon}, any wavepacket can be traced back in time to arbitrary thinness and arbitrary closeness to the horizon.  The modes are thus well-defined arbitrarily close to the horizon, and the region over which analytic continuation onto the complex $x$-plane is imposed can be made arbitrarily small.  Therefore, it does not matter how small is the region over which the linearization of the velocity profile is valid, so long as it is valid over \textit{some} small region.  Consequently, the Hawking temperature can depend only on the first derivative, $\alpha$, of $V$ precisely at the horizon.

This picture breaks down entirely as soon as dispersion is introduced.  If the wave speed is allowed to vary with wavelength, then the event horizon is not so well-defined as a single point and linearization of $V$ at the horizon is not generally justified, so that we cannot hone in on a single parameter which uniquely determines the Hawking temperature.  More drastically, a change of wave speed at some scale will tune wavepackets out of the grip of the horizon once they reach that scale.  It is no longer true that any outgoing wavepacket must originate from one captured by the creation of the horizon; instead, all wavepackets, when traced back in time, will blueshift to the dispersive scale and then propagate away from the horizon, originating from a fixed finite wavevector at spatial infinity!  Thus is the trans-Planckian problem resolved by dispersion -- but can such different wave propagation yield similar effects to the trans-Planckian model?  After all, given that a trans-sonic flow of dispersionless fluid can switch the sign of the norm by causing the wavevector to diverge at the horizon (see Figs. \ref{fig:Dispersion_comoving-frame} and \ref{fig:Dispersion_lab-frame}), it is far from obvious that a similar kind of norm switching will still be possible in a dispersive fluid, and we might reasonably doubt the existence of Hawking radiation in dispersive systems at all.

\newpage

\part{Dispersive model\label{part:Dispersive_model}}

\section{The dispersive wave equation\label{sub:The-wave-equation}}

Aiming to generalize the dispersion relation of waves in the medium, we are, due to the considerations of \S\ref{sub:Dispersion}, forced to abandon the spacetime metric as a starting point. We begin instead with the Lagrangian density, in which dispersion is modelled by the appearance of higher-order derivatives \cite{Unruh-1995,Corley-Jacobson-1996}:
\begin{equation}
L=\frac{1}{2}\left|\left(\partial_{t}+V\partial_{x}\right)\phi\right|^{2}-\frac{1}{2}\left|c\left(-i\partial_{x}\right)\partial_{x}\phi\right|^{2}\,.\label{eq:Lagrangian_with_dispersion}
\end{equation}
The only difference from Eq. (\ref{eq:scalar_massless_Lagrangian_moving_fluid}) is the replacement of the constant $c$ with the function $c(k)$; as a function of an operator, we consider it as a Taylor series in that operator.

Application of the Principle of Least Action to Eq. (\ref{eq:Lagrangian_with_dispersion}) leads to the generalized Euler-Lagrange equation:\begin{equation}
\frac{\partial L}{\partial\phi^{\star}}-\frac{\partial}{\partial t}\left(\frac{\partial L}{\partial\left(\partial_{t}\phi^{\star}\right)}\right)-\frac{\partial}{\partial x}\left(\frac{\partial L}{\partial\left(\partial_{x}\phi^{\star}\right)}\right)+\frac{\partial^{2}}{\partial x^{2}}\left(\frac{\partial L}{\partial\left(\partial_{x}^{2}\phi^{\star}\right)}\right)-\frac{\partial^{3}}{\partial x^{3}}\left(\frac{\partial L}{\partial\left(\partial_{x}^{3}\phi^{\star}\right)}\right)+\ldots=0\,.\label{eq:Euler_Lagrange_higher_derivatives}\end{equation}
Substituting $L$ from Eq. (\ref{eq:Lagrangian_with_dispersion}), we find the wave equation \cite{Unruh-1995}
\begin{equation}
\left(\partial_{t}+\partial_{x}V\right)\left(\partial_{t}+V\partial_{x}\right)\phi-c^{2}\left(-i\partial_{x}\right)\partial_{x}^{2}\phi=0\,.\label{eq:acoustic_wave_equation}
\end{equation}
In the geometrical optics approximation, the velocity profile $V(x)$ varies negligibly over a wavelength of the field $\phi$, so that the ``local wavevector'' $k(x)$ exists and is determined by the local value of $V$, analogously to Eqs. (\ref{eq:U_V_wavevectors}).  This leads to a dispersion relation between $\omega$, $k$ and $V$: neglecting derivatives of $V$ and $k$, we find
\begin{equation}
\left(\omega-V k\right)^{2}=c^{2}\left(k\right)k^{2}\,.\label{eq:acoustic_flow_dispersion}
\end{equation}
This is the generalization to dispersive media of Eq. (\ref{eq:Doppler_formula}).
$c\left(k\right)$ is the \textit{phase velocity} $\omega_{\mathrm{cm}}/k$ in the co-moving frame, which describes the propagation of the phase $kx-\omega_{cm}t$; in the lab frame, the phase velocity is rather $\omega/k$.  The \textit{group velocity} is the derivative $d\omega_{\mathrm{cm}}/dk=\pm d(c(k)k)/dk$ in the co-moving frame, or $d\omega/dk=V\pm d(c(k)k)/dk$ in the lab frame;
this gives the velocity of the envelope of a wavepacket strongly peaked at the wavevector $k$.  In the absence of dispersion, the phase and group velocities are identical.
Typically, there exists a low-wavevector regime where the phase and group velocities can be approximated as constant,
so that $c\left(k\right)\rightarrow c_{0}$ as
$k\rightarrow0$. For higher values of $k$, the dispersion can take
two basic forms: \textit{superluminal} and \textit{subluminal}
\footnote{Although these terms emphasise the Schwarzschild analogy, they have become standard terms to describe the dispersion in any context, much like the labels ``subsonic'' and ``supersonic'' to describe any type of flow.}.
These are defined according
to whether the magnitude of $c\left(k\right)$
becomes higher or lower than $c_{0}$ as $k$ increases.

As in \S\ref{sub:Stationary_modes}, there are two equivalent pictures for determining the possible solutions of the dispersion relation: from the point-of-view of the co-moving frame, they are the points of intersection of the dispersion profile $\pm|c(k)k|$ with the co-moving frequency $\omega_{\mathrm{cm}}=\omega-Vk$; or, from the point-of-view of the lab frame, they are the points of intersection of the frequency $\omega$ with the lab frame dispersion profile $Vk\pm|c(k)k|$.  Also as before, there are two branches of the dispersion, and two ways to define them:
\begin{itemize}
\item \textsc{The $u$- and $v$-branches}: the sign of the \textit{phase velocity} in the co-moving frame defines the counter-propagating $u$-branch ($\omega_{\mathrm{cm}}=\omega-Vk=c(k)k$) and the co-propagating $v$-branch ($\omega_{\mathrm{cm}}=\omega-Vk=-c(k)k$);
\item \textsc{The positive- and negative-norm branches}: the sign of the \textit{frequency} in the co-moving frame defines the positive-norm branch ($\omega_{\mathrm{cm}}=\omega-Vk=|c(k)k|$) and the negative-norm branch ($\omega_{\mathrm{cm}}=\omega-Vk=-|c(k)k|$).
\end{itemize}
The definitions of the positive- and negative-norm branches carry over because, like its dispersionless counterpart (\ref{eq:scalar_massless_Lagrangian_moving_fluid}), the Lagrangian density (\ref{eq:Lagrangian_with_dispersion}) remains invariant under phase rotations.  Therefore, the scalar product of Eq. (\ref{eq:scalar_product}) is conserved even for dispersive media.

\section{Field decomposition with dispersion\label{sec:Field_decomposition_with_dispersion}}

For a homogenous velocity profile, the wavevector $k$ is a conserved quantity, and the analysis of \S\ref{sub:k-representation} -- the definition of the orthonormal basis of $k$-modes, the decomposition of the field in this basis and the quantization of the field -- carries over exactly, the only difference being the more complicated functional form of the frequency, $\omega(k)$.

Frequency being the conserved quantity for a general \textit{in}homogeneous velocity profile, we wish rather to express the field operator in the form of an integral over $\omega$:
\begin{equation}
\hat{\phi}\left(x,t\right)=\int_{0}^{\infty} d\omega\,\left[\hat{\phi}_{\omega}\left(x\right)e^{-i\omega t}+\hat{\phi}_{\omega}^{\dagger}\left(x\right)e^{i\omega t}\right]\,,
\label{eq:w-repn_of_field}
\end{equation}
where the $\hat{\phi}_{\omega}\left(x\right)$ are the complete mode operators in the $\omega$-representation, combining all solutions of the same frequency.  The transformation to the $\omega$-representation performed in \S\ref{sub:w-representation} was written for a general dispersion relation $\omega(k)$, and therefore also carries over to the dispersive case.  Dispersion complicates the transformation only through the introduction of additional counter-propagating solutions, which are easily found with the aid of dispersion diagrams.  For simplicity, we shall restrict ourselves to \textit{subluminal} dispersion and examine its effects on the resulting transformation.  Superluminal dispersion behaves analogously; see reference \cite{Macher-Parentani-2009} for a detailed derivation of the transformation for both types of dispersion \footnote{Most of the analyses occurring in the literature (e.g., \cite{Balbinot-et-al-2008,Macher-Parentani-2009-ii,Recati-et-al-2009,Finazzi-Parentani-2011-JPCS,Finazzi-Parentani-2011}) use superluminal dispersion -- mainly because they are geared towards BECs, whose excitations follow a superluminal dispersion relation.}.  Realistic dispersion relations may be complicated enough to introduce even more solutions than are indicated here\footnote{For instance, in the first paper \cite{Jacobson-1991} to consider dispersive analogue systems, liquid helium is considered as an example, with a dispersion profile that includes a roton dip.}, but careful attention to the dispersion diagrams should make the required modes clear.

\subsection{Homogeneous flow\label{sub:Homogeneous flow}}

We learned from the dispersionless model that the set of plane-wave solutions to the dispersion relation depends crucially on the flow velocity -- in particular, on whether it is subsonic or supersonic.  Anticipating similar behaviour in the dispersive case, let us examine these two cases separately.

\paragraph{Supersonic flow}

\begin{figure}
\includegraphics[width=0.8\columnwidth]{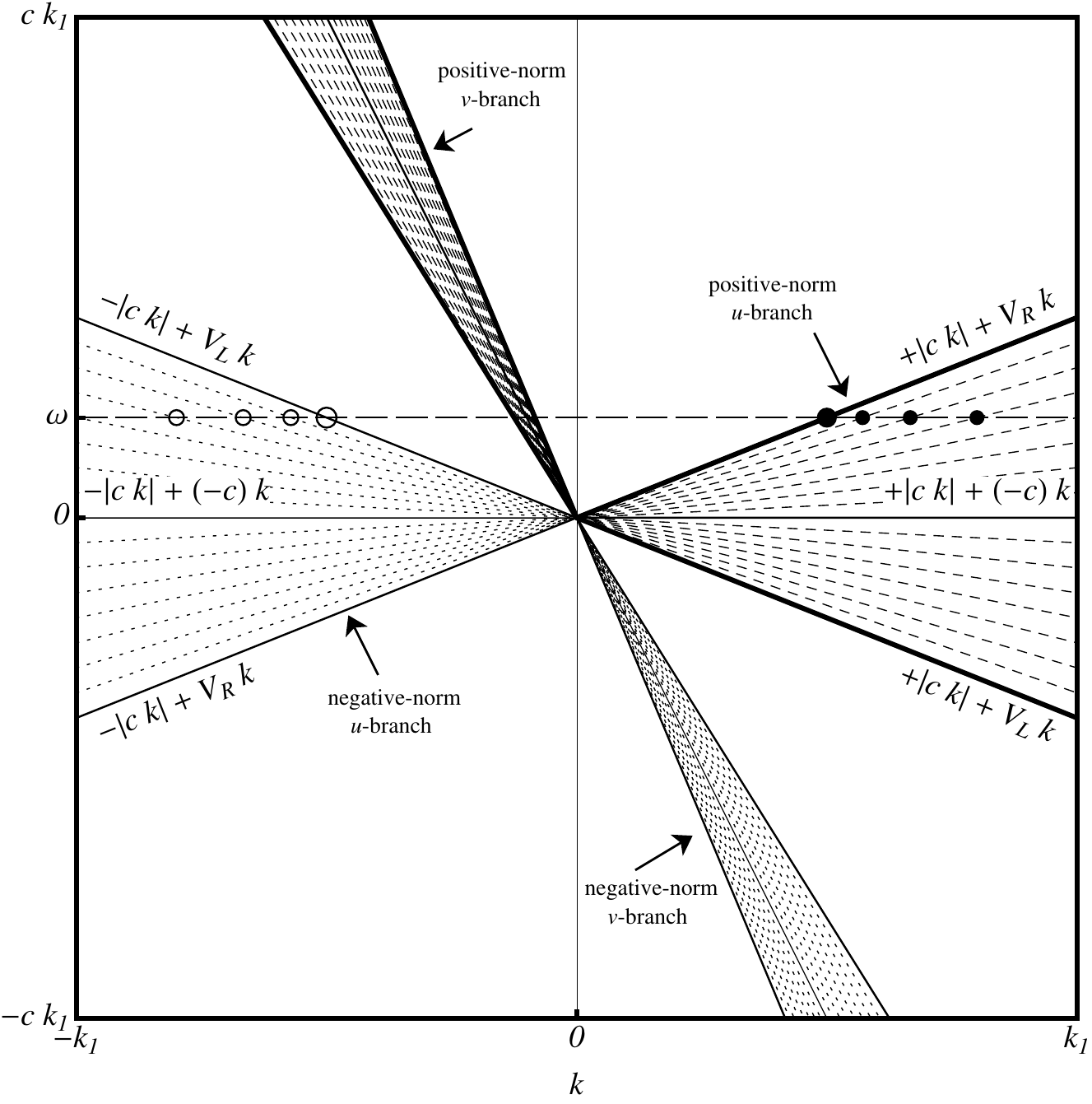}
\caption[\textsc{Dispersion profile in the lab frame}]{\textsc{Dispersion profile in the lab frame}:
Here is plotted the same information as in Fig. \ref{fig:Dispersion_comoving-frame}, but from the point of view of the lab frame.  The conserved lab-frame frequency $\omega$ appears as a horizontal line.  On the other hand, the dispersion profile of the medium in the lab frame is given by $Vk\pm|ck|$, where the plus and minus signs refer to the positive- and negative-norm branches, respectively.  The lab frame dispersion could also be written $Vk\pm ck$, where this time the plus and minus signs refer to the $u$- and $v$-branches, respectively.  As the flow velocity $V$ is varied, the dispersion profile tilts, and the possible wavevector solutions -- where the disperison profile is equal to $\omega$ -- vary accordingly.  Again, we observe critical behaviour when $V=-c$, for at this point the $u$-branch of the dispersion lies exactly along the horizontal $k$-axis, and is therefore parallel to the horizontal line at $\omega$.  For $V>-c$, it is the positive-norm $u$-branch which crosses $\omega$, while for $V<-c$ it is the negative-norm $u$-branch which does so.  In contrast, the $v$-branch -- which for $V=-c$ is shown as a solid line within the continuum of $V$-dependent $v$-branches -- experiences nothing of importance at $V=-c$.
\label{fig:Dispersion_lab-frame}}
\end{figure}

Suppose that the flow velocity is faster than $c_{0}$, the low-wavevector limit of the phase velocity.  The wavevector solutions of the dispersion are illustrated graphically in Figure \ref{fig:subluminal_supersonic}.  It is clear that, for all positive lab-frame frequencies $\omega>0$, there exist only two wavevector solutions: a co-propagating $v$-mode with positive norm and a counter-propagating $u$-mode with negative norm.  This is exactly what was found for the same (supersonic) flow in the dispersionless model (see Figs. \ref{fig:Dispersion_comoving-frame} and \ref{fig:Dispersion_lab-frame}), the only difference here being a different functional dependence of the wavevectors on frequency.  The transformation to the $\omega$-representation thus proceeds exactly as in \S\ref{sub:w-representation}, and the mode operators for $\omega>0$ are simply
\begin{equation}
\hat{\phi}_{\omega}\left(x\right)=\hat{a}^{v}_{\omega}\phi_{\omega}^{v}\left(x\right)+\hat{a}^{u\dagger}_{-\omega}\phi_{-\omega}^{u\star}\left(x\right)\,,
\label{eq:w-mode_subluminal_supersonic}
\end{equation}
where the mode on the $u$-branch has been written as a Hermitian conjugate on account of its negative norm.  For $\omega<0$, the mode operators are simply the Hermitian conjugates of those in Eq. (\ref{eq:w-mode_subluminal_supersonic}): $\hat{\phi}_{-\omega}(x)=\hat{\phi}_{\omega}^{\dagger}(x)$.

\paragraph{Subsonic flow}

Now consider a flow velocity slower than $c_{0}$.  The solutions are illustrated graphically in Figure \ref{fig:subluminal_subsonic}.  Here we observe behaviour significantly different from the dispersionless case.  Firstly, let us note that, for all positive lab-frame frequencies $\omega>0$, there exist a positive-norm $v$-mode and a \textit{negative}-norm $u$-mode, both being continuously connected to their counterparts in supersonic flow as $V$ is varied from one value to the other.  In the dispersionless model, the negative-norm $u$-mode experiences a diverging wavevector as the flow crosses from supersonic to subsonic (see Figs. \ref{fig:Dispersion_comoving-frame} and \ref{fig:Dispersion_lab-frame}), becoming a positive-norm $u$-mode in the latter type of flow.  Dispersion has completely removed this divergence, and consequently the negative-norm $u$-mode remains.

However, we see from Fig. \ref{fig:subluminal_subsonic} that removal of the wavevector divergence does not quite prohibit the emergence of a \textit{positive}-norm $u$-mode.  More precisely, it \textit{is} prohibited at high frequencies, above some threshold value $\omega_{\mathrm{max}}$ which depends on the value of the velocity (and vanishes in the limit where $V$ becomes supersonic, so that the two cases are continuously connected).  For frequencies below $\omega_{\mathrm{max}}$, there comes into existence a \textit{pair} of additional counter-propagating modes with positive norm.  For convenience, let us label these $u1$ and $u2$, where $|k^{u1}|<|k^{u2}|$; the negative-norm $u$-mode, since it exists for all frequencies and velocities, will continue to be labelled simply $u$.  The $u1$- and $u2$-waves can be considered as forming additional branches of solutions, of finite measure, which exist only at low frequencies and which vanish entirely when $V$ becomes supersonic.  Notice that, while the $u1$-, $u2$- and $u$-branches are all counter-propagating (i.e., they are right-moving in the co-moving frame), they are distinguished by their behaviour in the lab frame:
\begin{itemize}
\item $u1$ and $u2$ are distinct from $u$ in that they have oppositely-signed lab-frame \textit{phase velocities} $\omega/k$; and
\item $u1$ is distinct from $u2$ in that they have oppositely-signed lab-frame \textit{group velocities} $d\omega/dk$.
\end{itemize}
The first property regarding the phase velocity also applies to the dispersionless case, since it distinguishes the $u_{R}$- and $u_{L}$-modes that exist on opposite sides of the horizon: it enables the mixing of waves of opposite norm.  The second property, regarding the group velocity, is a purely dispersive effect: it will be found to regularize the behaviour at the horizon.

With these considerations, the mode operators can be written
\begin{equation}
\hat{\phi}_{\omega}\left(x\right)=\begin{cases}
\hat{a}_{\omega}^{v}\phi_{\omega}^{v}\left(x\right)+\hat{a}_{\omega}^{u1}\phi_{\omega}^{u1}\left(x\right)+\hat{a}_{\omega}^{u2}\phi_{\omega}^{u2}\left(x\right)+\hat{a}_{-\omega}^{u\dagger}\phi_{-\omega}^{u\star}\left(x\right)\, & \mathrm{for}\;0<\omega<\omega_{\mathrm{max}}\\
\hat{a}_{\omega}^{v}\phi_{\omega}^{v}\left(x\right)+\hat{a}_{-\omega}^{u\dagger}\phi_{-\omega}^{u\star}\left(x\right)\, & \mathrm{for}\;\omega>\omega_{\mathrm{max}}\end{cases}\,.
\label{eq:w-mode_subluminal_subsonic}
\end{equation}
Let us emphasise that the mode operators (\ref{eq:w-mode_subluminal_subsonic}) for subsonic flow connect continuously to the mode operators (\ref{eq:w-mode_subluminal_supersonic}) for supersonic flow, since $\omega_{\mathrm{max}}\rightarrow 0$ when $V$ becomes supersonic.

\subsection{Inhomogeneous flow: Mode mixing\label{sub:Inhomogeneous_flow}}

As remarked in \S\ref{sub:w-representation}, an inhomogeneous, asymptotically constant flow can be viewed as mixing or coupling the asymptotic plane waves to each other.  Physically, this concept of mode mixing is perhaps better understood by considering the motion of a wavepacket, strongly peaked at a certain value of $k$.  (Figures \ref{fig:out_mode} and \ref{fig:in_mode} include results of numerical wavepacket propagation, alongside space-time diagrams, to help make the scattering concept more concrete.)  In a region of constant $V$, the wavepacket will propagate in accordance with its group velocity, $v_{g}=d\omega/dk$.  A wavepacket incident on the inhomogeneous region will, after a complicated interaction, scatter into some combination of resultant wavepackets, with frequency $\omega$ equal to that of the original wavepacket and wavevectors $k$ corresponding to the possible solutions of the dispersion relation.  To use the terminology introduced in \S\ref{sub:In_and_out_modes}, there are two possible types of scattering mode: the \textit{out-mode}, which contains a single outgoing wavevector in the asymptotic future (illustrated in Fig. \ref{fig:out_mode}); and the \textit{in-mode}, which contains a single ingoing wavevector in the asymptotic past (illustrated in Fig. \ref{fig:in_mode}).  Throughout the scattering process, the total norm is always conserved; but, as discussed in \S\ref{sub:Spontaneous_creation}, some of the waves involved may carry opposite signs of norm, leading to a degree of amplification.

Dispersive effects result in a richer variety of possible scattering processes, due to the appearance of additional plane wave solutions and hence more possibilities for mode coupling.  Since the additional solutions exist only below the $V$-dependent critical frequency $\omega_{\mathrm{max}}$, the scattering possibilities are themselves frequency- and velocity-dependent.  The asymptotic flow velocities are generally different, yielding different critical frequencies for the left- and right-hand asymptotic regions.  For definiteness, we denote these $\omega_{\mathrm{max},1}$ and $\omega_{\mathrm{max},2}$, where $\omega_{\mathrm{max},1} \le \omega_{\mathrm{max},2}$, and where one or both of the critical frequencies may be zero.  These divide the spectrum into (at most) three regimes, in which quite different scattering behaviour is observed.  Figures \ref{fig:dispersion_sup_sup}, \ref{fig:dispersion_sub_sup} and \ref{fig:dispersion_sub_sub} show dispersion diagrams -- with varying $V$ to account for the inhomogeneous profile -- for the various arrangements of the frequency regimes described below.  Also note that $V$ is assumed to be single-valued, so that only a single horizon (if any) exists.

\subsubsection{$\omega>\omega_{\mathrm{max},2}$: suppression of Hawking radiation}

This regime always exists for high enough frequencies, and in the case where the flow is everywhere supersonic so that both $\omega_{\mathrm{max},1}$ and $\omega_{\mathrm{max},2}$ vanish, it encompasses the entire spectrum; this situation is illustrated in Figure \ref{fig:dispersion_sup_sup}.

\begin{figure}
\includegraphics[width=0.8\columnwidth]{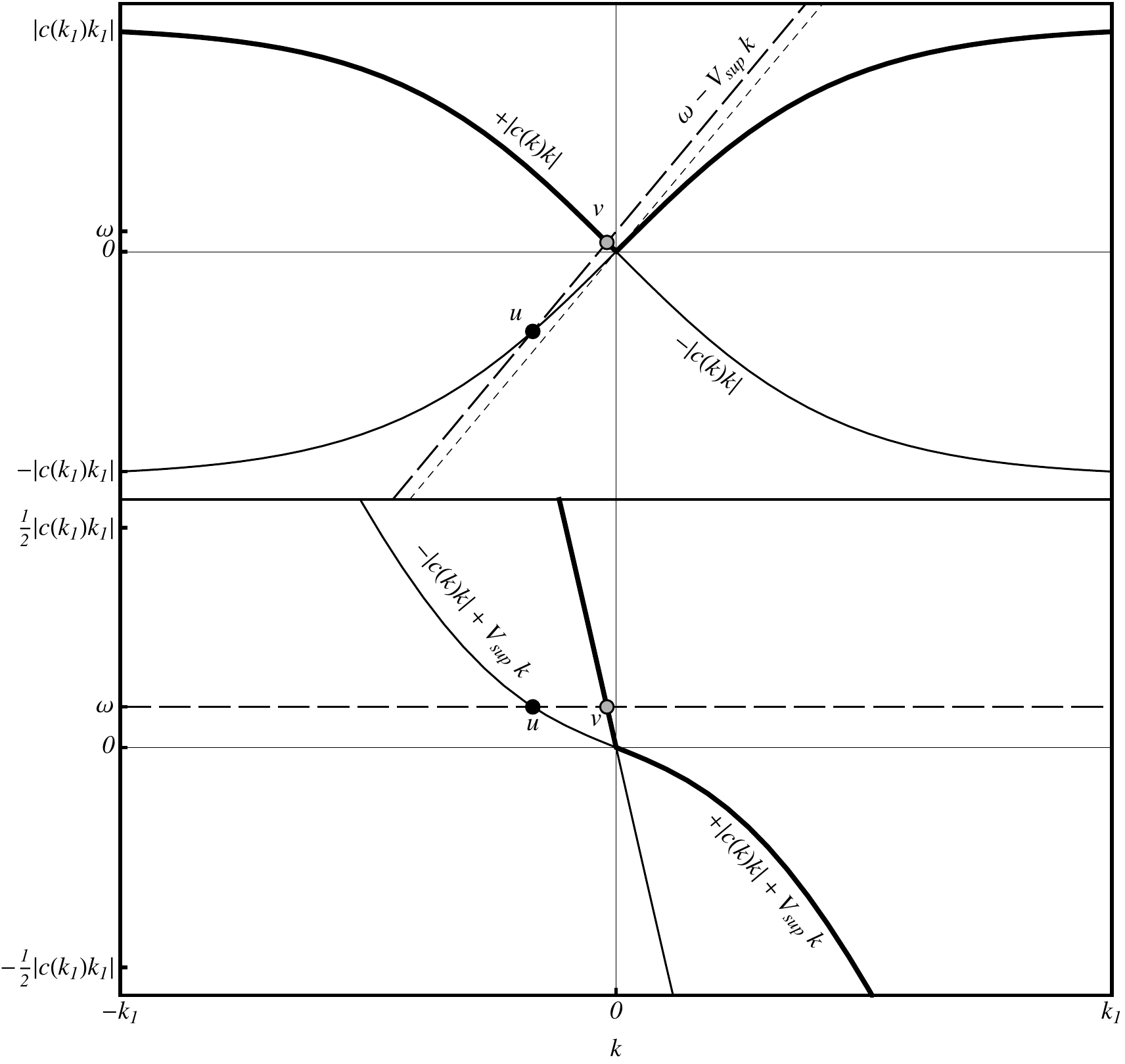}
\caption[\textsc{Subluminal dispersion with supersonic flow}]{\textsc{Subluminal dispersion with supersonic flow}:
Here are plotted subluminal dispersion diagrams from the points of view of both the co-moving frame (upper panel) and the lab frame (lower panel).  Positive- and negative-norm branches are shown as thick and thin curves, respectively, just as in Figs. \ref{fig:Dispersion_comoving-frame} and \ref{fig:Dispersion_lab-frame}.  The velocity is assumed to be supersonic: $V_{\mathrm{sup}}<-c_{0}$, where $c_{0}$ is the low-wavevector limit of the phase velocity, and therefore equal to the slope of the dispersion profile in the co-moving frame at $k=0$.  The line $\omega-V_{\mathrm{sup}}k$, then, has a greater slope than the dispersion profile at $k=0$, and since the slope of the dispersion profile decreases with increasing $k$, the two curves never cross for positive $k$.  Viewed from the lab frame, the slope of the dispersion profile $V_{\mathrm{sup}}k+c(k)k$ is negative at $k=0$ because of the supersonic flow, and the gradual decrease of $c(k)$ causes it to become more negative with increasing $k$, so that it can never equal $\omega$ for positive $k$.  There are thus no solutions on the positive-norm $u$-branch, and only the negative-norm $u$-wave and positive-norm $v$-wave remain.
\label{fig:subluminal_supersonic}}
\end{figure}

\begin{figure}
\includegraphics[width=0.8\columnwidth]{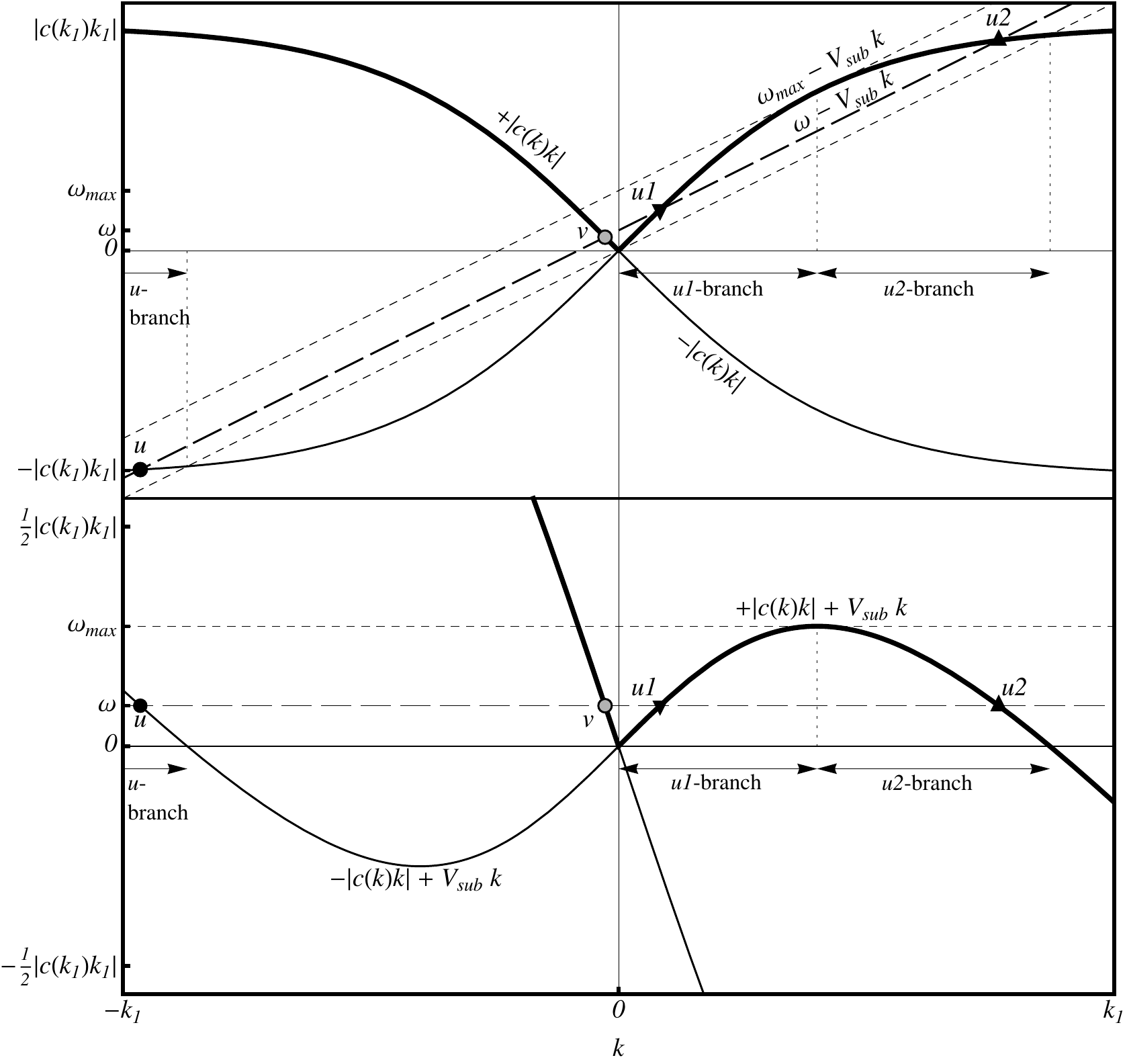}
\caption[\textsc{Subluminal dispersion with subsonic flow}]{\textsc{Subluminal dispersion with subsonic flow}:
Here are plotted subluminal dispersion diagrams from the points of view of both the co-moving frame (upper panel) and the lab frame (lower panel).  Positive- and negative-norm branches are shown as thick and thin curves, respectively, just as in Figs. \ref{fig:Dispersion_comoving-frame}, \ref{fig:Dispersion_lab-frame} and \ref{fig:subluminal_supersonic}.  The velocity is assumed to be subsonic: $V_{\mathrm{sub}}>-c_{0}$, where $c_{0}$ is the low-wavevector limit of the phase velocity.  As viewed from the co-moving frame, the slope of the line $\omega-V_{\mathrm{sub}}k$ is therefore \textit{less} than that of the dispersion profile at $k=0$; but, since $c(k)$ decreases with increasing $k$, the two curves become parallel at some point, and if $\omega$ is equal to $\omega_{\mathrm{max}}$ the line $\omega_{\mathrm{max}}-V_{\mathrm{sub}}k$ is actually tangent to the dispersion profile.  As viewed from the lab frame, the dispersion profile $Vk+c(k)k$ has positive slope at $k=0$, but the slope decreases so that it eventually turns round; at this maximum point, it is equal to $\omega_{\mathrm{max}}$.  If $\omega<\omega_{\mathrm{max}}$, the two curves cross twice for positive $k$, giving rise to the $u1$- and $u2$-branches.  The $u$-branch, then, does not extend to $k=0$, but only to the point at which the (complex conjugate of the) $u2$-branch ends.  On the other hand, if $\omega>\omega_{\mathrm{max}}$, only the $u$- and $v$-waves exist.
\label{fig:subluminal_subsonic}}
\end{figure}

\begin{figure}
\includegraphics[width=0.8\columnwidth]{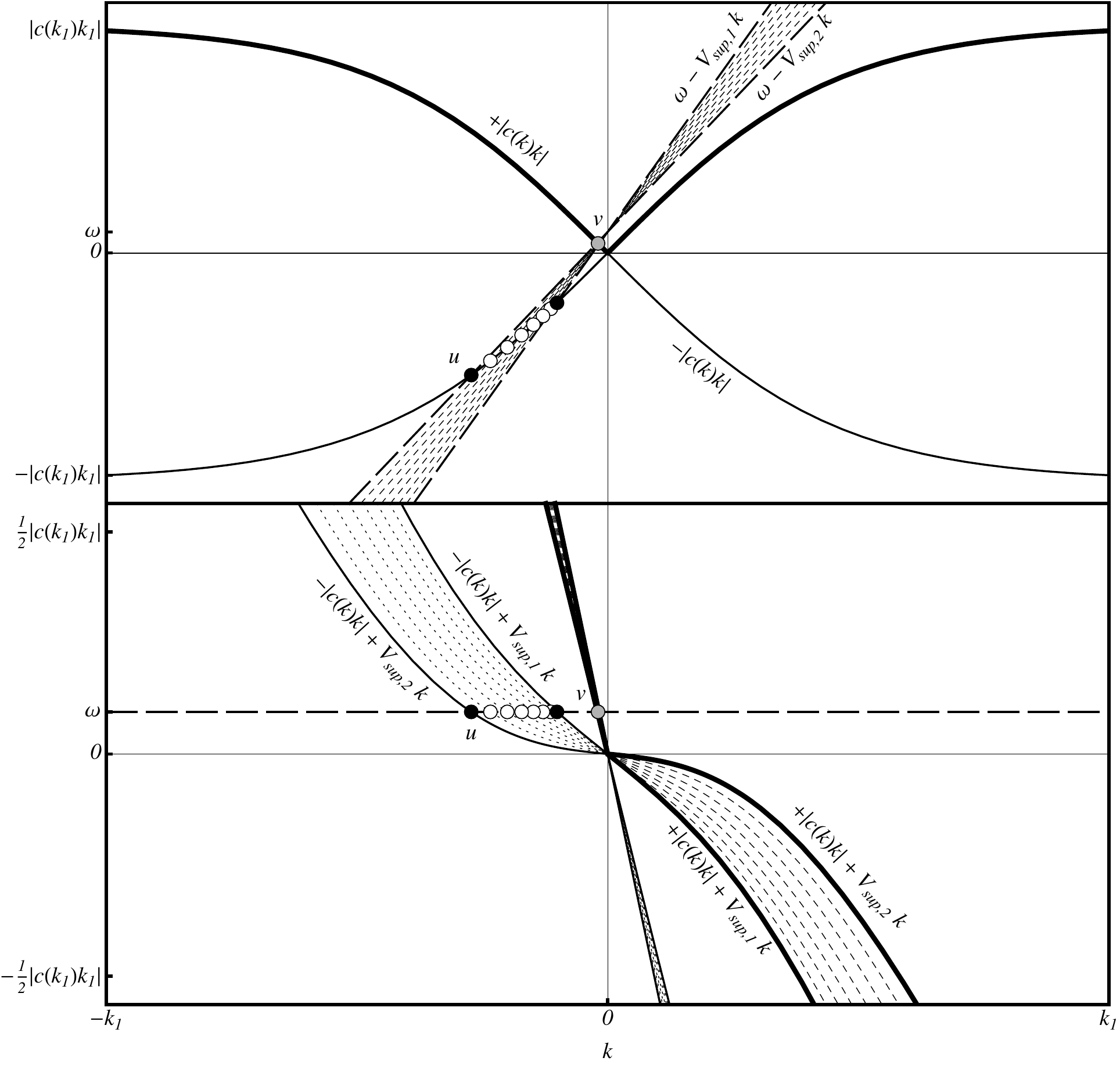}
\caption[\textsc{Dispersion with inhomogeneous $V$ (supersonic to supersonic)}]{\textsc{Dispersion with inhomogeneous $V$ (supersonic to supersonic)}:
Here are plotted subluminal dispersion diagrams from the points of view of both the co-moving frame (upper panel) and the lab frame (lower panel).  Positive- and negative-norm branches are shown as thick and thin curves, respectively, just as in Figs. \ref{fig:Dispersion_comoving-frame}-\ref{fig:subluminal_subsonic}.  The velocity is assumed to vary between two supersonic values, $V_{\mathrm{sup},1}$ and $V_{\mathrm{sup},2}$.  The entire spectrum is in the ``high'' frequency regime, where only the $u$- and $v$-waves -- indicated by circles -- exist.  These vary smoothly with varying $V$, having well-defined values at all points in space; they do not experience a horizon.  They do have opposite norm, however, so that they can still couple to form Hawking pairs.
\label{fig:dispersion_sup_sup}}
\end{figure}

In both asymptotic regions, the solutions of the dispersion are arranged similarly to Fig. \ref{fig:subluminal_supersonic}.  There exist only one positive-norm $v$-wave and one negative-norm $u$-wave.  These vary continuously on their respective branches, being perfectly well-behaved at any horizon which might be present.  This is quite different from the dispersionless case, where the $u$-wave experiencing divergence of its wavevector and a switch in its norm as the horizon is crossed.  It appears there is no channel for Hawking radiation.  It is not \textit{quite} non-zero, though, since $u$-$v$ mixing can occur, and since these have opposite norm they can form a Hawking pair \footnote{This is not true for superluminal dispersion, where the $u$- and $v$-waves both have positive norm \cite{Macher-Parentani-2009}.}.  An interesting aspect of this pair is that, as can be seen in the dispersion diagram of Fig. \ref{fig:subluminal_supersonic}, their group velocities are both negative, and both partners are emitted into the left-hand region.  This is not the usual Hawking channel, though, and the strength of the $u$-$v$ coupling is negligible for a slowly-varying velocity profile \cite{Schutzhold-Unruh-2008}.  This high-frequency regime therefore corresponds to strong suppression of the Hawking radiation.

Each of the $u$- and $v$-waves can form an ingoing or an outgoing wave.  In either the in- or out-basis, then, there are two independent modes in this regime.

\subsubsection{$\omega_{\mathrm{max},1}<\omega<\omega_{\mathrm{max},2}$: the group-velocity horizon}

\begin{figure}
\includegraphics[width=0.8\columnwidth]{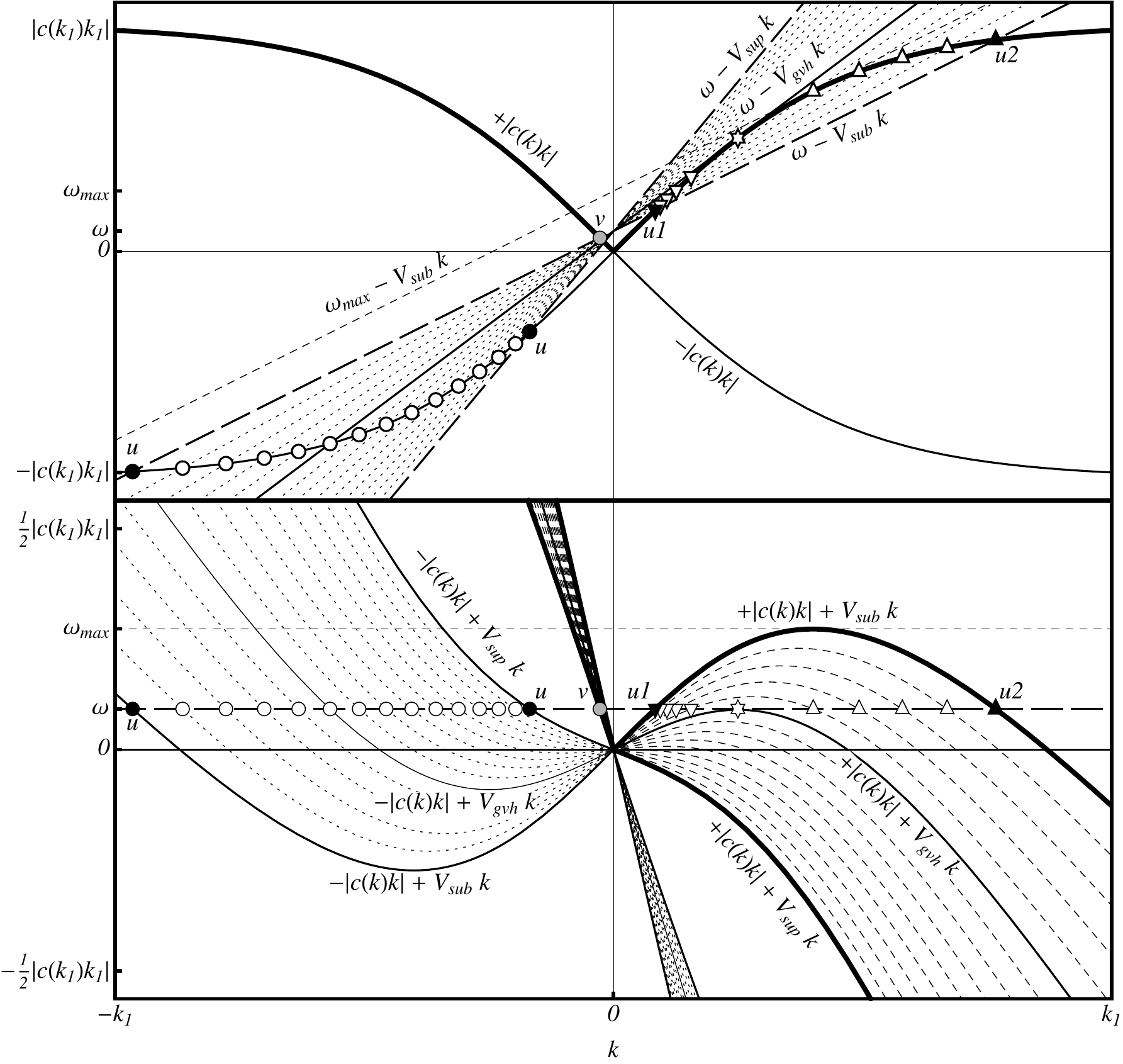}
\caption[\textsc{Dispersion with inhomogeneous $V$ (subsonic to supersonic)}]{\textsc{Dispersion with inhomogeneous $V$ (subsonic to supersonic)}:
Here are plotted subluminal dispersion diagrams from the points of view of both the co-moving frame (upper panel) and the lab frame (lower panel).  Positive- and negative-norm branches are shown as thick and thin curves, respectively, just as in Figs. \ref{fig:Dispersion_comoving-frame}-\ref{fig:dispersion_sup_sup}.  The velocity is assumed to vary between a subsonic value $V_{\mathrm{sub}}>-c_{0}$ and a supersonic value $V_{\mathrm{sup}}<-c_{0}$.  Therefore, of the critical frequencies, only $\omega_{\mathrm{max},2}$ is non-zero, and is labelled here simply as $\omega_{\mathrm{max}}$.  If $\omega>\omega_{\mathrm{max}}$, the situation is analogous to that described by Fig. \ref{fig:dispersion_sup_sup}.  If $\omega<\omega_{\mathrm{max}}$, we are in the mid-frequency regime where a group-velocity horizon exists.  There, the $u1$- and $u2$-waves -- indicated by inverted and upright triangles, respectively, in accordance with their lower and higher wavevectors -- exist in the subsonic region, but they cannot exist in the supersonic region.  Note their oppositely-directed group velocities -- this is indicated by the slope of the dispersion profile relative to that of the line $\omega-Vk$ in the co-moving frame, or simply by the slope of the dispersion curve itself in the lab frame.  If we track their evolution with varying $V$, we see that they vary towards each other, until, when $V=V_{\mathrm{gvh}}$ (indicated by a solid line among the continuum of curves), the straight line is tangent to the dispersion curve and the $u1$- and $u2$-wavevectors merge at the point indicated by a star.  This point corresponds to the group-velocity horizon: as $V$ varies further, the $u1$- and $u2$-waves cease to exist as real, propagating waves.  Instead, they cross into each other, reversing their group velocity and moving back in the direction from which they originally came.  They are thus not independent solutions, but degenerate into a single solution which ``bounces'' off the group-velocity horizon (see Fig. \ref{fig:in_mode}).  This positive-norm solution, as well as the positive-norm $v$-wave, can couple with the negative-norm $u$-wave to form Hawking pairs.\label{fig:dispersion_sub_sup}}
\end{figure}

\begin{figure}
\includegraphics[width=0.8\columnwidth]{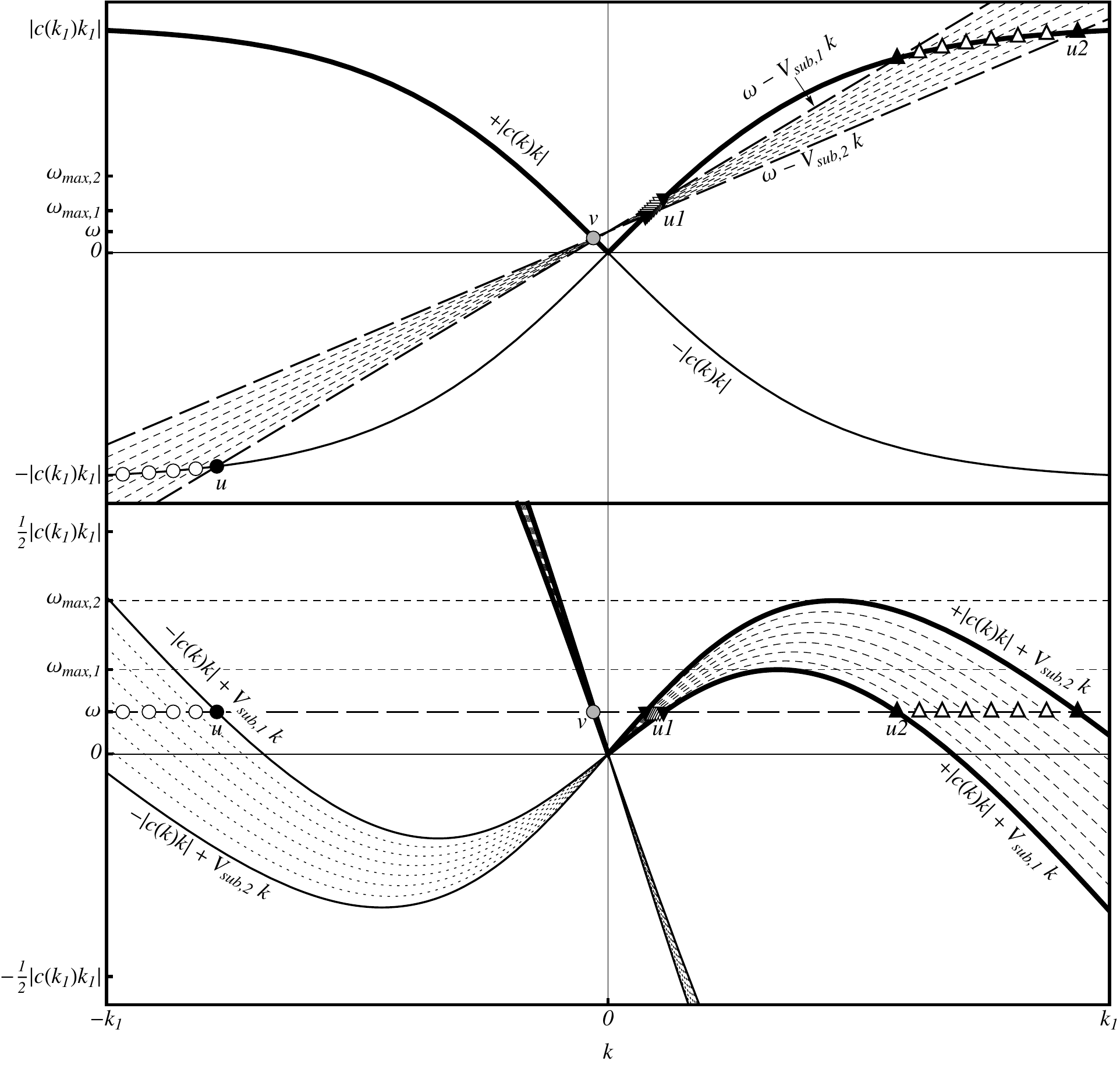}
\caption[\textsc{Dispersion with inhomogeneous $V$ (subsonic to subsonic)}]{\textsc{Dispersion with inhomogeneous $V$ (subsonic to subsonic)}:
Here are plotted subluminal dispersion diagrams from the points of view of both the co-moving frame (upper panel) and the lab frame (lower panel).  Positive- and negative-norm branches are shown as thick and thin curves, respectively, just as in Figs. \ref{fig:Dispersion_comoving-frame}-\ref{fig:dispersion_sub_sup}.  The velocity is assumed to vary between two subsonic values, $V_{\mathrm{sub},1}$ and $V_{\mathrm{sub},2}$.  The critical frequencies $\omega_{\mathrm{max},1}$ and $\omega_{\mathrm{max},2}$ are therefore both non-zero, so that all three velocity regimes described in \S\ref{sub:Inhomogeneous_flow} exist.  It is easiest to pick them out in the lab frame diagram: the critical frequencies are there simply the maxima of the dispersion curves corresponding to the asymptotic velocities.  If $\omega>\omega_{\mathrm{max},2}$, then only the $u$- and $v$-waves exist, so that the situation is analogous to that of Fig. \ref{fig:dispersion_sup_sup}.  If $\omega_{\mathrm{max},1}<\omega<\omega_{\mathrm{max},2}$, then the $u1$- and $u2$-waves only exist in one of the asymptotic regions, and the situation is similar to that described in Fig. \ref{fig:dispersion_sub_sup}.  If $\omega<\omega_{\mathrm{max},1}$, then the $u1$- and $u2$-waves exist in both asymptotic regions, varying smoothly as $V$ is varied.  They do not degenerate into a single solution, so that there are four independent solutions in total.  The $u1$-, $u2$- and $v$-waves all have positive norm, and can couple with the negative-norm $u$-wave to form Hawking pairs.
\label{fig:dispersion_sub_sub}}
\end{figure}

\begin{figure}
\subfloat{\includegraphics[width=0.45\columnwidth]{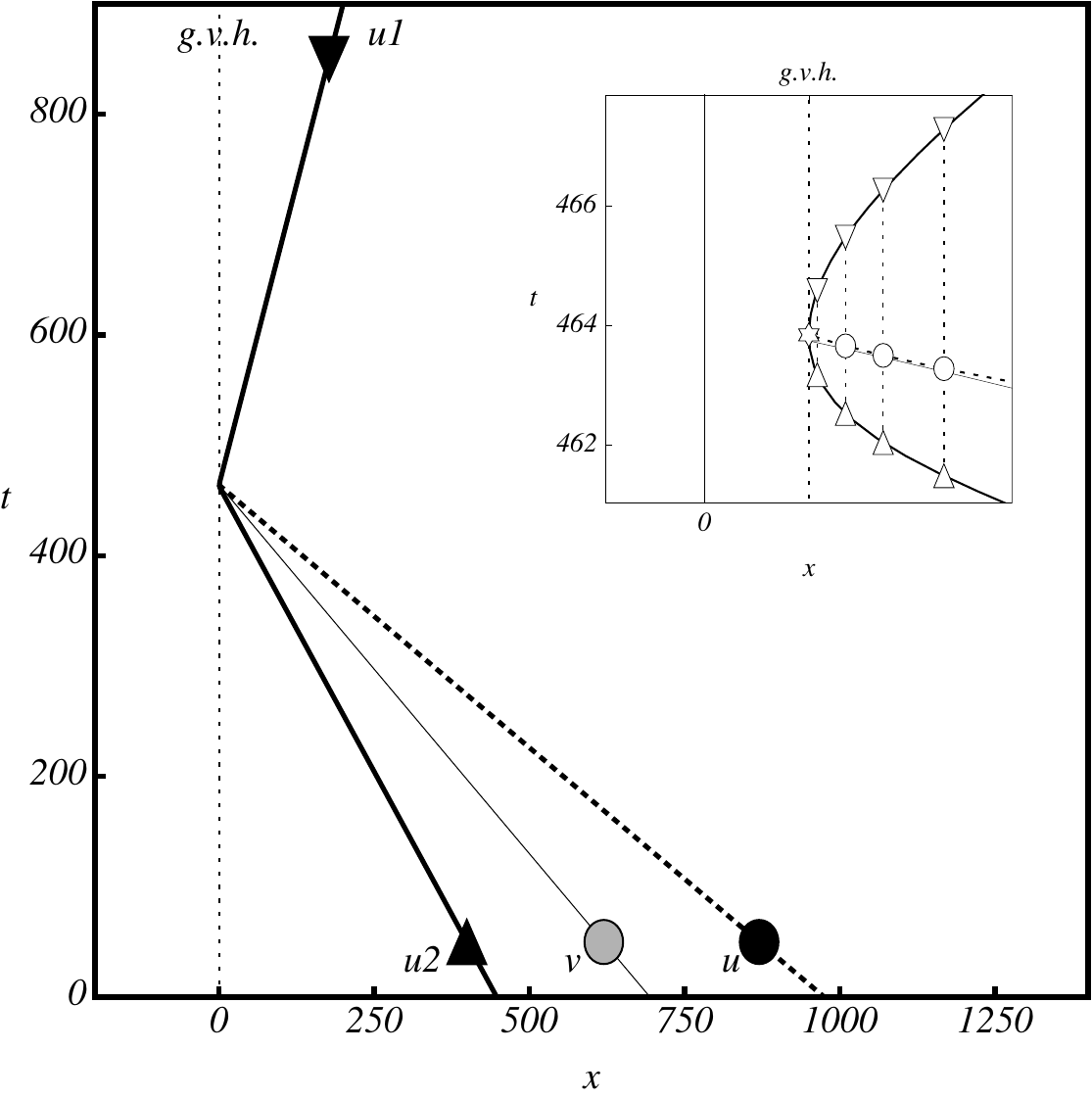}} \subfloat{\includegraphics[width=0.475\columnwidth]{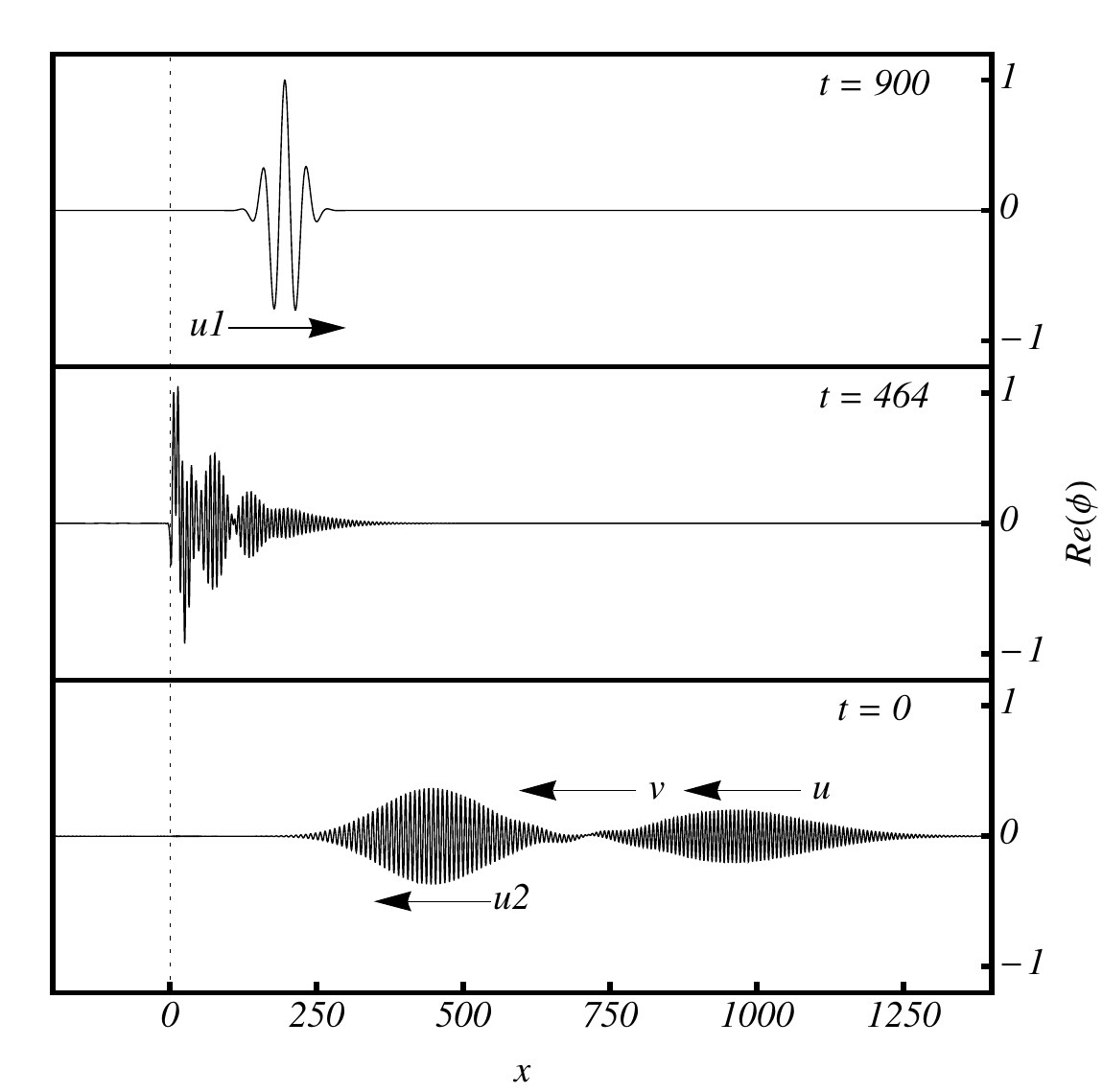}}
\caption[\textsc{$u1$-out mode in black hole configuration}]{\textsc{$u1$-out mode in black hole configuration}:
Here is illustrated the $u1$-out mode in the presence of a black hole horizon with subluminal dispersion.  The left panel shows a space-time diagram of the trajectories of the various wavepackets, which are labelled with the same symbols used in the dispersion diagram of Fig. \ref{fig:dispersion_sub_sup}, on which the evolution can be traced.  Included as an inset is a close-up of the group-velocity horizon region, where the incoming $u2$-wave crosses onto the $u1$-branch at the point marked by a star, ``bouncing'' off the group-velocity horizon.  Note that the simple geometric picture provided by the dispersion diagram cannot describe coupling into the $u$- and $v$-branches.  (The trajectories of the $u$- and $v$-waves are almost equal near the group-velocity horizon, making them difficult to distinguish).  In the right panel are shown the results of numerical wavepacket propagation at three representative times: the asymptotic past, the interaction with the near-horizon region, and the asymptotic future.  This was calculated by specifying the outgoing $u1$-wave and propagating it backwards in time.  (For these calculations was used the dimensionless velocity and dispersion profiles of \S\ref{sub:Specifying_dispersion_and_flow}: $C^{2}(K)=1-K^{2}$ and $U(X)=(U_{R}+U_{L})/2\,+\,(U_{R}-U_{L})/2\,\tanh(aX)$, with $U_{R}=-0.5$, $U_{L}=-1.5$ and $a=1$.)  While the outgoing $u1$-wave has positive norm, the $u$-wave is the only ingoing wave with negative-norm; so, according to \S\ref{sub:Spontaneous_creation-II}, the norm of the $u$-wave relative to the $u1$-wave gives the creation rate of the outgoing $u1$-wave.
\label{fig:out_mode}}
\end{figure}

\begin{figure}
\subfloat{\includegraphics[width=0.45\columnwidth]{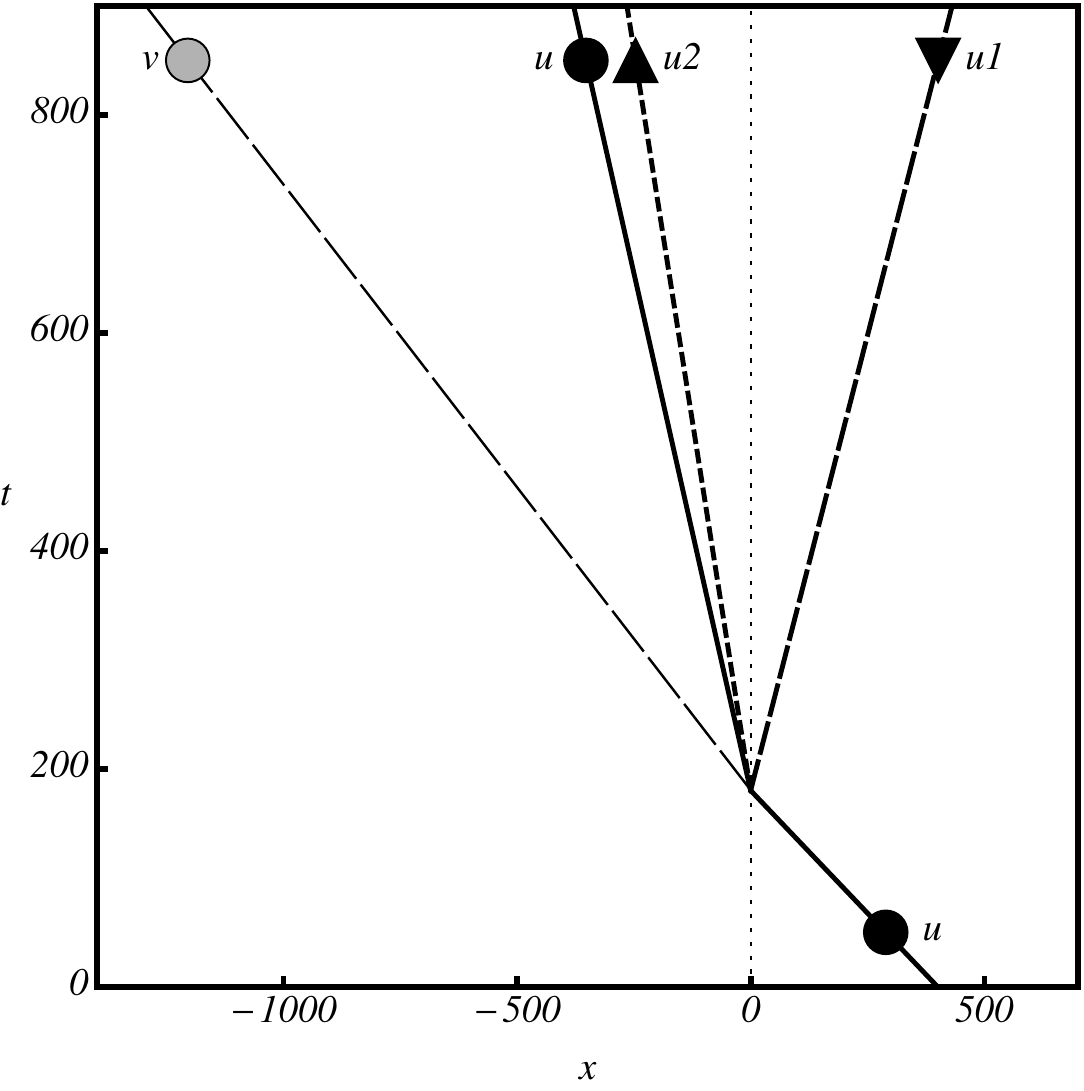}} \subfloat{\includegraphics[width=0.475\columnwidth]{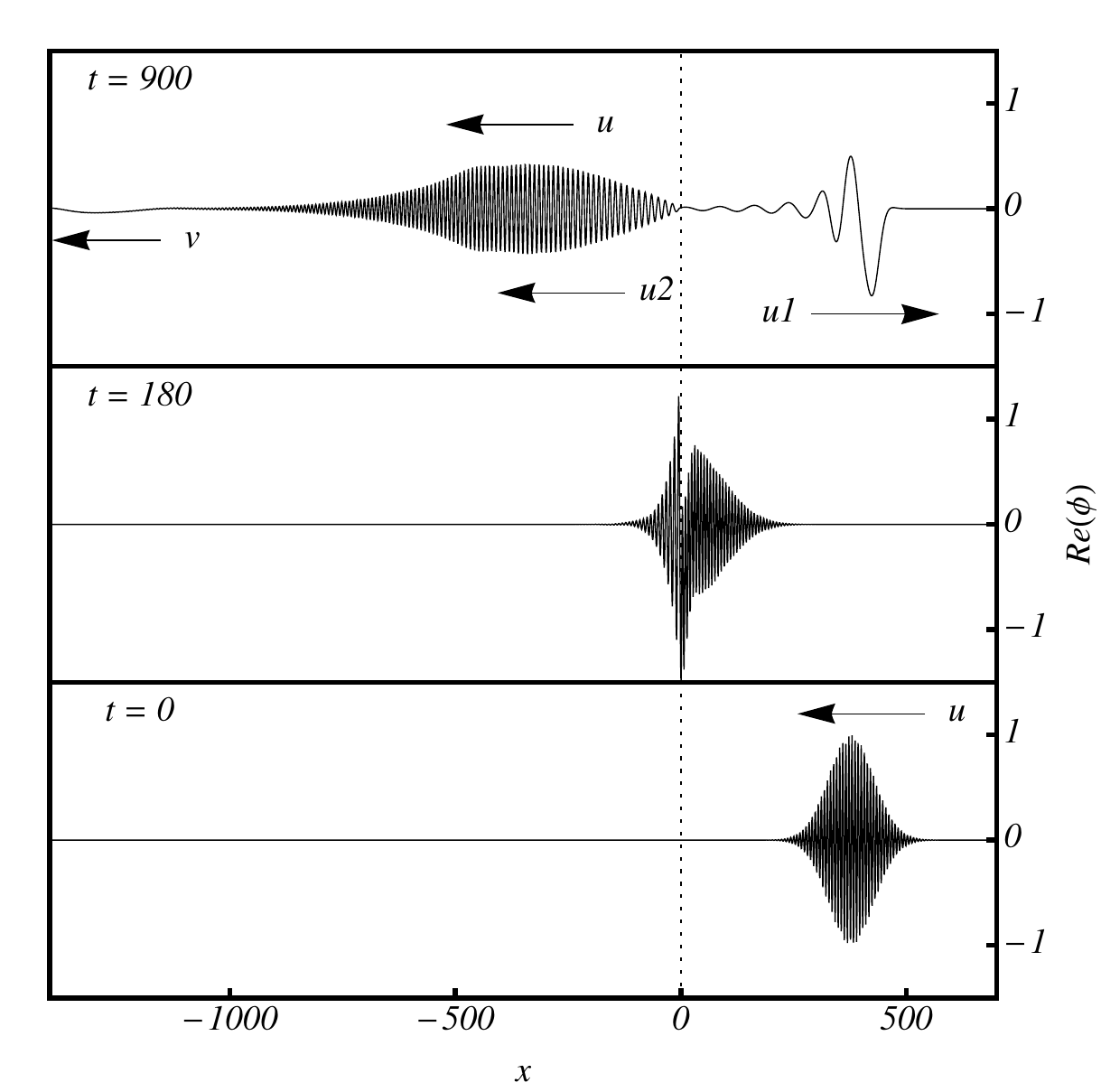}}
\caption[\textsc{$u$-in mode in horizonless configuration}]{\textsc{$u$-in mode in horizonless configuration}:
Here is plotted the $u$-in mode in the absence of a group-velocity horizon with subluminal dispersion.  The flow is entirely subsonic, and the frequency is in the low-horizon regime $\omega<\omega_{\mathrm{max},1}$, so that the $u1$- and $u2$-waves exist everywhere.  The situation is thus similar to that described in Fig. \ref{fig:dispersion_sub_sub}, although the simple geometric picture provided by wavevector evolution on the dispersion diagram cannot account for coupling between the various waves.  The left panel shows a space-time diagram of the trajectories of the various wavepackets, which are labelled with the same symbols used for the various wavevector solutions in Fig. \ref{fig:dispersion_sub_sub}.  The right panel shows the results of numerical wavepacket propagation at three representative times: the asymptotic past, the interaction with the inhomogeneity in $V$, and the future.  Note that the $u2$- and $u$-waves have almost the same group velocity, so that they lie on top of each other in the future and are difficult to distinguish.  (For these calculations was used the dimensionless velocity and dispersion profiles of \S\ref{sub:Specifying_dispersion_and_flow}: $C^{2}(K)=1-K^{2}$ and $U(X)=(U_{R}+U_{L})/2\,+\,(U_{R}-U_{L})/2\,\tanh(aX)$, with $U_{R}=-0.4$, $U_{L}=-0.8$ and $a=1$.)  The $u$-wave is the only one with negative norm; therefore, according to \S\ref{sub:Spontaneous_creation-II}, the norms of the outgoing $u1$-, $u2$- and $v$-waves relative to that of the ingoing $u$-wave give their respective creation rates.
\label{fig:in_mode}}
\end{figure}

$\omega_{\mathrm{max},2}$ must be strictly non-zero for this regime to exist, so that at least one of the asymptotic velocities must be subsonic.  If the other is supersonic, $\omega_{\mathrm{max},1}$ vanishes and this regime extends down to zero frequency, a situation described by Fig. \ref{fig:dispersion_sub_sup}.  On the other hand, if the flow is everywhere subsonic, $\omega_{\mathrm{max},1}$ is also non-zero, and this regime exists in a (possibly narrow) frequency window, as in Fig. \ref{fig:dispersion_sub_sub}.

In the asymptotic region corresponding to $\omega_{\mathrm{max},2}$, there are four wavevector solutions, while in the other asymptotic region there are only two.  The two common solutions are the $u$- and $v$-waves; the extra solutions in one of the regions are $u1$ and $u2$.  Following the evolution of the dispersion diagram as $V$ varies between the asymptotic velocities is particularly instructive; see Fig. \ref{fig:dispersion_sub_sup}.  It is seen that the additional solutions $u1$ and $u2$ vary towards each other, and, at some velocity $V_{\mathrm{gvh}}$, merge into a single wavevector; as $V$ is varied further, they cease to exist.  Mathematically, the $u1$- and $u2$-wavevectors cease to be real, but continue to exist as complex conjugates (as is most clearly seen in the complex $k$-plane diagram of Fig. \ref{fig:WKB_contour_plot}); that is, they become evanescent waves -- one exponentially damped, the other exponentially divergent -- in the opposite asymptotic regime.  Only the exponentially damped evanescent wave is allowed on physical grounds.  Thus are we led to the conclusion that the $u1$- and $u2$-waves do not form two, but rather just \textit{one} additional solution, for they must be combined in just the right combination to cancel the exponentially divergent wave \cite{Macher-Parentani-2009}.  We shall label this single additional solution $u12$.

The physical interpretation of this behaviour is again made clearer by considering the evolution of wavepackets.  In particular, we pay attention to their group velocities, $v_{g}=d\omega/dk=Vk\pm|c(k)k|$, which is given pictorially by the slope of the lab-frame dispersion profile.  As already noted, the $u1$- and $u2$-waves have oppositely-directed lab-frame group velocities: $u1$ is right-moving in the lab frame, while $u2$ is left-moving.  As the magnitude of $V$ increases, the $u1$- and $u2$-wavevectors vary in such a way that the magnitudes of their group velocities decrease, until, precisely at the merging point, \textit{the group velocity vanishes}.  So a wavepacket on either the $u1$- or $u2$-branch, when sent towards the inhomogeneous region, will experience a shift in its wavevector such that it slows down to a standstill at the point where the velocity is $V_{\mathrm{gvh}}$.  It does not stay there, however, as numerical simulations show (see reference \cite{Unruh-1995} and Fig. \ref{fig:out_mode}).  Instead, its wavevector continues to evolve, crossing from one of these branches to the other, and its group velocity changes sign.  (This is why they degenerate into a single branch: waves of one type evolve in time onto those of the other.)  So the wavepacket continues to shift in wavevector, but moves back in the direction from which it came.  The turning point where $V=V_{\mathrm{gvh}}$ is a \textit{group-velocity horizon}, a dispersive analogue of the event horizon beyond which the wavepacket cannot propagate.  It differs in that it does not cause the phase singularities of the dispersionless model, and hence does not give rise to the trans-Planckian problem.  But it differs also in that its position is frequency-dependent, so that it is impossible to speak unambiguously of ``the horizon''.  That said, in the limit $\omega\rightarrow 0$ where the group velocity approaches $c_{0}$, the group-velocity horizon also approaches a limiting point: the point where $V=-c_{0}$.  This low-frequency limiting horizon is the closest equivalent to the event horizon of the dispersionless model, since it ensures the existence of a horizon for all (counter-propagating) wavevectors in a spectral region around $k=0$.

Unfortunately, the simple geometrical picture just described does not capture the details of the full wave equation; in particular, it fails to capture the crucial coupling between the positive-norm $u12$- and the negative-norm $u$-branches.  Hawking radiation can thus occur in $u$-$u12$ pairs (as well as in $u$-$v$ pairs).  Whether the emitted quasiparticle has wavevector $k^{u1}$ or $k^{u2}$ depends on which of these is the \textit{outgoing} wave, which in turn depends on whether the velocity profile describes a black- or white-hole configuration:
\begin{itemize}
\item \textsc{Black hole}: In this case, since $V<0$, we must have $V_{R}>V_{L}$, and the $u1$- and $u2$-waves exist in the right-hand asymptotic region but not in the left.  Then $u1$, with low wavevector, is the outgoing wave.  The $u$-wave is always left-moving, and is thus emitted into the left-hand region.  The Hawking partners are thus emitted in opposite directions from the group-velocity horizon, exactly as in the case of a gravitational black hole.
\item \textsc{White hole}: In this case, $V_{R}<V_{L}$, so the $u1$- and $u2$-waves exist in the \textit{left}-hand region but not in the right.  So it is $u2$, with high wavevector, which is outgoing.  Moreover, it is emitted into the left-hand region along with the $u$-wave: the Hawking partners are emitted in the same direction!
\end{itemize}
As earlier remarked, we restrict our attention mainly to black holes, but the white hole is included here for completeness, since its behaviour in this respect is so different.  Using super- rather than subluminal dispersion, however, does not yield drastically different results, merely switching the group velocities of the waves and hence the directions in which the quasiparticles are emitted \cite{Macher-Parentani-2009}.

\subsubsection{$0<\omega<\omega_{\mathrm{max},1}$: horizonless Hawking radiation}

If \textit{both} asymptotic regions are subsonic (so that no low-frequency limiting horizon exists), then the critical frequencies are ordered thus: $\omega_{\mathrm{max},2}>\omega_{\mathrm{max},1}>0$.  The smaller critical frequency $\omega_{\mathrm{max},1}$ is strictly non-zero, and a third, low-frequency regime is available: $0<\omega<\omega_{\mathrm{max},1}$.  It is shown in Fig. \ref{fig:dispersion_sub_sub}.

The $u1$- and $u2$-waves exist in both asymptotic regions \cite{Finazzi-Parentani-2011-JPCS}.  Following their variation with $V$, their wavevectors never reach the merging point; they experience no group-velocity horizon, and wavepackets in either of the two modes can propagate from one spatial infinity to the other.

However, as we noted previously, the negative-norm $u$-wave exists regardless of the existence of the additional solutions.  There is no need, as in the dispersionless case, to make a transition between subsonic and supersonic flow in order to have two solutions of opposite norm -- they already exist, at a single value of $V$.  So they can couple with each other to produce Hawking radiation, even in the complete absence of a group-velocity horizon \cite{Finazzi-Parentani-2011-JPCS}!

Another novel aspect of this regime is that, there being no group-velocity horizon, $u1$ and $u2$ do not degenerate into a single solution but remain as two completely separate solutions.  They exist in both asymptotic regions, and so each of them can form an incoming wave or an outgoing wave.  So \textit{each} of them can couple to the negative-norm $u$-waves.  There are thus three channels for the radiation, corresponding to the three distinct Hawking pairs $u$-$v$, $u$-$u1$ and $u$-$u2$, each of which will be created according to its own spectrum.  One of the latter two spectra will be a low-frequency continuation of that in the group-velocity horizon regime ($\omega_{\mathrm{max},1}<\omega<\omega_{\mathrm{max},2}$), while the other exists only for $\omega<\omega_{\mathrm{max},1}$ and must vanish outside this region.  Generally, then, there are three Hawking spectra: the two $\omega_{\mathrm{max}}$ values mark the cut-off frequencies of those involving only the counter-propagating waves, whereas the $u$-$v$ pair has no cut-off.

\subsubsection{Form of the modes}

Taking the above considerations into account, the mode operators differ in the various frequency regimes by the total number of independent solutions they contain:
\begin{equation}
\hat{\phi}_{\omega}\left(x\right)=\begin{cases}
\hat{a}_{\omega}^{v}\phi_{\omega}^{v}\left(x\right)+\hat{a}_{\omega}^{u1}\phi_{\omega}^{u1}\left(x\right)+\hat{a}_{\omega}^{u2}\phi_{\omega}^{u2}\left(x\right)+\hat{a}_{-\omega}^{u\dagger}\phi_{-\omega}^{u\star}\left(x\right)\, & 0<\omega<\omega_{\mathrm{max,1}}\\
\hat{a}_{\omega}^{v}\phi_{\omega}^{v}\left(x\right)+\hat{a}_{\omega}^{u12}\phi_{\omega}^{u12}\left(x\right)+\hat{a}_{-\omega}^{u\dagger}\phi_{-\omega}^{u\star}\left(x\right)\, & \omega_{\mathrm{max,1}}<\omega<\omega_{\mathrm{max,2}}\\
\hat{a}_{\omega}^{v}\phi_{\omega}^{v}\left(x\right)+\hat{a}_{-\omega}^{u\dagger}\phi_{-\omega}^{u\star}\left(x\right)\, & \omega>\omega_{\mathrm{max,2}}\end{cases}\,,\label{eq:phi_w_inhomogeneous_subluminal}
\end{equation}
where we recall that $\omega_{\mathrm{max},1}$ or both $\omega_{\mathrm{max},1}$ and $\omega_{\mathrm{max},2}$ may vanish, depending on the nature of the asymptotic velocities.  Note that the modes can be interpreted either as in-modes or out-modes, which exist in one-to-one correspondence.

\section{Hawking radiation in dispersive media\label{sec:Hawking_radiation_in_dispersive_media}}

\subsection{Transforming between the in- and out-bases}

The $\omega$-representation of the stationary modes given in Eq. (\ref{eq:phi_w_inhomogeneous_subluminal}) can be expressed as a dot product between a vector of operators and a vector of modes:
\begin{equation}
\hat{\phi}_{\omega}\left(x\right)=\left(\begin{array}{ccc}
\cdots & \hat{a}_{\omega}^{\nu,\mathrm{in}} & \cdots\end{array}\right)\left(\begin{array}{c}
\vdots\\
\phi_{\omega}^{\nu,\mathrm{in}}\\
\vdots\end{array}\right)=\left(\begin{array}{ccc}
\cdots & \hat{a}_{\omega}^{\mu,\mathrm{out}} & \cdots\end{array}\right)\left(\begin{array}{c}
\vdots\\
\phi_{\omega}^{\mu,\mathrm{out}}\\
\vdots\end{array}\right)\,,\label{eq:mode_dot_product_form}
\end{equation}
where we have explicitly shown both the in- and out-mode representations.  The generic modes and operators $\phi_{\omega}^{\mu}$ and $\hat{a}_{\omega}^{\mu}$ stand for all the possible branches for any single value of $\omega$, including any complex-conjugate modes and Hermitian-conjugate operators that may appear.  (Recall that $\phi_{\omega}^{\mu}=\phi_{-\omega}^{\mu\star}$ and $\hat{a}_{\omega}^{\mu}=\hat{a}_{-\omega}^{\mu\dagger}$; these definitions take care of the negative-norm modes.)  Since both the in- and out-modes form a complete set, they are related via a linear transformation:
\begin{equation}
\left(\begin{array}{c}
\vdots\\
\phi_{\omega}^{\mu,\mathrm{out}}\\
\vdots\end{array}\right)=\mathcal{S}\left(\begin{array}{c}
\vdots\\
\phi_{\omega}^{\nu,\mathrm{in}}\\
\vdots\end{array}\right)\,.\label{eq:in_out_mode_transformation}
\end{equation}
Plugging this into Eq. (\ref{eq:mode_dot_product_form}), we find a corresponding transformation for the operators:
\begin{equation}
\left(\begin{array}{c}
\vdots\\
\hat{a}_{\omega}^{\nu,\mathrm{in}}\\
\vdots\end{array}\right)=\mathcal{S}^{T}\left(\begin{array}{c}
\vdots\\
\hat{a}_{\omega}^{\mu,\mathrm{out}}\\
\vdots\end{array}\right)\,,\label{eq:in_out_operator_transformation}
\end{equation}
where $\mathcal{S}^{T}$ is the transpose of $\mathcal{S}$.

$\mathcal{S}$ is simply the \textit{scattering matrix} which describes how the outgoing modes scatter into ingoing modes when propagated backwards in time into the infinite past.  Since the scalar product is bilinear and the in-modes are themselves orthogonal to each other, these coefficients can be written as a scalar product: $\mathcal{S}_{\mu\nu}=(\phi^{\nu}_{\omega},\phi^{\mu}_{\omega})$.  This also means that the inverse matrix -- which describes the in-modes as linear combinations of out-modes -- is easily obtained through use of the property $(\phi_{2},\phi_{1})=(\phi_{1},\phi_{2})^{\star}$.  In particular, the magnitudes of their elements are identical: $|\mathcal{S}_{\mu\nu}|=|\mathcal{S}^{-1}_{\nu\mu}|$.  In words: the absolute value of the amplitude of the $\nu$-in mode when the $\mu$-out mode is scattered backwards in time is equal to the absolute value of the amplitude of the $\mu$-out mode when the $\nu$-in mode is scattered forwards in time.  This identity can be useful when calculating the radiation spectra, as we shall see in \S\ref{sec:Application_to_a_simple_model}.

If some of the modes scatter into modes with oppositely-signed norm, then, when their amplitudes are substituted in Eq. (\ref{eq:in_out_operator_transformation}), this will result in a mixing of annihilation and creation operators between the in- and out-bases.

\subsection{Spontaneous creation\label{sub:Spontaneous_creation-II}}

As in \S\ref{sub:Spontaneous_creation}, we assume that the field is in the in-vacuum
$\left|0_{\mathrm{in}}\right\rangle $ -- that is, that there are no
incoming particles, and the quantum state is annihilated by all in-mode
annihilation operators.  From Eq. (\ref{eq:in_out_operator_transformation}), the expectation value of \textit{outgoing} particles in a particular mode $\mu$ is
\begin{equation}
\left\langle 0_{\mathrm{in}}\right|\left(\hat{a}_{\omega_{1}}^{\mu,\mathrm{out}}\right)^{\dagger}\hat{a}_{\omega_{2}}^{\mu,\mathrm{out}}\left|0_{\mathrm{in}}\right\rangle = \left\langle 0_{\mathrm{in}}\right|\left\{ \sum_{\nu} \mathcal{S}_{\nu\mu}^{\star} \left( \hat{a}_{\omega_{1}}^{\nu,\mathrm{in}} \right)^{\dagger} \right\} \left\{ \sum_{\nu} \mathcal{S}_{\nu\mu} \, \hat{a}_{\omega_{2}}^{\nu,\mathrm{in}} \right\} \left|0_{\mathrm{in}}\right\rangle \, .
\end{equation}
The subsequent algebra is entirely analogous to that in the derivation of Eq. (\ref{eq:thermal_spectrum}).  As there, the sum of operators that left-multiplies the in-vacuum state may be reduced to its creation operator terms, while the other is simply the Hermitian conjugate of this.  Using the fact that the operators of different states commute, we have simply
\begin{equation}
\left\langle 0_{\mathrm{in}}\right|\left(\hat{a}_{\omega_{1}}^{\mu,\mathrm{out}}\right)^{\dagger}\hat{a}_{\omega_{2}}^{\mu,\mathrm{out}}\left|0_{\mathrm{in}}\right\rangle = \sum_{\nu_{n}} \left| \mathcal{S}_{\nu_{n}\mu} \right| ^{2} \delta \left( \omega_{1} - \omega_{2} \right) \,,
\label{eq:out_expectation_value}
\end{equation}
where $\nu_{n}$ denotes all states of oppositely-signed norm to the state $\mu$.  As before (see \S\ref{sub:Spontaneous_creation}), the occurrence of the $\delta$ function shows that this is a number density.  The spectral flux density -- that is, the number of particles emitted per unit time per unit (angular) frequency interval -- is obtained on division by $2\pi$ \cite{Corley-Jacobson-1996}:
\begin{equation}
\frac{\partial^{2}N}{\partial \omega \, \partial t} = \frac{1}{2\pi} \sum_{\nu_{n}} \left| \mathcal{S}_{\nu_{n}\mu} \right| ^{2} \, .
\label{eq:spectral_flux_density}
\end{equation}
Recalling the definition of $\mathcal{S}$ in Eq. (\ref{eq:in_out_mode_transformation}), the spectral flux density of particles in the mode $\mu$ is proportional to
\begin{itemize}
\item the sum of the squared amplitudes of the opposite-norm in-modes into which the $\mu$-out mode scatters when propagated backwards in time; or
\item the sum of the squared amplitudes of the $\mu$-out mode over all opposite-norm in-modes when they are scattered forwards in time.
\end{itemize}

Equation (\ref{eq:spectral_flux_density}) is the generalization to dispersive media of Eq. (\ref{eq:thermal_spectrum}).  There are three important differences to note.  Firstly, due to the inability to factorize the dispersive wave equation (\ref{eq:acoustic_wave_equation}) into $u$- and $v$-parts as in Eq. (\ref{eq:U_V_wave_equation}), it is now possible for $u$- and
$v$-waves to couple to each other.  This coupling is negligible in the limit of a slowly-varying velocity profile \cite{Schutzhold-Unruh-2008}, but in general it should be taken into account (see references \cite{Carusotto-et-al-2008}, \cite{Recati-et-al-2009} and \cite{Macher-Parentani-2009-ii} for predicted $u$-$v$ coupling in Bose-Einstein condensates).
Secondly, pairs created in dispersive media are not necessarily localised to opposite sides of the horizon.  In the usual black hole case, it is certainly true, for the created $u1$- and $u$-waves have opposite group velocities.  But for a white hole \cite{Mayoral-et-al-2011}, or in the horizonless case where the $u1$- and $u2$-waves exist in either asymptotic region \cite{Finazzi-Parentani-2011-JPCS}, it is possible to create a $u$-$u2$ pair, with the same sign of the group velocity and hence emitted into the same asymptotic region \footnote{Studying the emission from a white hole in BEC, in \cite{Mayoral-et-al-2011} it was found that interference between the correlated $u$- and $u2$-quasiparticles leads to a very pronounced ``checkerboard'' pattern in the density-density correlations.}.  Finally, unlike the dispersionless case, the scattering amplitudes in Eq. (\ref{eq:spectral_flux_density}) are generally unknown, and must be found either approximately or numerically.

\section{Calculation methods\label{sec:Calculation_methods}}

The dispersive framework just developed does not lead to an exact analytical derivation of the Hawking spectrum, as in Eq. (\ref{eq:thermal_spectrum}).  There are no exact solutions akin to (\ref{eq:general_solution}), valid for arbitrary velocity profiles.  The main exact solutions we do have are the plane waves (\ref{eq:normalized_w-modes}) whenever $V$ is constant; finding how these are coupled by inhomogeneities in $V$ requires either numerical or approximate methods.

There are two arbitrary functions in the wave equation (\ref{eq:acoustic_wave_equation}): the velocity profile $V(x)$ and the dispersion profile $c(k)$.  These are defined in the dual spaces of $x$ and $k$; therefore, we should be mindful that, depending on their functional forms, it might be more convenient to work in one space than the other.  If one of these functions appears in the wave equation as a finite polynomial, for example, then it will form an exact finite-order differential equation in the dual space.  Since $V(x)$ is taken to be asymptotically constant, only $c(k)$ can be described as such, leading to an exact differential equation in $x$-space which can be solved numerically.

On the other hand, if $V(x)$ is monotonic, then the appearance of multiple solutions means that the position of a given wavevector is better-defined than the wavevector at a given position.  Moreover, the group-velocity horizon is also better-defined in terms of positions of wavevectors, which vary quite smoothly in contrast to the abrupt transition from real to complex wavevectors that the horizon engenders in $x$-space.  From an analytical point of view, then, the problem is better-suited to the $k$-space representation.

\subsection{Numerical method: differential equation in $x$-space\label{sub:General_velocity_profile}}

We shall first consider the intuitive method of solving the wave equation in position-space through numerical integration, thus yielding exact solutions.  This method is also used in \cite{Corley-Jacobson-1996,Macher-Parentani-2009,Corley-1997,Macher-Parentani-2009-ii,Finazzi-Parentani-2011}.

Assuming a stationary solution $\phi_{\omega}(x)e^{-i\omega t}$, the wave equation (\ref{eq:acoustic_wave_equation}) is transformed into its time-independent form
\begin{equation}
\left(-i\omega+\partial_{x}V\right)\left(-i\omega+V\partial_{x}\right)\phi_{\omega}-c^{2}\left(-i\partial_{x}\right)\partial_{x}^{2}\phi=0\,.
\label{eq:acoustic_steady_state_eqn}
\end{equation}
This is a linear ordinary differential equation.  If the phase velocity $c^{2}(k)$ forms a finite polynomial in $k$, then the differential equation is of finite degree and may be integrated numerically provided appropriate boundary conditions are specified.  Different boundary conditions lead to the various solutions associated with a given frequency.

As always, $V$ is taken to be asymptotically constant.  In the asymptotic regions, then, the normalized plane waves (\ref{eq:normalized_w-modes}) are particular solutions, and the general solution is a linear combination of these:
\begin{equation}
\phi_{\omega}\left(x\right) = \sum_{\mu} c^{\mu} \phi_{\omega}^{\mu}\left(x\right)\,,
\label{eq:sum_of_exps}
\end{equation}
where $\mu$ is a generic label for the various plane-wave solutions, and the $c^{\mu}$ are arbitrary complex constants.  Some of the $\phi_{\omega}^{\mu}$ may have negative norm, and can always be written as the complex conjugate $\phi_{-\omega}^{\mu\star}$ to emphasize this.  Others may have complex wavevectors, and since the dispersion relation is real, these always occur in complex-conjugate pairs.  Only that wavevector \textit{decreasing} towards infinity is physically allowed, so only it can appear with non-zero coefficient in the general solution (\ref{eq:sum_of_exps}).  Note that, since complex wavevectors do not correspond to propagating waves, their ``norm'' is physically meaningless (it is actually zero \cite{Leonhardt-et-al-2003}), and their amplitudes do not enter into equations of norm conservation.

\begin{figure}
\includegraphics[width=0.8\columnwidth]{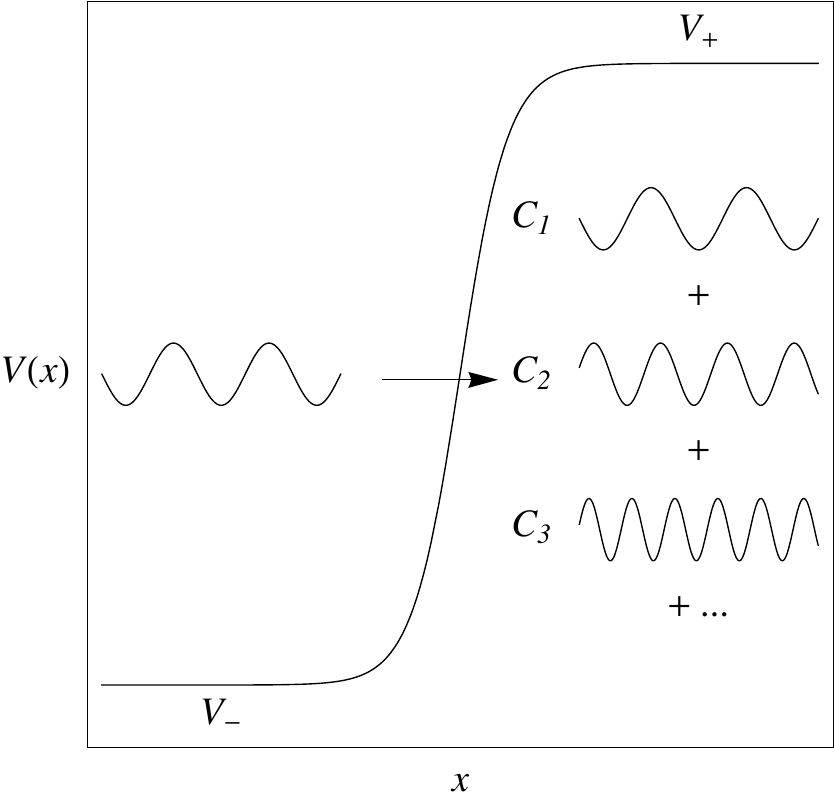}
\caption[\textsc{Integration of stationary solution}]{\textsc{Integration of stationary solution}: In one asymptotic region,
the solution is chosen to be a particular plane wave. Eq. (\ref{eq:acoustic_steady_state_eqn})
is then integrated through to the other asymptotic region, where the solution
is a linear combination of plane waves (\ref{eq:sum_of_exps}). This is done for all possible
plane wave solutions in the initial region. Thus we find the transfer matrix $\mathcal{T}$ appearing
in Eqs. (\ref{eq:left_right_mode_transformation}) and (\ref{eq:left_right_coefficient_transformation}).
\label{fig:Integration-of-Stationary-Solution}}
\end{figure}

The general solution (\ref{eq:sum_of_exps}) applies in each of the asymptotic regions \textit{separately}, with modes and coefficients particular to their own region.  The left- and right-hand solutions are not independent, but \textit{linearly} related through the linear differential equation (\ref{eq:acoustic_steady_state_eqn}).  We may think of the set of plane-wave solutions in one region as a basis of solutions, and the velocity profile as a linear operator, transforming between the left- and right-bases.  This is entirely analogous to the linear transformation between the in- and out-bases, Eq. (\ref{eq:in_out_mode_transformation}), except that the spatial form of the left- and right-modes makes them better-suited to solution by numerical integration.  Specifying boundary conditions appropriate to a single plane wave in one asymptotic region, Eq. (\ref{eq:acoustic_steady_state_eqn}) can be numerically integrated through to the other region, and its solution there fitted to the sum of plane waves (\ref{eq:sum_of_exps}); this is illustrated pictorially in Fig. \ref{fig:Integration-of-Stationary-Solution}.  Working through all possible initial plane waves, the transformation matrix -- a \textit{transfer matrix}, linking two regions in space -- is built up column by column.  This yields the matrix $\mathcal{T}$ such that

\begin{equation}
\left(\begin{array}{c}
\vdots\\
\phi_{\omega}^{\mu,\mathrm{left}}\\
\vdots\end{array}\right)\,=\,\mathcal{T}\left(\begin{array}{c}
\vdots\\
\phi_{\omega}^{\nu,\mathrm{right}}\\
\vdots\end{array}\right)\,,\label{eq:left_right_mode_transformation}
\end{equation}
while for the coefficients we have
\begin{equation}
\left(\begin{array}{c}
\vdots\\
c^{\mu,\mathrm{left}}\\
\vdots\end{array}\right)\,=\,\mathcal{T}^{T}\left(\begin{array}{c}
\vdots\\
c^{\nu,\mathrm{right}}\\
\vdots\end{array}\right)\,,\label{eq:left_right_coefficient_transformation}
\end{equation}
where $\mathcal{T}^{T}$ is the transpose of $\mathcal{T}$.

Any in- or out-mode can be solved for in terms of its scattering coefficients using Eq. (\ref{eq:left_right_coefficient_transformation}).  We simply set to zero all those coefficients corresponding to ingoing or outgoing plane waves other than the one we are considering, as well as the coefficients of any exponentially diverging solutions
\footnote{If the divergence of a complex wavevector is too strong, it may swamp the numerics and make the calculated solution unreliable \cite{Macher-Parentani-2009}.  The solution can be tested by checking whether the norm is conserved, and whether the solution is significantly affected by increasing the accuracy and precision of the numerical integration.}.
There remains the coefficient of the single incoming or outgoing wave, which is set to unity to normalize the complete mode.  Plugging these constraints into Eq. (\ref{eq:left_right_coefficient_transformation}) yields the scattering coefficients of the mode.  The norm of each individual scattered wavevector is simply the squared modulus of its scattering coefficient, with the inclusion of a minus sign for those with negative norm.  Conservation of norm implies that the sum of the norms of the scattered waves should equal unity, the norm of the single ingoing or outgoing wave; this is very useful as a numerical check.

\subsection{Analytic method: the step-discontinuous limit\label{sub:Discontinuous_velocity_profile}}
While numerical integration as just described can be applied to an arbitrary velocity profile, there exist two limiting regimes in which instructive approximation methods can be applied.  These regimes relate to the steepness of the velocity profile in relation to the fundamental length scale determined by the dispersion.  If the velocity changes over a distance much longer than this dispersive length scale, then the system is in the slowly-varying regime of geometrical optics; an analytical method applicable to this regime is described in detail in \S\ref{sub:WKB-type-analysis}.  On the other hand, if the velocity changes over a distance significantly shorter than the dispersive length scale, the waves of the system are unable to resolve the details of the change, so that shortening the transition region even further will have no significant effect.  In this regime, the system is well-approximated by a velocity profile with a step discontinuity \footnote{A detailed study in \cite{Finazzi-Parentani-2012-arXiv} shows that the relevant dispersive scale to be used in such a comparison is actually the cut-off frequency $\omega_{\mathrm{max}}$, and not -- as might be expected -- the wavevector at which the dispersion profile deviates from the dispersionless model.}.  This type of profile is studied in references \cite{Corley-1997, Recati-et-al-2009,Mayoral-et-al-2011,Finazzi-Parentani-2012-arXiv}.

The step-discontinuous velocity profile is exactly solved by a sum (\ref{eq:sum_of_exps}) of plane waves (and non-divergent evanescent waves) in each of its constant-velocity regions, and these solutions must be matched appropriately at each discontinuity.  The matching conditions are determined by the differential equation (\ref{eq:acoustic_steady_state_eqn}): as the discontinuous limit is approached by an ever-steepening profile, Eq. (\ref{eq:acoustic_steady_state_eqn}) must be satisfied throughout the steepening process.  So long as the system is dispersive so that Eq. (\ref{eq:acoustic_steady_state_eqn}) contains spatial derivatives of order greater than $2$, this limit is well-defined.  Suppose for definiteness that the differential equation is of order $2n$, the phase velocity being described by the order-$(2n-2)$ polynomial
\begin{equation}
c^{2}(k)=\sum_{j=0}^{n-1} C_{j} k^{2j} \,.
\label{eq:phase_velocity_polynomial}
\end{equation}
Then the stationary field $\phi$ and its first $2n-2$ spatial derivatives are continuous everywhere, while at a point of discontinuity where $V$ changes abruptly from $V^{-}$ to $V^{+}$, the $(2n-1)$-th derivative of $\phi$ has the discontinuity
\begin{equation}
\left(\partial_{x}^{2n-1}\phi\right)^{+}-\left(\partial_{x}^{2n-1}\phi\right)^{-} = \frac{(-1)^{n}}{C_{n-1}}\left[ i\omega\phi - (V^{+}+V^{-})\partial_{x}\phi \right] \left( V^{+}-V^{-} \right) \, .
\label{eq:discontinuity_condition}
\end{equation}
Since $\phi$ and $\partial_{x}\phi$ are assumed continuous everywhere, there is no ambiguity in their values at the point of discontinuity.  These conditions -- the continuity of $\phi$ through $\partial_{x}^{2n-2}\phi$ and the discontinuity condition (\ref{eq:discontinuity_condition}) for $\partial_{x}^{2n-1}\phi$ -- provide $2n$ independent relations between the $2n$ unknown amplitudes of the ingoing or outgoing waves (relative to the amplitude of the single outgoing or ingoing wave, typically set to unity).  This is then a solvable linear system.  Since there is no need to perform numerical integration or to build up the transfer matrix $\mathcal{T}$ by considering one asymptotic plane wave at a time, the solution can be found much faster than in the general case of \S\ref{sub:General_velocity_profile}.

Condition (\ref{eq:discontinuity_condition}) is particular to the system obeying wave equation (\ref{eq:acoustic_wave_equation}), though other systems will admit analogous discontinuity conditions \footnote{Reference \cite{Recati-et-al-2009} gives a similar such condition in a one-dimensional Bose-Einstein condensate, while \cite{Robertson-thesis} works out condition (\ref{eq:discontinuity_condition}) in detail and gives a similar derivation for nonlinear light interaction in optical fibres.}.  What is noteworthy is that, in the presence of dispersion, this infinite-steepness limit exists and can be calculated, for the reason outlined above: dispersion introduces a fundamental length scale beneath which the details of the velocity transition cannot be resolved.  The Hawking flux must therefore approach a finite limit in the limit of infinite steepness.  This is in stark contrast to the dispersionless case, where the Hawking temperature is simply proportional to the steepness at the horizon, and would thus diverge in the discontinuous limit.  From a physical viewpoint, the absence of dispersion allows infinite resolution, so that a decrease of the length of the transition region will always be ``visible'' to the waves of the system, and will therefore affect the Hawking flux.  This is yet another rewording of the trans-Planckian problem, and the existence of the infinite-steepness limit is another manifestation of the regularization of this problem brought about by dispersion.

\subsection{Analytical method: the slowly-varying limit\label{sub:WKB-type-analysis}}

The earliest analytical descriptions of Hawking radiation in dispersive media \cite{Brout-et-al-1995,Corley-1998} tackle the problem in $k$-space, where horizons are better-defined, and transform back to $x$-space via the saddle-point approximation.  The solution in $k$-space is approximated by linearizing the velocity profile in the vicinity of the horizon.  This is a severe limitation of the method, for even if the geometrical optics approximation holds, it is not necessarily true that higher-order derivatives at the horizon are negligible.  Other treatments also rely on linearization at the horizon \cite{Himemoto-Tanaka-2000,Saida-Sakagami-2000}, or on a specific form for $V$ \cite{Schutzhold-Unruh-2008}.

Here, an alternative theory is presented (as also in references \cite{Robertson-thesis,Leonhardt-Robertson-2012}).  Inspired by earlier treatments, it is developed mainly in $k$-space, and uses the saddle-point approximation to transform back to $x$-space.  But $V$ is not linearized; it is completely general, except that it is assumed to be slowly-varying in such a way that a WKB-type approximation is justified.

\subsubsection{Solution in $k$-space
\label{sub:Solution_in_k_space}}

We take the Fourier transform of Eq. (\ref{eq:acoustic_steady_state_eqn}) by making the substitutions $\phi \rightarrow \widetilde{\phi}$, $x \rightarrow i\partial_{k}$ and $\partial_{x} \rightarrow ik$.  Rearranging slightly, we find that the wave equation in $k$-space takes the form
\begin{equation}
\left[ \left( \frac{\omega}{k} - V(i\partial_{k}) \right)^2 - c^{2} (k) \right] \left(k \widetilde{\phi} \right) \, = \, 0 \, .
\label{eq:wave_eqn_mom}
\end{equation}
The velocity profile can be characterized by a scale, and we indicate this by including a parameter
$\epsilon$ in its argument: $V(i\partial_{k}) \rightarrow V(i\epsilon \partial_{k})$; this is later included in the definition of $V$ by formally setting $\epsilon=1$.  The purpose of $\epsilon$ is as a measure of the steepness of $V$, so that a ``slowly-varying'' velocity profile can be defined as that remaining in the limit of small $\epsilon$
\footnote{In a mathematical sense, $\epsilon$ plays here a role analogous to that played by $\hbar$ in the WKB approximation in Quantum Mechanics.  Just as the WKB approximation works well in the semi-classical limit, the present approximation works well in the geometrical optics limit; both cases refer to a ``slowly-varying'' background.}.
We make the ansatz $k\widetilde{\phi} = \exp \left( \widetilde{\varphi} \right)$, where $\widetilde{\varphi}$ may be Laurent-expanded in powers of $\epsilon$: $\widetilde{\varphi}=\epsilon^{-1}\left(\widetilde{\varphi}_{0} + \epsilon \widetilde{\varphi}_{1} + \epsilon^{2} \widetilde{\varphi}_{2} + ...\right)$.  
Plugging this ansatz into the left-hand side of Eq. (\ref{eq:wave_eqn_mom}) yields a Taylor series in $\epsilon$:
\begin{eqnarray*}
\left[ \left( \frac{\omega}{k} - V(i \epsilon \partial_{k}) \right)^2 - c^{2} (k) \right] \left(k \widetilde{\phi} \right) \, & = & \, \Bigg[ \left[ \left( \frac{\omega}{k} - V(i\partial_{k}\widetilde{\varphi}_{0}) \right)^{2} - c^{2}(k) \right] \\
& & +\, \epsilon \left[ \partial_{k}\widetilde{\varphi}_{1} + \frac{1}{2} \partial_{k} \ln \left( V^{\prime}(i\partial_{k}\widetilde{\varphi}_{0}) c(k) \right) \right] + \ldots \Bigg] \left( k \widetilde{\phi} \right) \,.
\end{eqnarray*}
Equation (\ref{eq:wave_eqn_mom}) is solved by setting the coefficient of each power of $\epsilon$ to zero, yielding an infinite sequence of coupled equations for the coefficients $\widetilde{\varphi}_{j}$.  We implement the slowly-varying approximation for $V$ by truncating the Taylor series at the first power of $\epsilon$.  This leaves only the two lowest-order equations:
\begin{eqnarray}
\left( \frac{\omega}{k} - V(i\partial_{k}\widetilde{\varphi}_{0}) \right)^{2} - c^{2}(k) & = & 0 \, ,
\label{eq:mom_zeroth_eqn} \\
\partial_{k}\widetilde{\varphi}_{1} + \frac{1}{2} \partial_{k} \ln \left( V^{\prime}(i\partial_{k}\widetilde{\varphi}_{0}) c(k) \right) & = & 0 \, .
\label{eq:mom_first_eqn}
\end{eqnarray}

Equation (\ref{eq:mom_zeroth_eqn}) is precisely the dispersion relation (\ref{eq:acoustic_flow_dispersion}), except that the flow velocity $V$ is not assumed to be constant.  The argument of $V$ is the position $x$, but imposing Eq. (\ref{eq:mom_zeroth_eqn}) selects a particular position for each value of $k$: that position at which $k$ is a solution of the dispersion relation for the local value of $V$.  Let us denote this wavevector-dependent position by $\chi(k)$, and note its distinctness from the position \textit{variable} $x$.  Explicitly, $\chi(k)$ is defined such that
\begin{equation}
\left( \frac{\omega}{k} - V(\chi(k)) \right) ^{2} - c^{2}(k) = 0\,,
\label{eq:position_dependent_dispersion_relation}
\end{equation}
or, inverting this relation,
\begin{equation}
\chi(k) = V^{-1}\left(\frac{\omega}{k} - c(k)\right)\,,
\label{eq:inverted_dispersion_relation}
\end{equation}
Note that we have chosen the sign of the square root of $c^{2}(k)$ to correspond with the counter-propagating $u$-modes; changing this sign gives another, independent, function corresponding to the $v$-modes, but we shall not consider this here \footnote{The independence of the position functions corresponding to the $u$- and $v$-modes shows that, to this level of approximation, these modes are decoupled from each other.  So, while we could perform the following analysis for $v$-modes, we would not predict any scattering from $u$- into $v$-modes or vice versa.  This corroborates the fact that the $u$-$v$ coupling becomes negligible in the slowly-varying regime \cite{Schutzhold-Unruh-2008}.}.  Equation (\ref{eq:mom_zeroth_eqn}) is solved by setting $i\partial_{k}\widetilde{\varphi}_{0}$ equal to $\chi(k)$, or, integrating,
\begin{equation}
\widetilde{\varphi}_{0} = -i \int^{k} \chi(k^{\prime}) dk^{\prime} + C_{0}\,,
\label{eq:varphi0_integral}
\end{equation}
where the lower limit of integration is omitted as it can be incorporated into the constant of integration $C_{0}$.

The second of our sequence of equations, Eq. (\ref{eq:mom_first_eqn}), may now be solved by substituting $\chi(k)$ for $i\partial_{k}\widetilde{\varphi}_{0}$, and it is easily seen that
\begin{equation}
\widetilde{\varphi}_{1} = -\frac{1}{2} \ln \left( V^{\prime}(\chi(k)) c(k) \right) + C_{1} \,.
\label{eq:varphi1}
\end{equation}
Plugging the solutions for $\widetilde{\varphi}_{0}$ and $\widetilde{\varphi}_{1}$ into the original ansatz, we find the approximate $k$-space solution
\begin{equation}
\widetilde{\phi} \left( k \right) \, \approx \, \frac{\Phi_{0}}{k\, \sqrt{V^{\prime}(\chi(k)) \, c(k)}} \exp \left( -i \int^{k} \chi(k^{\prime}) dk^{\prime} \right) \, ,
\label{eq:soln_mom}
\end{equation}
where the constants of integration $C_{0}$ and $C_{1}$ are incorporated into the overall prefactor $\Phi_{0}$.

The economical value of the $k$-space representation is demonstrated by Eq. (\ref{eq:soln_mom}).  Whereas, in the $x$-space representation, the number of independent solutions can (depending on the dispersion relation) be arbitrarily large, in $k$-space we have found only two independent solutions, corresponding to the counter- and co-propagating modes; neglecting the latter, the solution is uniquely determined up to an unimportant multiplicative prefactor.  The question inevitably arises: how can these dramatically different dimensionalities be reconciled, when the two representations ostensibly describe one and the same system?  The answer is that, in Fourier transforming back to $x$-space, the $k$-space solution (\ref{eq:soln_mom}) can yield different $x$-space solutions depending on the integration contour taken in the complex $k$-plane \cite{Brout-et-al-1995}.  Just as the precise linear combination of $x$-space solutions is selected by appropriate boundary conditions, so the integration contour in $k$-space is also selected by appropriate boundary conditions; but the latter -- especially in the presence of a horizon -- is much easier to identify.

\subsubsection{Transforming to $x$-space: Saddle-point approximation
\label{sub:Transforming_to_z_space}}

Inverse Fourier transforming Eq. (\ref{eq:soln_mom}) to find the $x$-space solution,
\begin{eqnarray}
\phi \left( x \right) & = & \int_{-\infty}^{+\infty} \widetilde{\phi} \left( k \right) e^{i k x} \, dk \nonumber \\
& \approx & \int_{-\infty}^{+\infty} \frac{\Phi_{0}}{k\,\sqrt{V^{\prime}(\chi(k)) \, c(k)}} \, \exp \left( i x k - i \int^{k} \chi(k^{\prime}) \, dk^{\prime} \right) dk \nonumber \\
& = & \int_{-\infty}^{+\infty} \mathcal{A}(k) \, \exp \left( i\theta(k) \right) \, dk \, ,
\label{eq:inverse_Fourier_transform}
\end{eqnarray}
where we have defined
\begin{alignat}{1}
\mathcal{A}(k)=\frac{\Phi_{0}}{k\sqrt{V^{\prime}(\chi(k))\,c(k)}}\,,\,\,\,\,\,\, & \theta(k)= x k - \int^{k}\chi(k^{\prime})\,dk^{\prime}\,.
\label{eq:amplitude_phase}
\end{alignat}
Its integrand taking the form of an amplitude multiplied by a phase factor, the integral in Eq. (\ref{eq:inverse_Fourier_transform}) seems to lend itself well to the saddle-point approximation.  This observation is corroborated by noticing that, keeping the scale parameter $\epsilon$, it would appear only as a factor of $1/\epsilon$ multiplying the phase $\theta(k)$ (with $x$ rescaled by $\epsilon$ so that $\epsilon x$ remains constant).  So, in the limit of small $\epsilon$ -- the limit in which our approximation for $\widetilde{\phi}(k)$ is valid -- the saddle-point approximation also becomes a valid means of evaluating the inverse Fourier transform in the asymptotic regions.

The saddle points of the phase $\theta(k)$ are simply those points in the complex $k$-plane at which the derivative $\partial_{k}\theta(k)$ vanishes.  From the second of
Eqs. (\ref{eq:amplitude_phase}), we see that such points occur where $x - \chi(k)=0$; recalling the definition of $\chi(k)$ in Eqs. (\ref{eq:position_dependent_dispersion_relation}) and (\ref{eq:inverted_dispersion_relation}), this condition simply inverts the relationship between position and wavevector so that, for a given $x$, the saddle points lie at those values of $k$ which are solutions of the dispersion relation for the local value of $V$.  As $x$ is varied from one asymptotic region to the other, the saddle points trace out all values of $k$ which solve the dispersion relation (\ref{eq:position_dependent_dispersion_relation}) for real $x$; in particular, they vary continuously between wavevector solutions in the left-hand region and those in the right, at which extremes $\chi(k)$ is singular.  We shall denote the various wavevector solutions by $\kappa^{\mu}(x)$, so as not to confuse them with the wavevector \textit{variable} $k$.

Consider the example of Figure \ref{fig:WKB_contour_plot}, which shows the imaginary part of $\chi(k)$ for the example velocity and dispersion profiles specified later in \S\ref{sub:Specifying_dispersion_and_flow}.  The locus of saddle points (equivalently, of wavevector solutions) for real $x$ is shown as blue dashed lines.  This forms two disconnected structures.  On the right, with real positive $k$, are the positive-norm solutions, intersected by complex-conjugate wavevector solutions.  This crossbow-like structure is symptomatic of the presence of a group-velocity horizon: at the point of intersection, where the group velocity vanishes, two real wavevector solutions merge and form complex-conjugate wavevector solutions beyond the horizon.  The real wavevectors on either side of the group-velocity-horizon point are precisely the $u1$- and $u2$-branches.  On the left, with real negative $k$, lies the $u$-branch, which does not experience a horizon and therefore exists throughout the real space, in agreement with \S\ref{sub:Homogeneous flow}.

The saddle-point approximation treats the integral of an amplitude times a phase factor -- as in Eq. (\ref{eq:inverse_Fourier_transform}) -- as a sum of contributions from the saddle points of the phase.  In the vicinity of a saddle point, the phase is Taylor-expanded to second order:
\begin{equation}
\theta(k) \approx \theta(\kappa^{\mu}(x)) + \frac{1}{2} \partial_{k}^{2} \theta(\kappa^{\mu}(x)) \left( k - \kappa^{\mu}(x) \right)^{2} \, .
\label{eq:phase_expansion}
\end{equation}
Substituting back into the integrand of Eq. (\ref{eq:inverse_Fourier_transform}), and assuming that the amplitude varies slowly enough so that it can be replaced by the constant value $\mathcal{A}(\kappa^{\mu}(x))$, the integrand takes the simple Gaussian form:
\begin{equation}
\mathcal{A}(k) \exp \left(i\theta(k)\right) \approx \, \mathcal{A}(\kappa^{\mu}(x))\exp\left(i\theta(\kappa^{\mu}(x))\right) \exp \left(i\frac{1}{2}\partial_{k}^{2} \theta(\kappa^{\mu}(x)) \left( k - \kappa^{\mu}(x) \right)^{2} \right) \, .
\label{eq:saddlepoint_Gaussian_approx}
\end{equation}
Upon integration, this is easily evaluated; but before we can do so, we must check that the resulting integral is convergent.  It will only be so if the integration contour can be deformed to pass through the saddle point in its direction of steepest descent, i.e., the direction in which the exponent $i\frac{1}{2}\partial_{k}^{2} \theta(\kappa^{\mu}(x)) \left( k - \kappa^{\mu}(x) \right)^{2}$ is negative.  In this direction, the integrand is indeed a convergent Gaussian function, at least in the vicinity of the saddle point; and, so long as the magnitude of the phase increases fast enough (equivalently, so long as $\epsilon$ is small enough, or the velocity profile is sufficiently slowly-varying), then the Gaussian function converges to zero quickly so that only the behaviour in the vicinity of the saddle point is important in the evaluation of the integral.  (Retrospectively, this justifies replacing the amplitude $\mathcal{A}(k)$ with its value $\mathcal{A}(\kappa^{\mu}(x))$ at the saddle point.)  Then the integral over this relatively small region can be treated as an integral over the entire real line, the differences far from the saddle point having no significant effect.  So a single saddle point, for which the contour can be deformed to lie along the direction of steepest descent, contributes approximately
\begin{multline}
\mathcal{A}(\kappa^{\mu}(x))\exp\left(i\theta(\kappa^{\mu}(x))\right) \int_{-\infty}^{+\infty} \exp \left(i\frac{1}{2}\partial_{k}^{2} \theta(\kappa^{\mu}(x)) \left( k - \kappa^{\mu}(x) \right)^{2} \right) \, dk \\
= \, \mathcal{A}(\kappa^{\mu}(x))\exp\left(i\,\theta(\kappa^{\mu}(x))\right) \sqrt{\frac{2\pi}{i\,\partial_{k}^{2}\theta(\kappa^{\mu}(x))}}
\label{eq:saddle_point_contribution}
\end{multline}
to the integral of Eq. (\ref{eq:inverse_Fourier_transform}).  If the integration contour passes through several such saddle points, these contributions are added together.

From the definition of $\chi(k)$ in Eqs. (\ref{eq:position_dependent_dispersion_relation}) and (\ref{eq:inverted_dispersion_relation}), and after some algebraic manipulation, the second derivative of the phase is given by
\begin{equation}
\partial_{k}^{2} \theta(\kappa^{\mu}(x)) = -\partial_{k} \chi (\kappa^{\mu}(x)) = \frac{v_{g}(x,\kappa^{\mu}(x))}{\kappa^{\mu}(x) V^{\prime} (x)} \, .
\label{eq:phase_second_derivative}
\end{equation}
Here, $v_{g}(x,k)=V(x)+\partial_{k}(c(k)k)$ is the group velocity of the wavevector $k$ at the position $x$.  Applying directly to Eq. (\ref{eq:inverse_Fourier_transform}), substituting the saddle point contributions (\ref{eq:saddle_point_contribution}) and replacing the amplitudes and phases of Eqs. (\ref{eq:amplitude_phase}), we find
\begin{eqnarray}
\phi \left( x \right) & \approx & \sum_{\mu} \mathcal{A}(\kappa^{\mu}(x)) \sqrt{\frac{2\pi}{i \, \partial_{k}^{2} \theta(\kappa^{\mu}(x))}} \, \exp \left( i \, \theta(\kappa^{\mu}(x)) \right) \nonumber \\
& = & \Phi_{0}\sqrt{2\pi} \sum_{\mu} e^{\pm i\frac{\pi}{4}} \frac{1}{\sqrt{|c(\kappa^{\mu}(x)) \, \kappa^{\mu}(x) \, v_{g}(x,\kappa^{\mu}(x))|}} \, \exp \left( i \, x \, \kappa^{\mu}(x) - i \int^{\kappa^{\mu}(x)} \chi(k^{\prime}) \, dk^{\prime} \right) \, \nonumber \\
& \rightarrow & \Phi_{0} \, \frac{\pi}{\sqrt{2}} \, \sum_{\mu} \, \phi^{\mu}_{\omega}\left(x\right) \, \exp\left(- i \int^{\kappa^{\mu}(\pm\infty)} \chi(k^{\prime}) \, dk^{\prime} \, \pm \, i\frac{\pi}{4}\right) \,. \label{eq:pos_soln_sum}
\end{eqnarray}
In the last line, the solution has been restricted to the asymptotic constant-velocity regions, $x\rightarrow \pm\infty$.  There, the solutions $\kappa^{\mu}$ become independent of $x$, and are equal to the wavevector solutions for the given asymptotic flow velocities.  The various plane wave solutions and their $k$-dependent amplitudes are found to be, up to a constant prefactor, the $\omega$-normalized plane waves of Eq. (\ref{eq:normalized_w-modes}).  So the solution in the asymptotic regions is, as one would expect, just a linear combination of these various plane wave solutions.  What is remarkable about Eq. (\ref{eq:pos_soln_sum}), though, is that the relative coefficients of these plane wave solutions are not arbitrary: they are determined by the phase integral of $\chi(k)$, which in turn encodes the effect of the full velocity profile on the scattering between the different wavevector solutions.  The real part of this integral gives the relative phases between the plane waves; and, since $\chi(k)$ can be analytically continued into the complex $k$-plane, the integral can also contain an imaginary part, yielding their relative \textit{amplitudes}.  The only ambiguity is in the path taken in the complex plane between the various wavevectors, or, more specifically, its topology with respect to the singularities and branch cuts of $\chi(k)$.  This is exactly as noted previously: in the $k$-space representation, the selection of a potentially large number of coefficients of $x$-space solutions is replaced by the selection of a single contour in the complex $k$-plane.

\subsubsection{Selection of the contour
\label{sub:Selection_of_the_contour}}
As previously stated, the integration contour is selected so as to satisfy appropriate boundary conditions.  These boundary conditions relate to the ``in'' or ``out'' nature of the mode we are aiming for, for this determines that certain plane waves -- those in- or outgoing waves which do not correspond to the single in- or outgoing wave of the mode -- must have zero amplitude.  We must therefore single out the particular mode that relates to Hawking radiation.  Neglecting the $v$-branch, the outgoing waves are $u1$ and $u$, so the radiation is produced in $u1$-$u$ pairs and the relevant modes are the $u1$-out and $u$-out modes.  Of these, the former is the easiest to calculate, for in the presence of a group-velocity horizon and for both subluminal and superluminal dispersion, it has a purely evanescent character in one of the asymptotic regions.  (This mode is illustrated by a space-time diagram and a wavepacket simulation in Fig. \ref{fig:out_mode}.)   The $u1$-out mode thus contains only an exponentially damped solution beyond the horizon.  This is the boundary condition which selects the required integration contour.

\begin{figure}
\includegraphics[width=0.8\columnwidth]{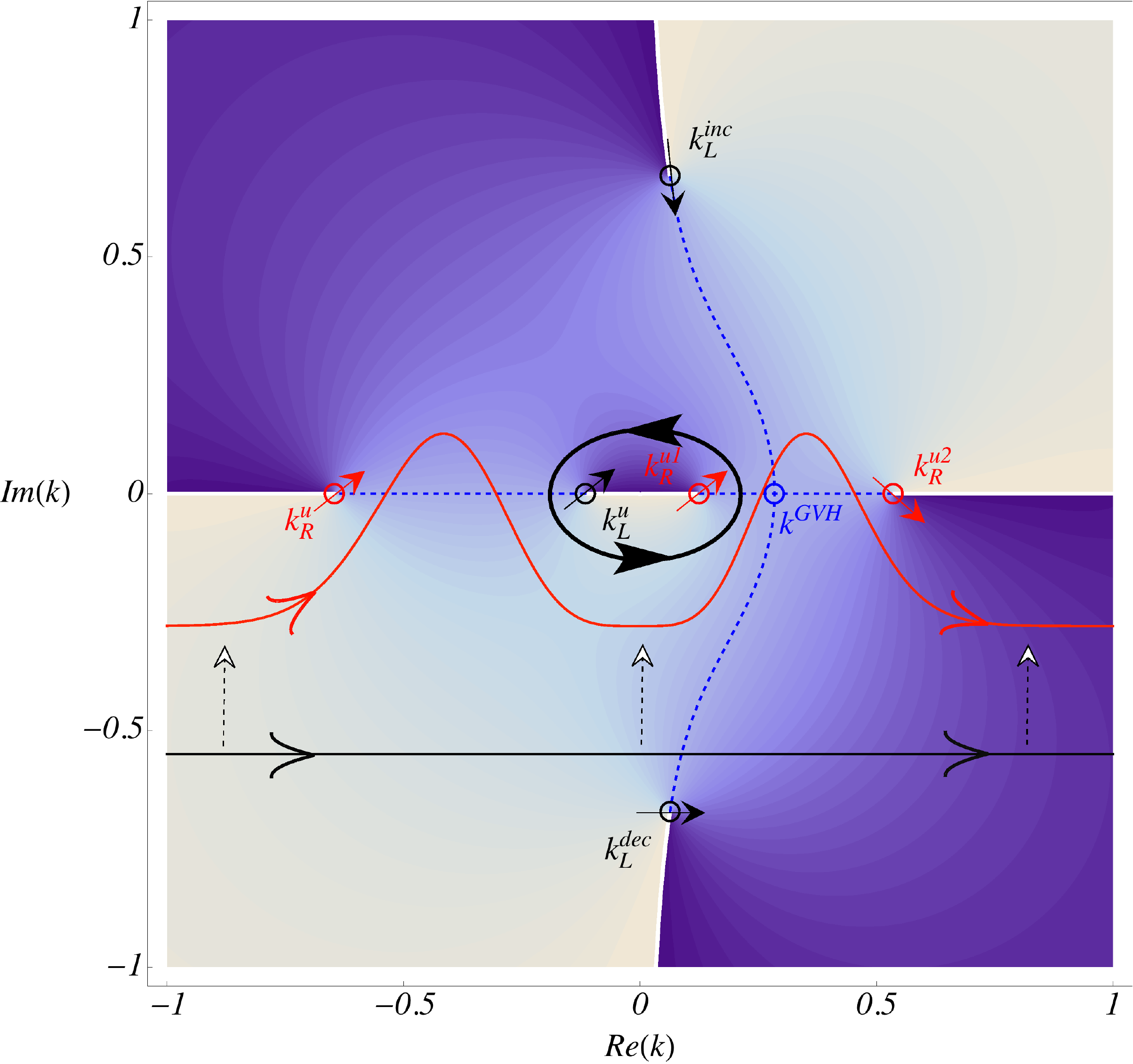}
\caption[\textsc{Phase integral in momentum space}]{\textsc{Phase integral in momentum space}:
Using the dispersion and velocity profiles of \S\ref{sub:Specifying_dispersion_and_flow} (plotted in Fig. \ref{fig:dispersion_and_velocity_profiles}), here the colour shading shows the imaginary part of the position $\chi(k)$ -- the darkest shade represents $-\pi/2a$, the lightest $+\pi/2a$ -- where the wavenumber $k$ is taken as a complex variable.  The dashed blue curves show where $\chi(k)$ is real, and split into two disconnected regions: the cross-bow shape on the right contains (on the real axis) the positive-norm $u1$- and $u2$-branches, as well as the complex conjugate wavevectors they connect to beyond the group-velocity horizon; the line on the left contains the negative-norm $u$-wavevectors.  The extremities of these curves correspond to the solutions in the asymptotic constant-velocity regions; those on the left are shown in black, those on the right in red, and the arrows show the directions of steepest descent through these saddle points.  When solving the wave equation, the boundary condition is that, beyond the horizon, only the exponentially \textit{decreasing} solution is allowed; therefore, for large negative $x$, the integration contour must be able to pass through the saddle point near $k_{L}^{dec}$ in its direction of steepest descent -- this contour is shown in black, and does not contain any contribution from the other left-hand (colour black) saddle points.  As $x$ varies continuously into the right-hand asymptotic constant-velocity region, this contour is continuously deformable to the red one, which is able to pass through all three of the right-hand (colour red) saddle points in their directions of steepest descent.  The relative phases between the various plane waves is given by the integral of $\chi(k)$ between their corresponding wavevectors, along the path selected by the integration contour.  When comparing the positive- and negative-norm components, this integral traverses a branch cut of $\chi(k)$ and hence contains an imaginary part, which can be isolated from the real part by taking the closed contour integral around the entire branch cut.  This gives their relative amplitudes -- and the Hawking temperature -- as described by Eqs. (\ref{eq:Boltzmann_factor}) and (\ref{eq:WKB_Boltzmann_factor}).
\label{fig:WKB_contour_plot}}
\end{figure}

Let us return to Fig. \ref{fig:WKB_contour_plot}.  Firstly, we note that, to ensure convergence, the integrand of Eq. (\ref{eq:inverse_Fourier_transform}) must tend to zero at the limits of the integral.  This restricts the possible directions in which the integration contour can tend to infinity.  The simplest (though by no means the only) way in which this condition can be satisfied is to have the contour tend from $-\infty$, where the imaginary part of $\chi(k)$ is positive, to $+\infty$, where the imaginary part of $\chi(k)$ is negative.  In Fig. \ref{fig:WKB_contour_plot}, we see then that there is a tendency of the contour to lie on the lower-half of the complex $k$-plane.

Now consider the boundary condition of pure exponential damping.  The wavevectors of interest are those in the asymptotic regions, which lie at the extremities of the locus of saddle points shown as blue dashed lines.  These have been colour-coded, with red representing the wavevector solutions in the right-hand subsonic region, and black representing the wavevector solutions in the left-hand supersonic region.  It is in the left-hand region, beyond the horizon, that the evanescent modes exist.  There, the exponentially damped solution has negative imaginary part: it is that complex wavevector lying on the lower-half complex plane, at the lower end of the crossbow.  Also shown in Fig. \ref{fig:WKB_contour_plot} are the directions of steepest descent through the extreme saddle points \footnote{The directions of steepest descent are actually calculated in the limit as $x\rightarrow \pm \infty$, for the extreme points themselves are singular and the directions of steepest descent are ill-defined precisely at those points.}.  For the exponentially damped wavevector, this direction is almost horizontal.  The contour, then, is that from $-\infty$ to $+\infty$ which lies entirely on the lower-half plane.  It is shown as a white curve in Fig. \ref{fig:WKB_contour_plot}.

The specification of the contour determines the complete solution in $x$-space through insertion into Eq. (\ref{eq:pos_soln_sum}), because the contour can be continuously deformed so long as its topology with respect to the singularities and branch cuts of $\chi(k)$ is not changed.  That is, we may move the contour in the complex $k$-plane, but so long as we do not cross any of the branch cuts where $\chi(k)$ is discontinuous (where light shading meets dark shading in Fig. \ref{fig:WKB_contour_plot}), the result of Eqs. (\ref{eq:inverse_Fourier_transform}) and (\ref{eq:pos_soln_sum}) do not change.  The final check, then, on whether the selected contour is the correct one is to see whether it can be deformed so as to pass through the other (colour black) saddle points in their directions of steepest descent.  It cannot; therefore, only the exponentially damped wavevector contributes to the solution in the left-hand supersonic region, and the selected contour is indeed that which corresponds to the $u1$-out mode.

\subsubsection{Extracting the Hawking temperature
\label{sub:Extracting_beta2}}

Recalling Eq. (\ref{eq:out_expectation_value}), the Hawking flux is determined by the squared amplitude of the oppositely-normed ingoing wave relative to that of the outgoing wave.  For the $u1$-out mode, as shown in Fig. \ref{fig:out_mode}, the scattering process takes place entirely in the right-hand subsonic region.  So, having selected the integration contour using boundary conditions at $x\rightarrow -\infty$, we now consider the solution at $x\rightarrow +\infty$.  The saddle points now lie at the wavevector solutions $u$, $u1$ and $u2$ coloured red in Fig. \ref{fig:WKB_contour_plot}.  Which of these wavevectors contribute to the solution is determined by whether the integration contour can be deformed to pass through these saddle points in their directions of steepest descent; it is easily seen that it can be deformed to pass through all three of them.  The $x$-space solution in this region is therefore a sum of the three plane waves, one outgoing ($u1$) and two ingoing ($u2$ and $u$), with relative amplitudes given by Eq. (\ref{eq:pos_soln_sum}).  The significance of the integration contour is not yet spent, for it determines the path taken between the three wavevector solutions in the complex $k$-plane when calculating the phase integral of $\chi(k)$.

Denoting the amplitudes of the ingoing waves $u2$ and $u$ as $\alpha_{\omega}$ and $\beta_{\omega}$, respectively, and assuming that the overall mode is normalized so that the outgoing $u1$-wave has unit norm, we have, by norm conservation,
\begin{equation}
|\alpha_{\omega}|^{2}-|\beta_{\omega}|^{2}=1\,.
\label{eq:norm_conservation}
\end{equation}
The Hawking flux is determined by the amplitude $\beta_{\omega}$ of the $u$-wave, the negative-norm ingoing component.  Defining the frequency-dependent temperature $T(\omega)$ in accordance with the form of the Planck spectrum of Eq. (\ref{eq:thermal_spectrum}),
\begin{equation}
|\beta_{\omega}|^{2}=\frac{1}{e^{\hbar \omega / k_{B} T(\omega)} - 1}\,,
\label{eq:temp_defn}
\end{equation}
and utilizing Eq. (\ref{eq:norm_conservation}), we have
\begin{equation}
\left| \frac{\alpha_{\omega}}{\beta_{\omega}} \right| ^{2} = \exp\left(\frac{\hbar\omega}{k_{B}T(\omega)}\right)\,.
\label{eq:Boltzmann_factor}
\end{equation}
Now, $|\alpha_{\omega}/\beta_{\omega}|$ -- the amplitude of the $u2$-wave relative to the $u$-wave -- is, according to Eq. (\ref{eq:pos_soln_sum}), given by the imaginary part \footnote{The real part of the integral gives their relative phase, which is unimportant in calculating the flux.} of the phase integral of $\chi(k)$ from $k^{u}_{R}$ to $k^{u2}_{R}$ along a path consistent with the integration contour selected in \S\ref{sub:Selection_of_the_contour} \footnote{Note that the integral taken between $k^{u1}_{R}$ and $k^{u2}_{R}$ is purely real, so that, according to this prediction, the incoming $u2$-mode and the outgoing $u1$-mode have \textit{equal} amplitude, and the exponentiated imaginary part of the integral taken between the positive- and negative-$k$ branches can equally well be interpreted as $|\alpha_{\omega}/\beta_{\omega}|$ or as simply $|\beta_{\omega}|$.  This is a defect of the approach, and we observe that of these two possibilities, the phase integral should give $|\alpha_{\omega}/\beta_{\omega}|$, since this relates the two high-magnitude wavevectors where the amplitude $\mathcal{A}(k)$ (defined in Eqs. (\ref{eq:amplitude_phase})) is better-behaved.  Other approaches combining WKB techniques in momentum- and position-space are able to resolve this issue -- see references \cite{Corley-1998}, \cite{Unruh-Schutzhold-2005} and \cite{Coutant-et-al-2012}.} .  In Fig. \ref{fig:WKB_contour_plot}, such a path lies in the lower-half complex $k$-plane, an important detail in that it determines how the central branch cut is to be circumvented.

Although the result could be left formulated in this way, it is made more elegant by noticing that the imaginary part of the phase integral can be isolated by closing the contour in the \textit{upper}-half complex plane.  Being an analytic continuation of an essentially real-valued quantity, the wavevector-dependent position $\chi(k)$ must obey the relation $\chi(k^{\star})=\chi^{\star}(k)$.  So, if we were to choose the contour to lie on the upper-half plane -- equivalently, transforming the integration variable $k$ to its complex conjugate $k^{\star}$ -- this would simply result in the complex conjugate of the original integral.  But, on closing the contour in this fashion, we subtract this value from the original integral, so that the real part is cancelled out while the imaginary part is doubled.  Since the phase integral appears as an exponent in Eq. (\ref{eq:pos_soln_sum}), this closed phase integral gives precisely the \textit{squared} relative amplitude $|\alpha_{\omega}/\beta_{\omega}|^{2}$, and by comparison with Eq. (\ref{eq:Boltzmann_factor}) we have
\begin{equation}
\frac{\hbar\omega}{k_{B}T(\omega)} = \left| \oint \chi(k) dk \right| \, ,
\label{eq:WKB_Boltzmann_factor}
\end{equation}
where the closed integration contour is taken around the central branch cut, as shown by the black curve in Fig. \ref{fig:WKB_contour_plot}.

Equation (\ref{eq:WKB_Boltzmann_factor}) is a generalization of the dispersionless result (\ref{eq:Hawking_temperature}), which, as we now show (see also \cite{Leonhardt-Robertson-2012}), it reproduces.  The phase integral of Eq. (\ref{eq:WKB_Boltzmann_factor}), instead of an integral of position over wavevector, may also be written as an integral of wavevector over position by integration by parts: $\oint \chi(k) dk = -\oint \kappa(x) dx$.  While in the general dispersive case the branch points and the integration contour take complicated forms in position space (see, e.g., references \cite{Leonhardt-et-al-2003,Leonhardt-et-al-2003-ii} for a position-space analysis of Bogoliubov modes in BEC), the dispersionless case is very simple there because it has only one singular point: the event horizon itself, where the wavevector diverges.  This is easily seen by examining the closed form of the position-dependent wavevector (recall Eq. (\ref{eq:U_V_wavevectors})): $\kappa^{u}(x)=\omega/(V(x)+c)$, which is analytic in the complex $x$-plane except at the point where $V(x)=-c$, where it has a simple pole.  The closed contour integral is equal to $2\pi$ times the residue of this pole; writing $V(x)=-c+\alpha x+...$, it is seen that the residue is $\omega/\alpha$.  Then we have
\begin{equation}
\frac{\hbar\omega}{k_{B}T}=\left|\oint \kappa^{u}(x) dx\right|=\left|\oint \frac{\omega}{\alpha x}dx\right|=\frac{2\pi\omega}{\alpha}\,,
\label{eq:dispersionless_phase_integral}
\end{equation}
from which we retrieve Eq. (\ref{eq:Hawking_temperature}).

Equation (\ref{eq:dispersionless_phase_integral}) also holds in the low-frequency limit of the dispersive model, for then the central branch cut closes in closer to the origin, and since the dispersion profile is assumed to approach $c(k)=c_{0}$ in this limit, we can then treat the phase integral as if the system were dispersionless.  So, as $\omega\rightarrow 0$, Eq. (\ref{eq:WKB_Boltzmann_factor}) reproduces precisely the dispersionless result (\ref{eq:Hawking_temperature}), where the horizon is understood as the low-frequency limiting horizon where $V(x)=-c_{0}$.  This can be taken as a measure of the applicability of Eq. (\ref{eq:WKB_Boltzmann_factor}): if the low-frequency temperature $T(\omega=0)$ is equal to that predicted by the dispersionless model, then the phase integral of Eq. (\ref{eq:WKB_Boltzmann_factor}) can be used to find the temperature for higher frequencies.  We shall examine a model in \S\ref{sec:Application_to_a_simple_model} in which this is found to be the case, and we shall see how the applicability of Eq. (\ref{eq:WKB_Boltzmann_factor}) is related to the criterion of a ``slowly-varying'' velocity profile.

\section{Application to a simple model\label{sec:Application_to_a_simple_model}}

Apart from restricting ourselves to dispersion relations of the subluminal type, the dispersive framework of previous sections has been kept fairly general, and we have noted only the main differences occurring when superluminal dispersion is used.  In this section, let us turn to a specific example, providing a quantitative account of, but mainly to get a feeling for the qualitative nature of the expected Hawking radiation.  The velocity and dispersion profiles used are kept as simple as possible, so that the qualitative properties discovered may still be expected to apply quite generally.

\subsection{Specifying the model\label{sub:Specifying_dispersion_and_flow}}

\subsubsection{Dispersion profile}

The dispersionless case corresponds to a wavelength-independent phase velocity: $c^{2}(k)=c_{0}^{2}$.  Since dispersive effects arise through the occurrence of higher-order derivatives in the wave equation -- and bearing in mind that $c^{2}(-k)=c^{2}(k)$ for an isotropic medium -- the simplest deviation from the dispersionless model is through the inclusion of a quadratic term in $c^{2}(k)$ \footnote{Dispersion relation (\ref{eq:quadratic_dispersion}) -- also used in \cite{Corley-Jacobson-1996} -- has a strict cut-off wavevector $k_{d}$, at which the free-fall frequency vanishes.  This differs from the subluminal dispersion relations shown in Figs. \ref{fig:subluminal_supersonic}-\ref{fig:dispersion_sub_sub}, where a limiting free-fall frequency is approached at high $k$, similarly to the dispersion considered in \cite{Unruh-1995}.  So long as the flow velocity does not become too small -- as will be the case here -- there is no significant difference in the behaviour of these two forms, since the wavevector remains well within the cut-off.  For low asymptotic velocity, the wavevector may reach $k_{d}$, at which point this dispersion relation breaks down \cite{Corley-Jacobson-1996}.  However, even the flat dispersion relation becomes problematic in the low-velocity limit; see ``Conceptual issues'' of \S\ref{sub:Overview}, and \S VI of \cite{Corley-Jacobson-1996}.}:
\begin{equation}
c^{2}\left(k\right)=c_{0}^{2}\left(1-\frac{k^{2}}{k_{d}^{2}}\right)\,.\label{eq:quadratic_dispersion}
\end{equation}
This dispersion relation is illustrated in Figure \ref{fig:dispersion_and_velocity_profiles}$(a)$.  Note that it reduces to the dispersionless form when $k^{2}/k_{d}^{2} \ll 1$.  The parameter $k_{d}$, then, indicates how large $k$ must be before dispersive effects begin to be felt.

In line with the framework we have so far considered, we have chosen the dispersion relation (\ref{eq:quadratic_dispersion}) to be subluminal: the phase velocity of short wavelengths drops below the long-wavelength limit $c_{0}$.  It is also possible to include the dispersive term with a plus sign, in which case the dispersion would be superluminal.  Indeed, a quadratic superluminal dispersion relation -- exactly of the type (\ref{eq:quadratic_dispersion}) with the sign switched -- is obeyed by the excitations of Bose-Einstein condensates \cite{Leonhardt-et-al-2003}, and has been numerically studied quite extensively in \cite{Macher-Parentani-2009,Macher-Parentani-2009-ii,Finazzi-Parentani-2011-JPCS,Finazzi-Parentani-2012-arXiv}.  We shall find that the results obtained from subluminal dispersion are qualitatively very similar.

\subsubsection{Velocity profile}

\begin{figure}
\subfloat{\includegraphics[width=0.45\columnwidth]{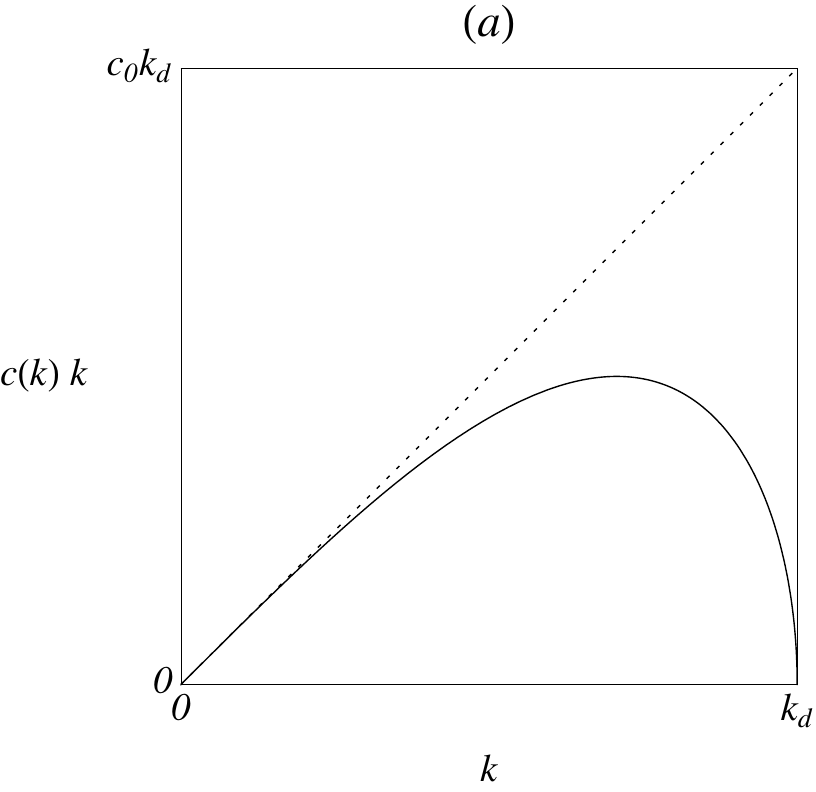}} \subfloat{\includegraphics[width=0.45\columnwidth]{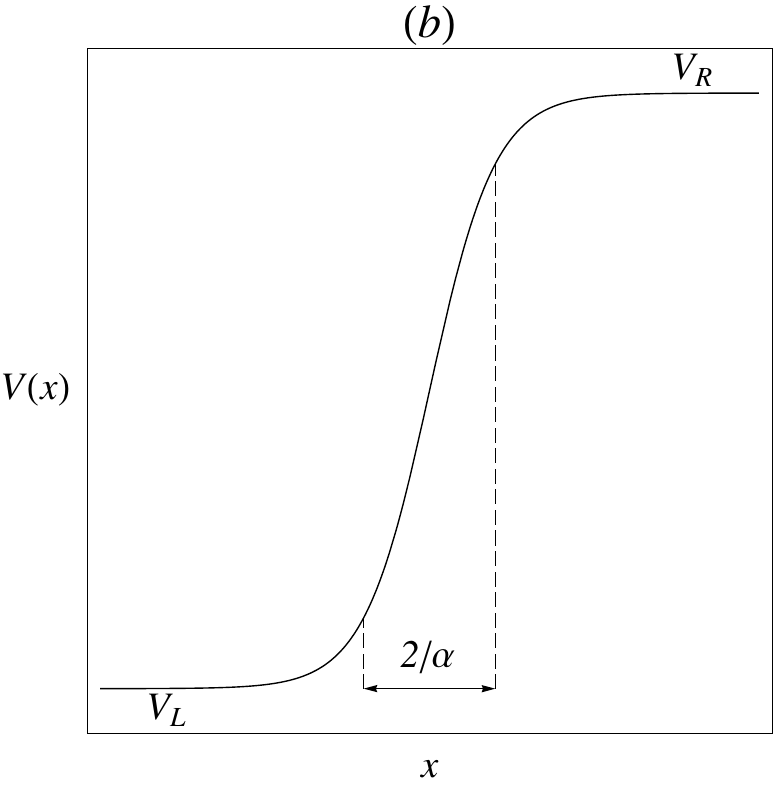}}
\caption[\textsc{Specifying dispersion and velocity profiles}]{\textsc{Specifying dispersion and velocity profiles}:

$(a)$ The solid line plots $c(k) k$, with $c(k)$ as in Eq. (\ref{eq:quadratic_dispersion}).  The dotted line plots the corresponding dispersionless curve, $c_{0} k$.

$(b)$ The velocity profile, given by Eq. (\ref{eq:hyperbolic_tangent_velocity_profile}), is monotonic and approaches asymptotically constant values.  The parameter $\alpha$ is inversely related to the length of the transition region.
\label{fig:dispersion_and_velocity_profiles}}
\end{figure}

Endeavouring to keep the physical system as simple as possible, we
shall use a velocity profile that varies monotonically between two
asymptotic values; such a profile shall then be invertible, and can be used in the WKB-type analysis of \S\ref{sub:WKB-type-analysis}. We choose
\begin{equation}
V\left(x\right)=\frac{1}{2}\left(V_{R}+V_{L}\right)+\frac{1}{2}\left(V_{R}-V_{L}\right)\tanh\left(\alpha x\right)\,.\label{eq:hyperbolic_tangent_velocity_profile}
\end{equation}
This is illustrated in Figure \ref{fig:dispersion_and_velocity_profiles}$(b)$.  $V_{R}$ and $V_{L}$ are the asymptotic values of $V\left(x\right)$
as $x$ tends to $+\infty$ (the right-hand region) or $-\infty$ (the left-hand region), respectively. The parameter
$\alpha$ characterizes the steepness of the velocity profile, but
this also depends on the difference $V_{R}-V_{L}$; a more direct
interpretation of $\alpha$ is that it governs the length of the region
of transition between $V_{L}$ and $V_{R}$, which is of order $2/\alpha$
as shown in Figure \ref{fig:dispersion_and_velocity_profiles}$(b)$.
As $\alpha\rightarrow\infty$, the velocity approaches the discontinuous step-profile
\begin{equation}
V_{\alpha\rightarrow\infty}\left(x\right) = V_{L}\,\theta\left(-x\right) + V_{R}\,\theta\left(x\right)\,,
\label{eq:step_velocity_profile}
\end{equation}
where $\theta(x)$ is the Heaviside step function, equal to zero for $x<0$ and unity for $x>0$.  In this limiting case, the treatment described in \S\ref{sub:Discontinuous_velocity_profile} may be applied.

We restrict our attention to a left-moving and accelerating flow, i.e. $V_{L}<V_{R}<0$.  Therefore, if a horizon is present, it corresponds to a black hole horizon.  White hole spectra are studied in \cite{Macher-Parentani-2009}.

\subsubsection{Normalizing the wave equation}

With the dispersion and velocity profiles given by Eqs. (\ref{eq:quadratic_dispersion})
and (\ref{eq:hyperbolic_tangent_velocity_profile}), the
wave equation (\ref{eq:acoustic_wave_equation}) contains five parameters: $c$ and $k_{d}$ from the
dispersion profile; and $V_{L}$, $V_{R}$ and $\alpha$ from the
velocity profile. However, two of these correspond to the scaling
of space and time; in redefining the variables to make them dimensionless,
we can reduce the number of independent parameters to three.

Let us proceed by defining a dimensionless velocity, which is the flow velocity divided by the low-wavevector phase velocity: $U \equiv V/c_{0}$; and also by defining a dimensionless distance, which is the ``phase'' of a wave with wavevector $k_{d}$: $X \equiv k_{d} x$.  Then a natural unit of time is $1/c_{0}k_{d}$, and we define the dimensionless time $T \equiv c_{0}k_{d}t$.  Dimensionless wavevectors and frequencies are defined accordingly: $K \equiv k/k_{d}$ and $\Omega \equiv \omega/c_{0}k_{d}$.  With these definitions, phases, which are anyway dimensionless quantities, are unchanged: $kx=KX$ and $\omega t=\Omega T$.  In terms of the dimensionless variables, the wave equation (\ref{eq:acoustic_wave_equation}) becomes
\begin{equation}
\left(\partial_{T}+\partial_{X}U\right)\left(\partial_{T}+U\partial_{X}\right)\phi-C^{2}\left(-i\partial_{X}\right)\partial_{X}^{2}\phi=0\,,\label{eq:normalized_wave_equation}
\end{equation}
where the rescaled dispersion relation is
\begin{equation}
C^{2}\left(K\right)=1-K^{2}\,.\label{eq:normalized_dispersion}
\end{equation}
and the rescaled velocity profile is
\begin{equation}
U\left(X\right)=\frac{1}{2}\left(U_{R}+U_{L}\right)+\frac{1}{2}\left(U_{R}-U_{L}\right)\tanh\left(a X\right)\,.\label{eq:normalized_velocity}
\end{equation}
The wave equation now contains only three parameters, and these relate only to the velocity profile $U(X)$: 
\begin{itemize}
\item $U_{R} \equiv V_{R}/c_{0}$ and $U_{L} \equiv V_{L}/c_{0}$, the asymptotic values of the flow velocity
relative to the low-wavevector wave speed; and
\item $a \equiv \alpha/k_{d}$, a parameter that combines steepness of the flow velocity profile with the wavevector characterising dispersion. Since the length of the region over which $V$ changes is on the order of $2/\alpha$, we see that $1/a$ is approximately the phase difference of the wavevector $k_{d}$ across this transition region.  So, if $a$ is small, this transition region is many cycles of $k_{d}$ wide; whereas, if $a$ is large, $V$ varies quickly over only a few cycles of $k_{d}$.  The geometrical optics and step function limits of \S\S \ref{sub:WKB-type-analysis} and \ref{sub:Discontinuous_velocity_profile} correspond to the limits of small and large $a$, respectively; the transition between these regimes will be discussed in \S\ref{sub:Varying-a} (see also \cite{Finazzi-Parentani-2012-arXiv,Robertson-thesis}).
\end{itemize}

Although in principle the values of these parameters can be set arbitrarily, there will typically be practical limits.  This is particularly relevant for the steepness parameter $a$, since the same dispersive effects which regulate the singular nature of the horizon will also tend to smooth out the sharpness of the velocity transition.  In BEC, for example, they are regulated by the so-called \textit{healing length}, which determines both the scale of short distance dispersion and the minimum length of the transition region.  $a=1$, then, represents a value which would be difficult to exceed in practice\footnote{In principle, however, arbitrarily high steepness can be achieved by suitably combining an external potential with a spatially-varying interaction constant \cite{Carusotto-et-al-2008,Recati-et-al-2009,Mayoral-et-al-2011}.}.

The difference in the asymptotic velocities can, in some contexts, also be limited.  Nonlinear optics provides a good example: a light pulse induces an effective velocity difference on the order of the change in refractive index, typically no more than $10^{-3}$ \cite{Philbin-et-al-2008,Robertson-thesis}.  There would seem to be no generally applicable limits to the velocity values, however.

\subsubsection{Required modes}

Before we can use the $\mathcal{S}$-matrix of \S\ref{sec:Hawking_radiation_in_dispersive_media} to calculate the quasiparticle creation rates, we must first select the appropriate modes whose scattering amplitudes encode their values.  For the case of a black-hole velocity profile with group-velocity horizon, we have actually already found such a mode in \S\ref{sub:Selection_of_the_contour}, since it is the boundary conditions appropriate to the desired mode which there determine the integration contour in momentum space.  This was the $u1$-out mode -- illustrated in Fig. \ref{fig:out_mode} -- which happened to have a single exponentially damped solution in the left-hand supersonic region.  Now, whereas the analytical approximation of \S\ref{sub:WKB-type-analysis} treats $u$- and $v$-modes as decoupled, this is not true in general, and through numerical integration the coupling is included automatically.  We should thus also account for the $v$-mode.  As can be seen in Fig. \ref{fig:out_mode}, the ingoing $v$-wave appears in the right-hand subsonic region along with all the other propagating waves.  So it remains true that the full $u1$-out mode is exactly equal to the left-mode which is purely a damped exponential in the left-hand region \footnote{This is \textit{not} true for superluminal dispersion, where the ingoing $v$-wave appears in the same region as the evanescent wave \cite{Macher-Parentani-2009}.}, and it may be found by straightforward integration, without any subsequent linear transformation.  Since the $u$-wave is the only one with opposite norm to $u1$, the creation rate of the $u1$-$u$ pair is (as found in \S\ref{sub:Spontaneous_creation-II}) given by the relative squared amplitude of the ingoing $u$-wave compared to that of the outgoing $u1$-wave.

There is another possible creation channel for this setup: the $u$- and $v$-waves have opposite norm, so $u$-$v$ pairs can be created.  Their creation rate is given by the relative squared amplitude of the ingoing $u$-wave to that of the outgoing $v$-wave in the $v$-out mode.  However, since the $v$-wave is always left-moving, the outgoing $v$-wave is in the left-hand supersonic region, where the evanescent waves also exist.  There can be a problem with numerical stability when we try to isolate the left-hand $v$-wave, since numerical integration in the left-hand region -- no matter what the initial conditions -- will tend to lead to exponential growth \cite{Corley-Jacobson-1996,Macher-Parentani-2009}.  That said, in the limiting step function profile there is no such problem in calculating the $v$-out mode, since this involves only matching at a single point and no numerical integration.

Finally, consider the case when there is no group-velocity horizon.  Then all three of the waves $u1$, $u2$ and $v$ can be outgoing, so all three of them can be created by pairing with the $u$-wave.  As we have just been considering, their creation rates could be calculated by calculating each of their out-modes and singling out the relative squared amplitude of the ingoing $u$-wave.  However, recall from \S\ref{sec:Hawking_radiation_in_dispersive_media} that this is equivalent to summing their relative squared amplitudes over all \textit{in}-modes of \textit{negative} norm.  This latter approach is the most economical because there is only one such in-mode, corresponding to the $u$-wave.  It is shown in Fig. \ref{fig:in_mode}, where it can be seen that it contains only the $u$- and $u1$-waves in the right-hand region, and no $u1$-wave in the left-hand region.  Two integrations are required to calculate it: we find first the $u1$-right and $u$-right modes, determine the amplitudes of their left-hand components, then form the linear combination which exactly cancels the $u1$-wave in the left-hand region.  From the resulting mode, the creation rates are the squared amplitudes of the outgoing $u1$-, $u2$- and $v$-waves relative to that of the ingoing $u$-wave.

\subsubsection{Predictions}

Before turning to numerical results, let us first extract the analytical predictions of the techniques of \S\ref{sub:Discontinuous_velocity_profile} and \S\ref{sub:WKB-type-analysis}.

In the high-steepness limit where the velocity profile approaches that of Eq. (\ref{eq:step_velocity_profile}), we simply need to solve a linear system involving derivatives of the plane waves, and hence powers of $K$ for each possible wavevector solution.  Since the dispersion $(\Omega-UK)^{2}=C^{2}(K)K^{2}$ is of fourth order, there will be four such solutions and the linear system is $4\times4$.  The resulting expressions are unwieldy; but, in the low-frequency limit where the Planck spectrum becomes the simple pole $T/\Omega$ \footnote{The normalized frequency-dependent temperature is defined by analogy with Eq. (\ref{eq:thermal_spectrum}) such that the relative squared amplitude of the corresponding mode is given by the expression for the Planck spectrum: $|\beta_{\Omega}|^{2} = \left(\exp(\Omega/T(\Omega))-1\right)^{-1}$.}, the expression for the $u$-$u1$ Hawking spectrum also approaches this form, and the low-frequency temperature can be read off.  After some straightforward but tedious algebra, we find
\begin{equation}
T_{\Omega\rightarrow 0}^{\mathrm{high}\,a} = \left(1-U_{R}^{2}\right)^{1/2} \frac{\left(1+U_{R}\right)\left(-1-U_{L}\right)\left(-U_{R}-U_{L}\right)}{\left(1-U_{R}\right)\left(1-U_{L}\right)\left(U_{R}-U_{L}\right)}\,.
\label{eq:step_temperature}
\end{equation}
This result was first stated in reference \cite{Corley-1997}.  A more detailed derivation can be found in Appendix A of reference \cite{Robertson-thesis}.

In the low-steepness limit where $V$ can be treated as slowly-varying, the temperature at any frequency should be well-approximated by the phase integral of Eq. (\ref{eq:WKB_Boltzmann_factor}).  Moreover, this phase integral takes a very simple analytic form when the velocity profile is of the hyperbolic tangent form (\ref{eq:hyperbolic_tangent_velocity_profile}), as was used to plot Fig. \ref{fig:WKB_contour_plot}: at the branch cuts, the imaginary part of $\chi(k)$ jumps between $-\pi/2a$ and $+\pi/2a$, while the central branch cut around which the integral is taken lies exactly on the real axis.  We can deform the contour to lie exactly along the branch cut, in which case it becomes simply the real integral of a constant.  After some trivial rearrangement, Eq. (\ref{eq:WKB_Boltzmann_factor}) then gives (in units where $\hbar=k_{B}=1$)
\begin{equation}
T^{\mathrm{low}\,a}\left(\Omega\right) = \frac{a}{\pi}\,\frac{\Omega}{K_{R}^{u1}-K_{L}^{u}}\,,
\label{eq:tanh_temperature}
\end{equation}
where $K_{R}^{u1}$ and $K_{L}^{u}$ are the (normalized) $u1$- and $u$-wavevectors calculated in the right- and left-hand regions, respectively.  It is easily shown (see \S4.6 of reference \cite{Robertson-thesis}) that, in the low-frequency limit, this becomes
\begin{equation}
T^{\mathrm{low}\,a}_{\Omega\rightarrow0} = \frac{a}{\pi}\,\frac{\left(1+U_{R}\right)\left(-1-U_{L}\right)}{\left(U_{R}-U_{L}\right)}\,,
\label{eq:tanh_temperature_low-frequency}
\end{equation}
which, as can be checked by evaluating the derivative of $V$ at the horizon, is exactly Hawking's prediction, Eq. (\ref{eq:Hawking_temperature}).  As observed at the end of \S\ref{sub:WKB-type-analysis}, Eq. (\ref{eq:WKB_Boltzmann_factor}) always yields this result in the low-frequency limit.

\subsection{Calculated spectra\label{sub:Numerical-results}}

The richness of the dispersive spectra is perhaps best appreciated by fixing one of the asymptotic velocities and varying the other \textit{through} $U=-1$.  This allows us to examine spectra from the various frequency regimes described in \S\ref{sub:Inhomogeneous_flow} and the manner in which they vary between these regimes.

\subsubsection{Varying $U_{L}$: the subsonic side\label{sub:Varying-Ul}}

In Figure \ref{fig:temp_varying-ul_Hawking-regime}, the parameter $a$ is fixed at $0.1$, $U_{R}$ is fixed at $-0.5$ and $U_{L}$ is varied from $-0.51$ to $-1.5$; Figure \ref{fig:temp_varying-ul_step-regime}, on the other hand, shows spectra for the same values of $U_{L}$ and $U_{R}$ but in the discontinuous limit $a\rightarrow\infty$.  (We call this the ``subsonic side'' because the central value of the velocity is greater than $-1$.)  When $U_{L}$ is close to $U_{R}$, the spectral region that experiences a group-velocity horizon (i.e., $\omega_{\mathrm{max},1}<\omega<\omega_{\mathrm{max},2}$) is narrow, and most of the spectrum is in the horizonless ($\omega<\omega_{\mathrm{max},1}$ regime.  As $U_{L}$ drifts away from $U_{R}$, $\omega_{\mathrm{max},1}$ decreases so the spectral region experiencing a group-velocity horizon widens at the expense of the horizonless regime; the group-velocity horizon regime encompasses the entire spectrum when $U_{L}$ reaches $-1$ and $\omega_{\mathrm{max},1}$ vanishes.  As $U_{L}$ varies further, we are in a true ``black hole'' regime, where a low-frequency limiting horizon ($V=-c_{0}$) exists.

In Figures \ref{fig:temp_varying-ul_Hawking-regime} and \ref{fig:temp_varying-ul_step-regime}, the solid lines plot the frequency-dependent temperature of the $u1$-wave, while the dashed lines do so for the $u2$-wave, which is only emitted where there is no group-velocity horizon (i.e., where $\Omega<\Omega_{\mathrm{max},1}$).  The dotted lines of Figure \ref{fig:temp_varying-ul_step-regime} plot the temperature of the $v$-wave, which is difficult to calculate for finite $a=0.1$ and, where it is possible, is found to be vanishingly small.

A list of the noteworthy features of the spectra follows.  For the $u1$-waves:
\begin{itemize}
\item The $u1$ spectra show a general increase in temperature as $U_{L}$ is decreased (i.e., as the left-hand flow is made faster).  Since the derivative of the velocity profile increases in this direction, this is to be expected.  As the limit of infinite steepness, the discontinuous limit shows temperatures higher by about an order of magnitude compared with the finite value $a=0.1$.
\item The low-frequency temperature of the $u1$ spectra increases in the same fashion, but it is finite only when the low-frequency limiting horizon exists (i.e., when $U_{L}<-1$).  On the other hand, when the flow is everywhere subsonic, the low-frequency temperature vanishes.  This is the main effect of the presence or otherwise of the horizon; in particular, the spectrum at higher frequencies varies quite smoothly over $U_{L}=-1$.
\item The spectra typically show significant dependence of temperature on frequency, and are therefore far from Planckian.  Interestingly, the most Planckian spectrum (i.e., that with the flattest temperature curve) occurs for $U_{L}=-1.5$, where the velocity profile is symmetric around the horizon $U_{L}=-1$.  Indeed, the more asymmetric the velocity profile, the greater the deviations from the Planck spectrum seem to be.
\item In the low-steepness regime and when $U_{L}>-1$ (so that $U$ is everywhere subsonic), the $u1$ spectra vary smoothly between the horizonless and group-velocity horizon regimes (i.e., at $\omega_{\mathrm{max},1}$).  In the discontinuous limit, however, the transition between these regimes at $\omega_{\mathrm{max},1}$ is more pronounced.
\item There is a cut-off in the $u1$ spectra at $\omega_{\mathrm{max},2}$, at which point the outgoing $u1$-waves cease to exist.  In the low-steepness regime, this cut-off is very sharp; in the discontinuous limit, it becomes more rounded, the fall-off beginning at lower frequencies.  Since the $u1$-waves are emitted into the right-hand region, their cut-off frequency depends only on $U_{R}$, and since this is held fixed all $u1$ spectra are seen to vanish at the same frequency.
\end{itemize}
For the $u2$-waves:
\begin{itemize}
\item At a fixed frequency, the $u2$ spectra also show increasing temperature with decreasing $U_{L}$, and temperatures higher by about an order of magnitude in the discontinuous limit.  Since the $u2$ spectra exist only when there is no low-frequency limiting horizon, their temperature always vanishes in the low-frequency limit.
\item The $u2$ spectra experience a cut-off at $\omega_{\mathrm{max},1}$, at which point the outgoing $u2$-wave ceases to exist and a group-velocity horizon comes into existence.
\end{itemize}
For the $v$-waves:
\begin{itemize}
\item The $v$ spectra, shown only in the discontinuous case, generally have lower temperatures than the $u1$- and $u2$-waves (for equal frequency and $U_{L}$).  Following the general trend, their temperature increases with decreasing $U_{L}$, though only significantly so when $U_{L}$ is close to $U_{R}$.  Their low-frequency temperature always vanishes, in accordance with the fact that the $v$-waves never experience a horizon.
\item Since the $u$- and $v$-waves exist even when the $u1$- and $u2$-waves do not, the $v$-spectra do not experience any cut-off.
\end{itemize}

\subsubsection{Varying $U_{R}$: the supersonic side\label{sub:Varying-Ur}}

Consider now the case where $U_{L}$ is fixed at a supersonic value and $U_{R}$ is varied through $-1$.  (The central velocity will always be less than $-1$, hence we call this the ``supersonic side''.)  Since the $u1$- and $u2$-waves do not exist when the flow velocity is supersonic, there can be no mixing and hence no radiation when $U_{R}<-1$, and the spectrum is expected to vanish in this limit.  Moreover, as $U_{R}$ decreases (i.e., the flow in the right-hand region becomes faster), the cut-off frequency of the $u1$-waves decreases.  Therefore, the spectrum should become narrower with decreasing $U_{R}$, approaching zero width in the limit $U_{R} \rightarrow -1$ where the $u1$-waves cease to exist.

\begin{figure}
\includegraphics[width=0.8\columnwidth]{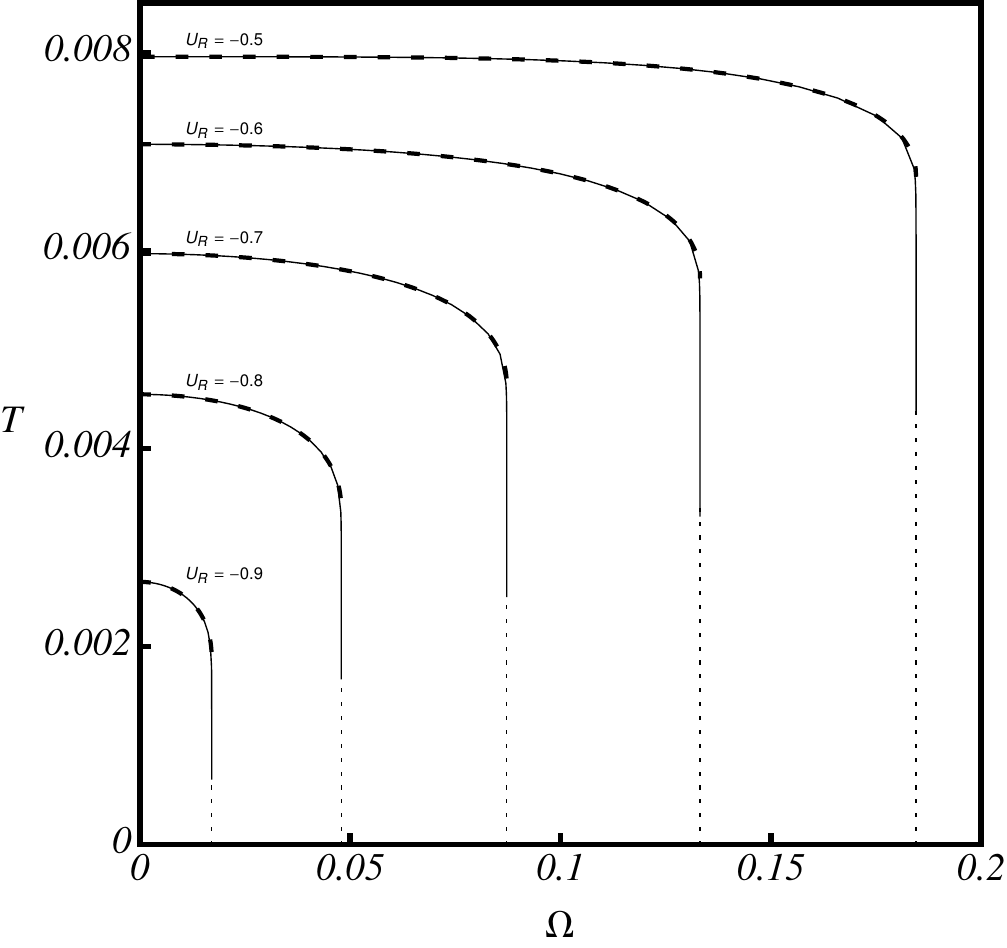}
\caption[\textsc{Temperature -- varying $U_{R}$ (low-$a$ regime)}]{\textsc{Temperature -- varying $U_{R}$ (low-$a$ regime)}: The (normalized) Hawking temperature, for the dispersion profile (\ref{eq:normalized_dispersion}) and velocity profile (\ref{eq:normalized_velocity}), is shown for various values of $U_{R}$ while $U_{L}$ is fixed at $-1.5$ and $a$ is fixed at $0.1$.  Since $U_{L}<-1$, there is no ``horizonless'' regime and hence no emission of $u2$-waves.  The solid curves show the numerically calculated $u1$-wave spectra, while the dashed curves correspond to the analytical prediction of Eq. (\ref{eq:tanh_temperature}).  The maximum frequency (shown as dotted vertical lines) decreases with $U_{R}$, eventually vanishing in the limit $U_{R}\rightarrow -1$.
\label{fig:temp_varying-ur_Hawking-regime}}
\end{figure}

In Figure \ref{fig:temp_varying-ur_Hawking-regime}, $a$ is fixed at $0.1$, $U_{L}$ is fixed at $-1.5$ and $U_{R}$ is varied from $-0.5$ to $-0.9$; in Figure \ref{fig:temp_varying-ur_step-regime}, $U_{L}$ and $U_{R}$ take the same values while the spectra are calculated in the discontinuous limit $a\rightarrow\infty$.  There is always a point at which $U=-1$, so there is no outgoing $u2$-wave and only the $u1$ and $v$ spectra are non-zero.

These spectra behave exactly as predicted: they both become narrower and experience an overall decrease in temperature as $U_{R}$ is decreased.  Notice that, as before, the discontinuous limit shows temperatures about an order of magnitude greater than for $a=0.1$, and that the spectra become very rounded so that the fall-off takes hold at lower frequencies than it does for $a=0.1$.

In the discontinuous limit (Fig. \ref{fig:temp_varying-ur_step-regime}), we also include the spectra of the $v$-waves.  As noted previously, these have generally lower temperatures than the $u1$-waves, with zero temperature in the low-frequency limit, and do not experience a cut-off.

\subsubsection{Varying $a$: the low- and high-steepness regimes\label{sub:Varying-a}}

Finally, let us examine the transition between the low- and high-steepness regimes by continuously varying $a$.  Rather than consider the evolution of a whole spectrum, let us consider the evolution of the low-frequency temperature in the limit $\Omega\rightarrow 0$.  As evidenced in Figs. \ref{fig:temp_varying-ul_Hawking-regime} and \ref{fig:temp_varying-ul_step-regime}, for this to be finite requires the existence of a low-frequency horizon where $U=-1$, and so $U_{L}<-1$ and $U_{R}>-1$.

\begin{figure}
\includegraphics[width=0.8\columnwidth]{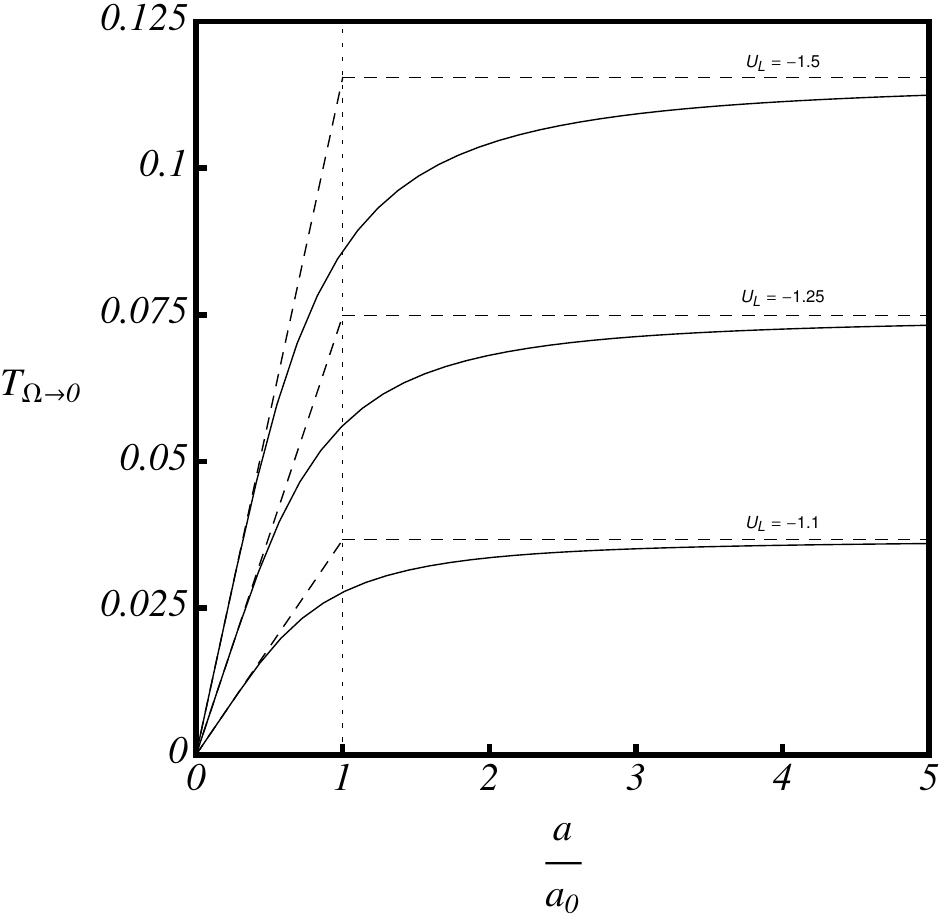}
\caption[\textsc{Low-frequency temperature -- varying $a$}]{\textsc{Low-frequency temperature -- varying $a$}: Here is shown the (normalized) low-frequency temperature with increasing steepness $a$, for various values of $U_{L}$ while $U_{R}$ is fixed at $-0.5$.  The parameter $a$ is normalized with respect to its value at which the linear Hawking prediction is exactly equal to the temperature in a step-discontinuous flow; they are equal at the dotted line, and the value of $a$ in relation to this gives a good indication of the point of transition between the two regimes.  Dashed lines correspond to the linear Hawking prediction for low $a$ and the limiting step-function temperature for high $a$.
\label{fig:temp_varying-a}}
\end{figure}

Figure \ref{fig:temp_varying-a} plots the low-frequency temperature against $a/a_{0}$, where $a_{0}$ is the value of $a$ at which the low-steepness prediction, Eq. (\ref{eq:tanh_temperature_low-frequency}), is exactly equal to the high-steepness prediction, Eq. (\ref{eq:step_temperature}):
\begin{equation}
a_{0} = \pi \left(1-U_{R}^{2}\right)^{1/2} \frac{\left(-U_{R}-U_{L}\right)}{\left(1-U_{R}\right)\left(1-U_{L}\right)}\,.
\label{eq:defining_a0}
\end{equation}
The high- and low-steepness predictions are also shown in Fig. (\ref{fig:temp_varying-a}), and by construction they meet at $a/a_{0}=1$.  It can been seen that $a/a_{0}$ provides a very good indicator for the regimes of low steepness (where the low-frequency temperature obeys Hawking's formula and the spectrum as a whole obeys Eq. (\ref{eq:WKB_Boltzmann_factor})) and of high steepness (where the step-function approximation of \S\ref{sub:Discontinuous_velocity_profile} is valid).

\subsubsection{Comparison with superluminal dispersion\label{sub:Comparison_with_superluminal_dispersion}}

\begin{figure}
\includegraphics[width=0.8\columnwidth]{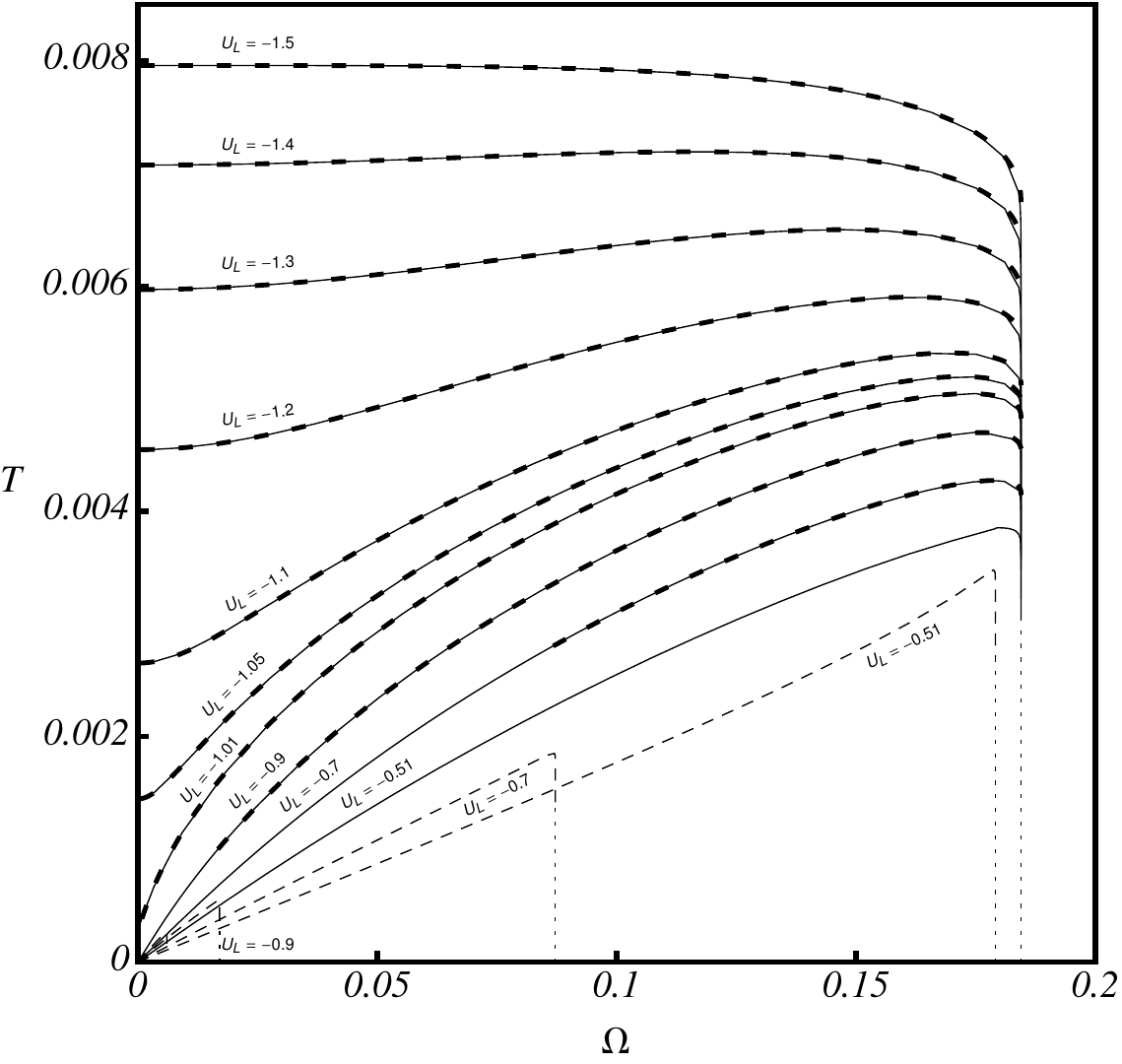}
\caption[\textsc{Temperature -- varying $U_{L}$ (low-$a$ regime)}]{\textsc{Temperature -- varying $U_{L}$ (low-$a$ regime)}: The (normalized) Hawking temperature, for the dispersion profile (\ref{eq:normalized_dispersion}) and the velocity profile (\ref{eq:normalized_velocity}), is shown as a function of frequency for various values of $U_{L}$, while $U_{R}$ is fixed at $-0.5$ and $a$ is fixed at $0.1$.  The solid curves correspond to the numerically calculated spectra of the $u1$-wave, while the thin dashed curves show the spectra for the $u2$-wave, with a maximum frequency (indicated by vertical dotted lines) that vanishes when $|U_{L}| \ge 1$.  The thick dashed curves show the analytic prediction of Eq. (\ref{eq:tanh_temperature}), derived from the phase-integral method described in \S\ref{sub:WKB-type-analysis}; this is applied only when a group-velocity horizon is present, and so only for the $u1$-wave spectra.  The agreement is seen to be very good for $a=0.1$ and for $U_{L}\le-0.7$.  (It is not so good for $U_{L}=-0.51$, where the group-velocity horizon exists only over a very narrow frequency range.)
\label{fig:temp_varying-ul_Hawking-regime}}
\end{figure}

\begin{figure}
\includegraphics[width=0.8\columnwidth]{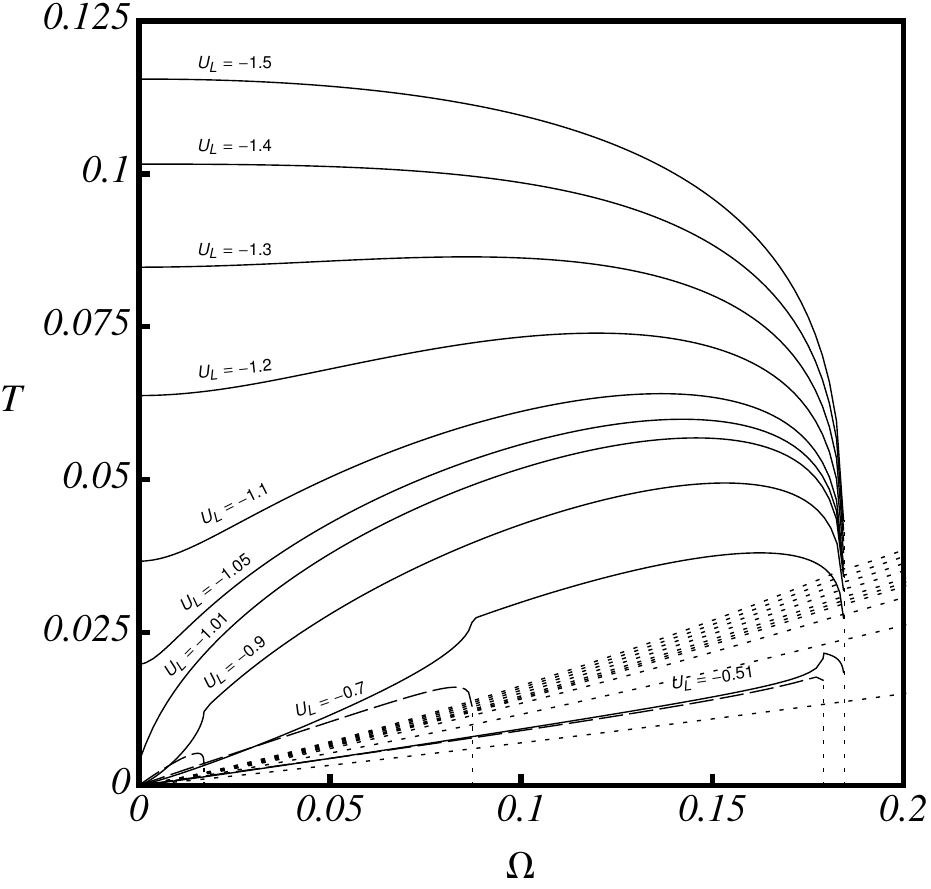}
\caption[\textsc{Temperature -- varying $U_{L}$ (discontinuous limit)}]{\textsc{Temperature -- varying $U_{L}$ (discontinuous limit)}: The (normalized) Hawking temperature, for the dispersion (\ref{eq:normalized_dispersion}), in a step-discontinuous flow is shown as a function of frequency for various values of $U_{L}$, with $U_{R}$ fixed at $-0.5$.  The solid curves correspond to the $u1$-wave, the dashed curves to the $u2$-wave (coupling into which only occurs in the purely subsonic regime where $|U_{L}|<1$), and we have also included dotted lines showing the temperature of the $v$-wave, which is easily calculated in the step-discontinuous limit.  (The highest $v$-curve corresponds to $U_{L}=-1.5$ and the lowest to $U_{L}=-0.51$, with monotonic variation in between.)  The most significant difference with respect to the low-steepness regime of Fig. \ref{fig:temp_varying-ul_Hawking-regime} is in the temperature values, which are here (for $u1$ and $u2$) an order of magnitude greater.  The shapes of the spectra are essentially the same as in Fig. \ref{fig:temp_varying-ul_Hawking-regime}, except that the fall-off on the approach to $\Omega_{\mathrm{max,2}}$ begins to take hold at lower frequencies, and the connection between the regimes in which a group-velocity horizon does and does not exist is a more clearly visible feature in the spectra of the $u1$-wave.  The $v$-wave spectra do not go to zero at $\Omega_{\mathrm{max}}$ because the $u$-$v$ coupling (between positive and negative norm) exists even above this value; this can be seen in Figs. \ref{fig:subluminal_supersonic} and \ref{fig:dispersion_sub_sup}.
\label{fig:temp_varying-ul_step-regime}}
\end{figure}

\begin{figure}
\includegraphics[width=0.8\columnwidth]{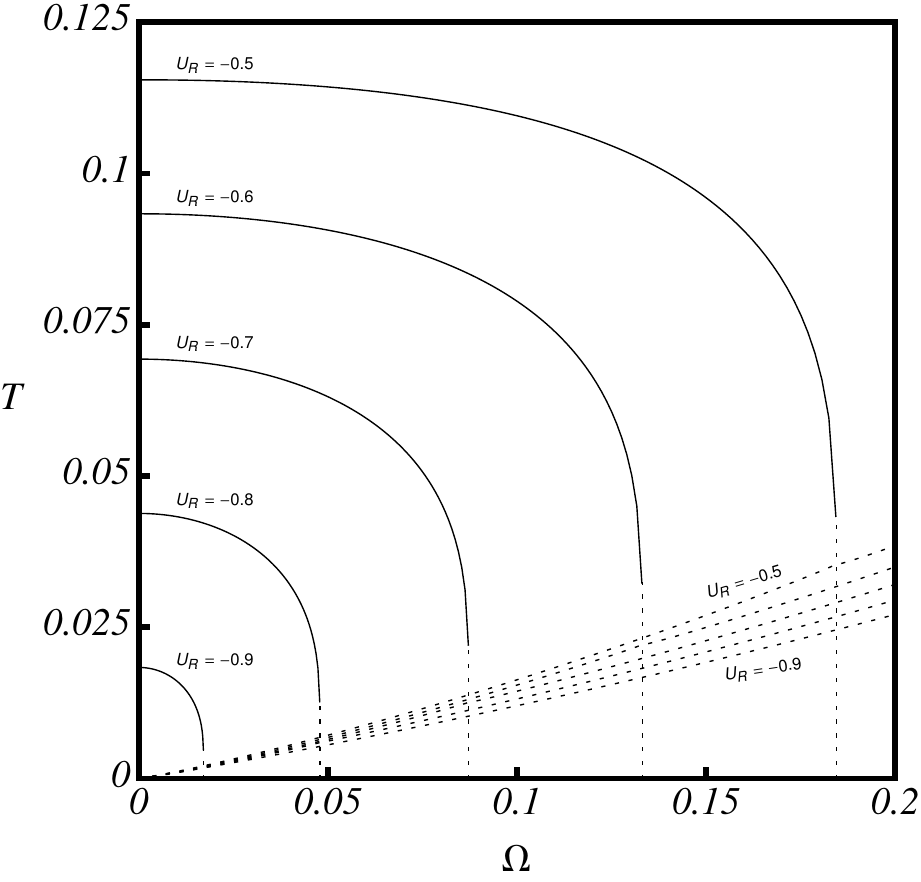}
\caption[\textsc{Temperature -- varying $U_{R}$ (discontinuous limit)}]{\textsc{Temperature -- varying $U_{R}$ (discontinuous limit)}: The (normalized) Hawking temperature, for the dispersion profile (\ref{eq:normalized_dispersion}) and a step-discontinuous flow, is shown as a function of frequency for various values of $U_{R}$ while $U_{L}$ is fixed at $-1.5$.  Solid lines show spectra for the $u1$-wave, and dotted lines for the $v$-wave; since $U_{L}<-1$, there is no ``horizonless'' regime and no emission of $u2$-waves.  Vertical dotted lines show the maximum frequencies at which the $u1$-$u$ coupling ceases and the spectrum vanishes; since the $u$-$v$ coupling can occur at all frequencies, the $v$-wave spectra do not experience such a cut-off.  As in the case of varying $U_{L}$ (see Figs. \ref{fig:temp_varying-ul_Hawking-regime} and \ref{fig:temp_varying-ul_step-regime}), we find in comparison with the low-steepness regime of Fig. \ref{fig:temp_varying-ur_Hawking-regime} that the temperatures are greater by an order of magnitude and the fall-off is more noticeable at lower frequencies.}
\label{fig:temp_varying-ur_step-regime}
\end{figure}

It is instructive to compare these observations with what is reported in the literature for superluminal dispersion\footnote{The reader is reminded that subluminal and superluminal dispersion for the wave equation (\ref{eq:acoustic_wave_equation}) are studied and compared in \cite{Macher-Parentani-2009}.  Also, \S IV E of \cite{Coutant-et-al-2012} details an approximate symmetry between subluminal and superluminal dispersion, helping to explain many of the observations made here.}.  The most relevant paper is \cite{Finazzi-Parentani-2011-JPCS}, since it also considers highly asymmetric profiles and includes the regime of all-supersonic flow with no low-frequency limiting horizon (although it considers fixed velocity difference $U_{R}-U_{L}$ and a change of the central velocity, so the actual temperature values are not directly comparable).  The resulting $u1$ spectra are qualitatively very similar to those of Fig. \ref{fig:temp_varying-ul_Hawking-regime}: when a low-frequency limiting horizon where $U=-1$ is present, the spectra approach finite temperatures in the low-frequency limit $\Omega\rightarrow 0$; and all spectra show a cut-off at some maximum frequency.  An important difference due to the reversed behaviour of the $u$-modes for sub- and superluminal dispersion with respect to sub- and supersonic flow (see \S\ref{sec:Field_decomposition_with_dispersion}) is in the reversed behaviour with respect to asymmetry towards the sub- and supersonic sides:
\begin{itemize}
\item When the velocity profile is asymmetric and towards the \textit{supersonic} side, the spectrum shows strong enhancement towards the high-frequency end.  This is analogous to what we have found for subluminal dispersion in Fig. \ref{fig:temp_varying-ul_Hawking-regime} for velocity profiles asymmetric towards the \textit{subsonic} side.
\item When the velocity profile is asymmetric and towards the \textit{subsonic} side, the spectrum narrows and is suppressed at higher frequencies, just as we have seen in Fig. \ref{fig:temp_varying-ur_Hawking-regime} for velocity profiles asymmetric towards the \textit{supersonic} side.
\end{itemize}

Reference \cite{Finazzi-Parentani-2011-JPCS} does not calculate temperatures for the case of no group-velocity horizon, but it does find non-vanishing spectra for the case of an entirely \textit{supersonic} flow in the mid-frequency regime where a group-velocity horizon exists, just as we did in Fig. \ref{fig:temp_varying-ul_Hawking-regime} for the case of an entirely \textit{subsonic} flow.

The $v$ spectra are not calculated in \cite{Finazzi-Parentani-2011-JPCS}.  However, we can predict that these will behave very differently in the cases of subluminal and superluminal dispersion:
\begin{itemize}
\item For subluminal dispersion, the $v$-wave has opposite norm to the $u$-wave; so, it can form a Hawking pair with the $u$-wave in the same way that the $u1$- and $u2$-waves can, and since there is no cut-off frequency for the $v$modes, this $v$ spectrum exists in principle for all frequencies.
\item For superluminal dispersion, the $v$-wave has opposite norm to the $u1$- and $u2$-waves; it thus forms Hawking pairs with these in the same way as the $u$-wave, and since the $u1$- and $u2$-waves have cut-off frequencies above which they no longer exist, so these $v$ spectra will vanish at these cut-off frequencies.
\end{itemize}

The transition between the low- and high-steepness regimes was recently studied in detail in \cite{Finazzi-Parentani-2012-arXiv} in the context of Bose-Einstein condensates.  A smooth transition between the two regimes is also observed there.  It is found that, to determine in which regime the system lies, one should compare the steepness of the transition (the ``surface gravity'' $\kappa$) with $\omega_{\mathrm{max}}$, the cut-off frequency of the spectrum.  This suggests that $\omega_{\mathrm{max}}/\kappa$ is simply related to $a/a_{0}$, the normalized steepness used to plot Fig. \ref{fig:temp_varying-a}.  Work on this particular aspect is still ongoing, however.

\section{Discussion\label{sec:Final_discussion}}

\subsection{Summary\label{sub:Final_summary}}

Generalizing the Lagrangian (\ref{eq:scalar_massless_Lagrangian}) derived from the spacetime metric to the Lagrangian (\ref{eq:Lagrangian_with_dispersion}) exhibiting dispersion, we find a wave equation (\ref{eq:acoustic_wave_equation}) and a dispersion relation (\ref{eq:acoustic_flow_dispersion}) analogous to the dispersionless model but with a $k$-dependent phase velocity $c(k)$, which we take to approach $c_{0}$ as $k\rightarrow 0$ and which might be superluminal or subluminal at higher wavevectors.  The scalar product (\ref{eq:scalar_product}) is also applicable with dispersion, and on writing the total field as an integral over stationary modes and quantizing, it is still true that positive- and negative-norm modes are multiplied by bosonic annihilation and creation operators, respectively.  However, the switch in the norm of the counter-propagating mode in the dispersionless model is replaced by a different kind of behaviour.  There is no divergence of $k$ anywhere, so there is a counter-propagating wave which exists for all values of $V$, never changing the sign of its norm.  However, for either subsonic or supersonic flow (depending on the type of dispersion), there comes into existence a pair of counter-propagating waves of opposite norm to the already existing one.  This allows opposite-norm pairing as before, though -- thanks to shifting and reflection at a group-velocity horizon -- without trans-Planckian frequencies in the past (though see \S\ref{sub:Overview} for a caveat).  More surprisingly, it also allows opposite-norm pairing in the complete absence of a group-velocity horizon, and since this can occur in two possible pairings, there are two types of Hawking spectra in such a situation.  The spectra can be calculated numerically by solving for the appropriate out- or in-modes via numerical integration; or, when the steepness is sufficiently large, by solving a linear system with matching conditions at a discontinuity; or, when there is a group-velocity horizon and the velocity gradient is not too great, by a phase integral in $k$-space.  Assuming a simple model of fluid flow and dispersion, we have found such Hawking spectra, confirming their theoretical existence and that they agree with the phase-integral formula in the appropriate regime.

\subsection{Overview of current research\label{sub:Overview}}

The field of analogue Hawking radiation has recently experienced its own inflationary period, the number of interested groups and proposed experimental setups having increased dramatically.  Here I offer a brief overview of the current efforts and challenges in this field.  For a review and complete historical overview of the general field of analogue gravity, see \cite{LivingReview}.

\paragraph{Analogue systems}

The earliest system to be considered for analogue Hawking radiation was superfluid helium \cite{Jacobson-1991,Jacobson-Volovik-1998}, being the main experimentally-accessible liquid in which quantum effects are important \footnote{Superfluid helium-3 provides an entire array of cosmological analogies \cite{Volovik-2001,HeliumDroplet}, in which analogues of the field theories familiar to us appear as low-frequency limits of more fundamental theories.}.  Once dilute gas Bose-Einstein condensates became experimentally accessible, their use as analogue systems was soon realised \cite{Garay-et-al-2000,Garay-et-al-2001}; their theoretical simplicity \cite{Leonhardt-et-al-2003-ii} and the relative ease with which they can be controlled in laboratory conditions have made them the most widely studied of all the analogue systems.  That said, due to flow instabilities induced by supersonic flow, it is not so easy to actually create a black hole configuration in superfluid systems, and it was only recently that this was finally achieved in BEC \cite{Lahav-et-al-2010}.

A change of reference frame can give rise to a black hole analogue in a \textit{stationary} medium, so long as there is an effective ``velocity profile'' which is \textit{moving} with respect to the medium.  This was first considered in the context of solitons in superfluid helium \cite{Jacobson-Volovik-1998}, and has also been considered for solitons in BECs \cite{Wuster-et-al-2007}.  It forms the basis of the analogue model in nonlinear optics: a strong pulse of light induces an effective change in the refractive index of the material, and this perturbed index travels through the medium with the pulse \cite{Philbin-et-al-2008,Belgiorno-et-al-2010,Schutzhold-2011,Finazzi-Carusotto-2012}.  In the co-moving frame of the pulse, the effective index is stationary while the material is moving -- exactly the type of situation studied in this tutorial, with the labels ``co-moving'' and ``lab'' switched.  This can give rise to a group-velocity horizon, and the resultant shift between the $u1$- and $u2$-waves has been measured in optical fibres \cite{Philbin-et-al-2008}.  Advantages of these systems are the high effective velocity -- which Doppler shifts the emitted photons into the UV range, making them easier to detect -- and the self-steepening effect, which can form optical shocks with very high steepness \cite{Philbin-et-al-2008,Robertson-thesis}.  Their main disadvantage is the smallness of the effective velocity change induced by the pulse \cite{Robertson-thesis,Schutzhold-2011}.  Given the complicated dispersion relations of optical media, the details of the Hawking process in these systems exhibit a rich phenomenology, as recently studied in \cite{Finazzi-Carusotto-2012}.

Although they are far from the quantum regime with no prospect of emitting detectable Hawking radiation, analogue black holes have been formed by trans-sonic flows of water \cite{Rousseaux-et-al-2008,Rousseaux-et-al-2010,Weinfurtner-et-al-2011}, and the scattering behaviour of incident waves has been observed.  In particular, the \textit{stimulated} Hawking effect -- the classical scattering between positive- and negative-norm waves -- can and has been observed in this system \cite{Rousseaux-et-al-2008,Weinfurtner-et-al-2011}.

We have taken only a glimpse at the wealth of proposed analogue systems, which also includes: ions moving on a ring \cite{Horstmann-et-al-2010,Horstmann-et-al-2011}; ``photon fluids'' in microcavities \cite{Marino-2008} and in an analogue Laval nozzle \cite{Fouxon-et-al-2010}; and ``fluids'' of exciton-polaritons \cite{Solnyshkov-et-al-2011,Gerace-Carusotto-2012}.

\paragraph{Detection of Hawking radiation}

Creating an analogue system is one thing; detecting Hawking radiation is quite another, and has so far proved elusive.  The predicted Hawking temperature of the setup realised in reference \cite{Lahav-et-al-2010} is on the orders of nanokelvin -- on the same order, but still somewhat less, than the temperature of the condensate itself.  The Hawking spectrum will then be practically indistinguishable from the thermal background.  This problem is reinforced when losses are taken into account \cite{Wuster-Savage-2007,Wuster-2008}, with quasiparticles induced by losses typically overwhelming the Hawking quasiparticles \cite{Wuster-2008}.

A promising proposal to overcome this thermal swamping is to exploit the entanglement between the Hawking partners, which leads to correlations between them.  These correlations have been studied extensively in BECs \cite{Balbinot-et-al-2008,Carusotto-et-al-2008,Balbinot-et-al-2010,Mayoral-et-al-2011}, where they show up as clear patterns in density-density correlations.  Remarkably, the correlation signatures remain strong even after the imposition of a finite condensate temperature higher than the Hawking temperature.

Another technique to increase the visibility of the Hawking spectrum is to make it highly non-thermal by inducing sharp resonance peaks.  This requires a several-horizon setup so that interference between them occurs.  If the asymptotic regions are one subsonic and one supersonic -- as studied in the BEC context in \cite{Zapata-et-al-2011} -- the system is (usually) dynamically stable, so that direct observation of spontaneous Hawking radiation is possible, its non-thermal character making it easier to distinguish from actual thermal radiation.  The black hole laser (previously mentioned in \S\ref{sub:White_holes}) -- of experimental relevance since many practical setups include a black-hole white-hole pair -- also yields strong resonant peaks at well-defined frequencies \cite{Corley-Jacobson-1999,Leonhardt-Philbin-2008,Coutant-Parentani-2010,Finazzi-Parentani-2010}; however, this radiation is strongly coherent and essentially classical in nature, forming from a runaway amplification process of \textit{stimulated} emission (as in a laser) that drowns out any spontaneous component of the radiation.

A signal claimed to represent Hawking radiation has been measured emitted from laser filaments in bulk nonlinear media, at a wavelength conforming to theoretical predictions \cite{Belgiorno-et-al-2010-ii}.  The interpretation of the signal, however, is controversial, with claims that the results are not consistent with Hawking radiation and that they are signs of another kind of creation process related to the non-stationarity of the background \cite{Schutzhold-Unruh-2011-comment,Belgiorno-et-al-2011-comment}.

\paragraph{Theoretical methods}

Similarly to the theoretical results presented in \S\ref{sec:Application_to_a_simple_model}, there are a number of works which investigate Hawking radiation by mode analysis in order to infer the spectrum for given dispersion and velocity profiles;  references \cite{Corley-Jacobson-1996,Corley-1997} provide the earliest examples.  Recent avenues of investigation include: comparison of black and white hole radiation for sub- and superluminal dispersion \cite{Macher-Parentani-2009}; spectrum and density-density correlations from a step-discontinuity in BEC \cite{Recati-et-al-2009}; the influence of asymmetric velocity profiles \cite{Finazzi-Parentani-2011-JPCS}; the dependence of the spectrum on multiple length scales \cite{Finazzi-Parentani-2011}; the transition between the low- and high-steepness regimes \cite{Finazzi-Parentani-2012-arXiv}; and mode analyses of black hole lasers \cite{Coutant-Parentani-2010,Finazzi-Parentani-2010}.

In a manner similar to that of \S\ref{sub:WKB-type-analysis}, there are several works which attempt purely analytic derivations of the Hawking spectrum in dispersive system.  The earliest examples are \cite{Brout-et-al-1995,Corley-1997}, which recover the Hawking result to leading order.  This was subsequently generalized: to general preferred reference frames \cite{Himemoto-Tanaka-2000}; to high-frequency corrections \cite{Saida-Sakagami-2000}; to more general dispersion relations \cite{Unruh-Schutzhold-2005}; to correlation functions \cite{Coutant-et-al-2012}; and, as presented in \S\ref{sub:WKB-type-analysis}, to more general velocity profiles \cite{Leonhardt-Robertson-2012}.  The latter approach was also considered in \cite{Robertson-thesis} for the ``horizonless'' case, where $\chi(k)$ has two central branch cuts in the complex $k$-plane rather than just one as in Fig. \ref{fig:WKB_contour_plot}; there are signs that this case does obey some kind of generalized phase-integral method, possibly involving linear combinations of various contours.

Analytical methods are almost always approximate to some degree, while mode analyses focus on stationary states (unless unstable modes are included, as for black hole lasers).  In order both to check our analytical results and to gain insight into the time-dependence of the system, \textit{ab initio} simulations are invaluable.  These have been used in the context of BECs, in the evaluation of the density-density correlation pattern \cite{Carusotto-et-al-2008,Mayoral-et-al-2011} and on the dynamics of the periodic configuration with a black hole-white hole pair \cite{Jain-et-al-2007}; the growth of an oscillating ``checkerboard'' pattern in the white hole case \cite{Mayoral-et-al-2011} is an interesting example of an unexpected feature revealed by simulations.

\paragraph{Conceptual issues}

It was early recognised \cite{Jacobson-1996,Corley-Jacobson-1996} that, in the case of a Schwarzschild black hole (or indeed any velocity profile with vanishing asymptotic velocity), dispersion does not quite solve the trans-Planckian problem.  Though it regularizes the wavevector divergence at the horizon by the shifting of high-wavevector ingoing waves to low-wavevector outgoing waves, tracing the ingoing waves back in time to spatial infinity sends them to arbitrarily high wavevectors \cite{Jacobson-1996,Corley-Jacobson-1996}.  So the position of the divergence has been shifted to infinity, but the divergence still exists in principle.  A very closely related point is that of conservation of the lab frequency when $V\rightarrow0$ at spatial infinity, since then the lab and co-moving frames coincide in the asymptotic region.  Thus, conservation of the lab frequency is equivalent there to conservation of co-moving frequency; and, since the sign of the co-moving frequency gives the sign of the norm (see Eq. (\ref{eq:free-fall-freq_pos-and-neg})), it follows there can be no conversion between positive- and negative-norm waves and hence no particle creation \cite{Jacobson-1996,Corley-Jacobson-1996}.  An interesting approach to this problem is to consider a discretized space whose points are freely-falling, so that the spacetime is not strictly stationary and the lab frequency not necessarily conserved \cite{Corley-Jacobson-1998,Jacobson-Mattingly-1999}.

\subsection{Final comments}

Despite some tantalizing experimental results -- the realization of an analogue black hole in BEC \cite{Lahav-et-al-2010}, the detection of a signal from laser pulse filaments in nonlinear optical media \cite{Belgiorno-et-al-2010-ii} and the observation of the classical stimulated Hawking effect for surface water waves \cite{Rousseaux-et-al-2008,Weinfurtner-et-al-2011} -- Hawking radiation remains stubbornly in the realms of theory.  But the \textit{idea} of its experimental realization is flourishing.  Its concepts are constantly being applied to newer analogue systems.  The goal is no longer a deeper understanding of gravity -- and perhaps, given the surprising emergence of the ``horizonless'' regime, not so much about general black or white holes either.  Instead, we are aiming to paint a picture of the quantum vacuum -- of the content of nothing, of physics at its most fundamental \footnote{From this perspective, we can take encouragement from recent experiments on the related phenomenon of the \textit{dynamical Casimir effect} (DCE), which predicts particle creation from the vacuum due to time-varying boundary conditions rather than a spatially-varying metric.  Analogue systems in superconducting circuits of variable electrical length have been shown, both in theory \cite{Johansson-et-al-2009,Johansson-et-al-2010} and experiment \cite{Wilson-et-al-2010,Wilson-et-al-2011}, to produce analogue DCE radiation.}.  For example: the heuristic picture of particle-antiparticle pairs arising from vacuum fluctuations and being torn apart by the event horizon might seem to give some insight into the gravitational black hole.  But how is this picture to be reconciled with analogue white hole radiation, or the ``horizonless'' regime emerging from dispersion, where Hawking partners can be emitted in the same direction?  Has our insight or our theory failed -- and if so, what should replace it?   For now, the vacuum and its contents might remain out of sight -- but I hope I, and the many others in this field, can offer a tantalizing glimpse beyond the horizon!

\newpage

\section*{Acknowledgements}

I am grateful to Dario Gerace, Steve Hill, Simon Horsley, Susanne Kehr, Friedrich K\"{o}nig, Chris Kuklewicz, Ulf Leonhardt, Joanna McLenaghan, Renaud Parentani, Thomas Philbin and Sahar Sahebdivan for valuable discussions.  In particular, I wish to thank Renaud Parentani for guidance on the WKB method and especially for pointing out that the negative-norm component gives directly $\beta_{\omega}/\alpha_{\omega}$.  Special thanks are due to Ulf Leonhardt, for putting me on this path and shedding light on the way ahead.  I am also grateful to Lucio Andreani for his support of later work and most of the writing process.  This tutorial is based on my PhD studies at the University of St Andrews, which was supported by EPSRC.  Work at the University of Pavia was supported by Fondazione Cariplo under project no. 2010-0523.

\newpage

\clearpage

\begin{thebibliography}{99}

\bibitem{Hawking-1974}
Hawking S W (1974) {\it Nature} {\bf 248} 30

\bibitem{Hawking-1975}
Hawking S W (1975) {\it Commun. Math. Phys.} {\bf 43} 199

\bibitem{Bekenstein-1973}
Bekenstein J D (1973) {\it Phys. Rev. D} {\bf 7} 2333

\bibitem{Bekenstein-1974}
Bekenstein J D (1974) {\it Phys. Rev. D} {\bf 9} 3292

\bibitem{Hartle-Hawking-1976}
Hartle J B and Hawking S W (1976) {\it Phys. Rev. D} {\bf 13} 2188

\bibitem{Unruh-1976}
Unruh W G (1976) {\it Phys. Rev. D} {\bf 14} 870

\bibitem{Damour-Ruffini-1976}
Damour T and Ruffini R (1976) {\it Phys. Rev. D} {\bf 14} 332

\bibitem{Sanchez-1978}
Sanchez N (1978) {\it Phys. Rev. D} {\bf 18} 1030

\bibitem{Fredenhagen-Haag-1990}
Fredenhagen K and Haag R (1990) {\it Commun. Math. Phys.} {\bf 127} 273

\bibitem{Parikh-Wilczek-2000}
Parikh M K and Wilczek F (2000) {\it Phys. Rev. Lett.} {\bf 85} 5042

\bibitem{Carr-Hawking-1974}
Carr B J and Hawking S W (1974) {\it Mon. Not. R. Astr. Soc.} {\bf 168} 399

\bibitem{Dimopoulos-Landsberg-2001}
Dimopoulos S and Landsberg G (2001) {\it Phys. Rev. Lett.} {\bf 87} 161602

\bibitem{Giddings-Thomas-2002}
Giddings S B and Thomas S (2002) {\it Phys. Rev. D} {\bf 65} 056010

\bibitem{Jacobson-1991}
Jacobson T (1991) {\it Phys. Rev. D} {\bf 44} 1731

\bibitem{Brout-et-al-Primer}
Brout R, Massar S, Parentani R and Spindel Ph (1995) {\it Phys. Rep.} {\bf 260} 329

\bibitem{ArtificialBlackHoles}
Novello M, Visser M and Volovik G (editors) (2002) {\it Artificial Black Holes} (Singapore: World Scientific)

\bibitem{Schutzhold-Unruh}
Sch\"{u}tzhold R and Unruh W G (editors) (2007) {\it Quantum Analogues: From Phase Transitions to Black Holes and Cosmology} (Berlin: Springer)

\bibitem{LivingReview}
Barcel\'{o} C, Liberati S and Visser M (2011) {\it Living Rev. Relativity} {\bf 14} 3

\bibitem{Unruh-1981}
Unruh W G (1981) {\it Phys. Rev. Lett} {\bf 46} 1351

\bibitem{Visser-1993}
Visser M (1993) arXiv:gr-qc/9311028

\bibitem{Unruh-1995}
Unruh W G (1995) {\it Phys. Rev. D} {\bf 51} 2827

\bibitem{Brout-et-al-1995}
Brout R, Massar S, Parentani R and Spindel Ph (1995) {\it Phys. Rev. D} {\bf 52} 4559

\bibitem{Garay-et-al-2000}
Garay L J, Anglin J R, Cirac J I and Zoller P (2000) {\it Phys. Rev. Lett.} {\bf 85} 4643

\bibitem{Garay-et-al-2001}
Garay L J, Anglin J R, Cirac J I and Zoller P (2001) {\it Phys. Rev. A} {\bf 63} 023611

\bibitem{Barcelo-Liberati-Visser-2001-arXiv}
Barcel\'{o} C, Liberati S and Visser M (2001) arXiv:gr-qc/0110036

\bibitem{Barcelo-Liberati-Visser-2001}
Barcel\'{o} C, Liberati S and Visser M (2001) {\it Class. Quantum Grav.} {\bf 18} 1137

\bibitem{Giovanazzi-et-al-2004}
Giovanazzi S, Farrell C, Kiss T and Leonhardt U (2004) {\it Phys. Rev. A} {\bf 70} 063602

\bibitem{Giovanazzi-2005}
Giovanazzi S (2005) {\it Phys. Rev. Lett.} {\bf 94} 061302

\bibitem{Jacobson-Volovik-1998}
Jacobson T A and Volovik G E (1998) {\it Phys. Rev. D} {\bf 58} 064021

\bibitem{Volovik-1999}
Volovik G E (1999) {\it JETP Lett.} {\bf 69} 705

\bibitem{Fischer-Volovik-2001}
Fischer U R and Volovik G E (2001) {\it Int. J. Mod. Phys.} {\bf 10} 57

\bibitem{Schutzhold-Unruh-2002}
Sch\"{u}tzhold R and Unruh W G (2002) {\it Phys. Rev. D} {\bf 66} 044019

\bibitem{Reznik-1997}
Reznik B (1997) arXiv:gr-qc/9703076

\bibitem{Schutzhold-Unruh-2005}
Sch\"{u}tzhold R and Unruh W G (2005) {\it Phys. Rev. Lett.} {\bf 95} 031301

\bibitem{Leonhardt-Piwnicki-1999}
Leonhardt U and Piwnicki P (1999) {\it Phys. Rev. A} {\bf 60} 4301

\bibitem{Leonhardt-Piwnicki-2000}
Leonhardt U and Piwnicki P (2000) {\it Phys. Rev. Lett.} {\bf 84} 822

\bibitem{Leonhardt-2002}
Leonhardt U (2002) {\it Nature} {\bf 415} 406

\bibitem{Unruh-1977}
Unruh W G (1977) {\it Phys. Rev. D} {\bf 15} 365

\bibitem{Jacobson-1996}
Jacobson T (1996) {\it Phys. Rev. D} {\bf 53} 7082

\bibitem{Schutzhold-Unruh-2008}
Sch\"{u}tzhold R and Unruh W G (2008) {\it Phys. Rev. D} {\bf 78} 041504

\bibitem{Misner-Thorne-Wheeler}
Misner C W, Thorne K S and Wheeler J A (1973) {\it Gravitation} (San Francisco: W H Freeman)

\bibitem{Painleve-1921}
Painlev\'{e} P (1921) {\it C. R. Acad. Sci. (Paris)} {\bf 173} 677

\bibitem{Gullstrand-1922}
Gullstrand A (1922) {\it Arkiv. Mat. Astron. Fys.} {\bf 16} 1

\bibitem{Lemaitre-1933}
Lema\^{i}tre G (1933) {\it Ann. Soc. Sci. Brux.} {\bf A53} 51

\bibitem{Jacobson-1999}
Jacobson T (1999) {\it Prof. Theor. Phys. Supp.} {\bf 136} 1

\bibitem{Hamilton-Lisle-2008}
Hamilton A J S and Lisle J P (2008) {\it Am. J. Phys.} {\bf 76} 519

\bibitem{Rousseaux-et-al-2008}
Rousseaux G, Mathis C, Ma\"{i}ssa P, Philbin T G and Leonhardt U (2008) {\it New J. Phys.} {\bf 10} 053015

\bibitem{Rousseaux-et-al-2010}
Rousseaux G, Ma\"{i}ssa P, Mathis C, Philbin T G and Leonhardt U (2010) {\it New J. Phys.} {\bf 12} 095018

\bibitem{Weinfurtner-et-al-2011}
Weinfurtner S, Tedford E W, Penrice M C J, Unruh W G and Lawrence G A (2011) {\it Phys. Rev. Lett.} {\bf 106} 021302

\bibitem{Macher-Parentani-2009-ii}
Macher J and Parentani R (2009) {\it Phys. Rev. A} {\bf 80} 043601

\bibitem{Leonhardt-et-al-2003}
Leonhardt U, Kiss T and \"{O}hberg P (2003) {\it Phys. Rev. A} {\bf 67} 033602

\bibitem{Barcelo-et-al-2006}
Barcel\'{o} C, Cano A, Garay L J and Jannes G (2006) {\it Phys. Rev. D} {\bf 74} 024008

\bibitem{Mayoral-et-al-2011}
Mayoral C, Recati A, Fabbri A, Parentani R, Balbinot R and Carusotto I (2011) {\it New J. Phys.} {\bf 13} 025007

\bibitem{Jain-et-al-2007}
Jain P, Bradley A S and Gardiner C W (2007) {\it Phys. Rev. A} {\bf 76} 023617

\bibitem{Corley-Jacobson-1999}
Corley S and Jacobson T (1999) {\it Phys. Rev. D} {\bf 59} 124001

\bibitem{Leonhardt-Philbin-2008}
Leonhardt U and Philbin T G (2008) arXiv:0803.0669

\bibitem{Coutant-Parentani-2010}
Coutant A and Parentani R (2010) {\it Phys. Rev. D} {\bf 81} 084042

\bibitem{Finazzi-Parentani-2010}
Finazzi S and Parentani R (2010) {\it New J. Phys.} {\bf 12} 095015

\bibitem{Macher-Parentani-2009}
Macher J and Parentani R (2009) {\it Phys. Rev. D} {\bf 79} 124008

\bibitem{Corley-Jacobson-1996}
Corley S and Jacobson T (1996) {\it Phys. Rev. D} {\bf 54} 1568

\bibitem{Leonhardt-Philbin-Invisibility}
Leonhardt U and Philbin T (2010) {\it Geometry and Light: The Science of Invisibility} (Dover Publications Inc.)

\bibitem{Birrell-Davies}
Birrell N D and Davies P C W (1982) {\it Quantum Fields in Curved Space} (Cambridge: Cambridge University Press)

\bibitem{Noether-1918}
Noether E (1918) {\it Nachr. d. K\"{o}nig. Gesellsch. d. Wiss. zu G\"{o}ttingen, Math-phys. Klasse} 235;
\,English translation: Tavel M A (1971) arXiv:physics/0503066

\bibitem{Castin-2000}
Castin Y (2000) arXiv:cond-mat/0105058

\bibitem{Leonhardt-QO}
Leonhardt U (2010) {\it Essential Quantum Optics} (Cambridge: Cambridge University Press)

\bibitem{Unruh-2011}
Unruh W G (2011) arXiv:1107.2669

\bibitem{Caves-1982}
Caves C M (1982) {\it Phys. Rev. D} {\bf 26} 1817

\bibitem{Fulling-1973}
Fulling S A (1973) {\it Phys. Rev. D} {\bf 7} 2850

\bibitem{Barnett-Phoenix-1989}
Barnett S M and Phoenix S J D (1989) {\it Phys. Rev. A} {\bf 40} 2404

\bibitem{Balbinot-et-al-2008}
Balbinot R, Fabbri A, Fagnocchi S, Recati A and Carusotto I (2008) {\it Phys. Rev. A} {\bf 78} 021603

\bibitem{Carusotto-et-al-2008}
Carusotto I, Fagnocchi S, Recati A, Balbinot R and Fabbri A (2008) {\it New J. Phys.} {\bf 10} 103001

\bibitem{Balbinot-et-al-2010}
Balbinot R, Carusotto I, Fabbri A and Recati A (2010) {\it Int. J. Mod. Phys. D} {\bf 19} 2371

\bibitem{Giovanazzi-2011}
Giovanazzi S (2011) {\it Phys. Rev. Lett.} {\bf 106} 011302

\bibitem{Rinaldi-2011}
Rinaldi M (2011) {\it Phys. Rev. D} {\bf 84} 124009

\bibitem{Corley-1997}
Corley S (1997) {\it Phys. Rev. D} {\bf 55} 6155

\bibitem{Recati-et-al-2009}
Recati A, Pavloff N and Carusotto I (2009) {\it Phys. Rev. A} {\bf 80} 043603

\bibitem{Finazzi-Parentani-2011}
Finazzi S and Parentani R (2011) {\it Phys. Rev. D} {\bf 83} 084010

\bibitem{Corley-1998}
Corley S (1998) {\it Phys. Rev. D} {\bf 57} 6280

\bibitem{Himemoto-Tanaka-2000}
Himemoto Y and Tanaka T (2000) {\it Phys. Rev. D} {\bf 61} 064004

\bibitem{Saida-Sakagami-2000}
Saida H and Sakagami M (2000) {\it Phys. Rev. D} {\bf 61} 084023

\bibitem{Leonhardt-Robertson-2012}
Leonhardt U and Robertson S (2012) {\it New J. Phys.} {\bf 14} 053003

\bibitem{Unruh-Schutzhold-2005}
Unruh W G and Sch\"{u}tzhold R (2005) {\it Phys. Rev. D} {\bf 71} 024028

\bibitem{Coutant-et-al-2012}
Coutant A, Parentani R and Finazzi S (2012) {\it Phys. Rev. D} {\bf 85} 024021

\bibitem{Leonhardt-et-al-2003-ii}
Leonhardt U, Kiss T and \"{O}hberg P (2003) {\it J. Opt. B} {\bf 5} S42

\bibitem{Finazzi-Parentani-2011-JPCS}
Finazzi S and Parentani R (2011) {\it J. Phys. Conf. Ser.} {\bf 314} 012030

\bibitem{Finazzi-Parentani-2012-arXiv}
Finazzi S and Parentani R (2012) {\it Phys. Rev. D} {\bf 85} 124027

\bibitem{Robertson-thesis}
Robertson S J (2011) arXiv:1106.1805 (PhD thesis)

\bibitem{Philbin-et-al-2008}
Philbin T G, Kuklewicz C, Robertson S, Hill S, K\"{o}nig F and Leonhardt U (2008) {\it Science} {\bf 319} 1367

\bibitem{Volovik-2001}
Volovik G E (2001) {\it Phys. Rep.} {\bf 351} 195

\bibitem{HeliumDroplet}
Volovik G E (2003) {\it The Universe in a Helium Droplet} (Oxford: Clarendon)

\bibitem{Lahav-et-al-2010}
Lahav O, Itah A, Blumkin A, Gordon C, Rinott S, Zayats A and Steinhauer J (2010) {\it Phys. Rev. Lett.} {\bf 105} 240401

\bibitem{Wuster-et-al-2007}
W\"{u}ster S and D\c{a}browska-W\"{u}ster B J (2007) {\it New J. Phys.} {\bf 9} 85

\bibitem{Belgiorno-et-al-2010}
Belgiorno F, Cacciatori S L, Ortenzi G, Sala V G and Faccio D (2010) {\it Phys. Rev. Lett.} {\bf 104} 140403

\bibitem{Schutzhold-2011}
Sch\"{u}tzhold R (2011) arXiv:1110.6064

\bibitem{Finazzi-Carusotto-2012}
Finazzi S and Carusotto I (2012) arXiv:1204.3603 (preprint)

\bibitem{Horstmann-et-al-2010}
Horstmann B, Reznik B, Fagnocchi S and Cirac J I (2010) {\it Phys. Rev. Lett.} {\bf 104} 250403

\bibitem{Horstmann-et-al-2011}
Horstmann B, Sch\"{u}tzhold R, Reznik B, Fagnocchi S and Cirac J I (2011) {\it New J. Phys.} {\bf 13} 045008

\bibitem{Marino-2008}
Marino F (2008) {\it Phys. Rev. A} {\bf 78} 063804

\bibitem{Fouxon-et-al-2010}
Fouxon I, Farberovich O V, Bar-Ad S and Fleurov V (2010) {\it Europhys. Lett.} {\bf 92} 14002

\bibitem{Solnyshkov-et-al-2011}
Solnyshkov D D, Flayac H and Malpuech G (2011) {\it Phys. Rev. B} {\bf 84} 233405

\bibitem{Gerace-Carusotto-2012}
Gerace D and Carusotto I (2012) arXiv:1206.4276 (preprint)

\bibitem{Wuster-Savage-2007}
W\"{u}ster S and Savage C M (2007) {\it Phys. Rev. A} {\bf 76} 013608

\bibitem{Wuster-2008}
W\"{u}ster S (2008) {\it Phys. Rev. A} {\bf 78} 021601

\bibitem{Zapata-et-al-2011}
Zapata I, Albert M, Parentani R and Sols F (2011) {\it New J. Phys.} {\bf 13} 063048

\bibitem{Belgiorno-et-al-2010-ii}
Belgiorno F, Cacciatori S L, Clerici M, Gorini V, Ortenzi G, Rizzi L, Rubino E, Sala V G and Faccio D (2010) {\it Phys. Rev. Lett.} {\bf 105} 203901

\bibitem{Schutzhold-Unruh-2011-comment}
Sch\"{u}tzhold R and Unruh W G (2011) arXiv:1012.2686

\bibitem{Belgiorno-et-al-2011-comment}
Belgiorno F, Cacciatori S L, Clerici M, Gorini V, Ortenzi G, Rizzi L, Rubino E, Sala V G and Faccio D (2011) arXiv:1012.5062

\bibitem{Corley-Jacobson-1998}
Corley S and Jacobson T (1998) {\it Phys. Rev. D} {\bf 57} 6269

\bibitem{Jacobson-Mattingly-1999}
Jacobson T and Mattingly D (1999) {\it Phys. Rev. D} {\bf 61} 024017

\bibitem{Johansson-et-al-2009}
Johansson J R, Johansson G, Wilson C M and Nori F (2009) {\it Phys. Rev. Lett.} {\bf 103} 147003

\bibitem{Johansson-et-al-2010}
Johansson J R, Johansson G, Wilson C M and Nori F (2010) {\it Phys. Rev. A} {\bf 82} 052509

\bibitem{Wilson-et-al-2010}
Wilson C M, Duty T, Sandberg M, Persson F, Shumeiko V and Delsing P (2010) {\it Phys. Rev. Lett.} {\bf 105} 233907

\bibitem{Wilson-et-al-2011}
Wilson C M, Johansson G, Pourkabirian A, Simoen M, Johansson J R, Duty T, Nori F and Delsing P (2011) {\it Nature} {\bf 479} 376

\end{thebibliography}
\end{document}